\documentclass[english,a4paper,notitlepage,aps,prl,groupedaddress,reprint,pdftex]{revtex4-1}
\usepackage[pdftex]{graphicx}
\usepackage{tabularx}
\usepackage{multirow}
\usepackage{longtable}
\usepackage{dcolumn}
\usepackage{bm}
\usepackage{amsmath}
\usepackage{amssymb}
\usepackage{txfonts}
\usepackage{url}
\usepackage{epstopdf}
\usepackage[usenames,dvipsnames]{color}
\definecolor{Gray}{gray}{0.0}
\definecolor{lightGray}{gray}{0.35}

\begin{document}
\title{
High-$T_c$ ternary metal hydrides, YKH$_{12}$ and LaKH$_{12}$, discovered by machine learning
}
\author{
  Peng Song$^{1}$,
  Zhufeng Hou$^{2}$,
  Pedro Baptista de Castro$^{3,4}$,
  Kousuke Nakano$^{1,5}$,
  Kenta Hongo$^{6}$,
  Yoshihiko Takano$^{3,4}$,
  Ryo Maezono$^{1}$ \\}

\affiliation{\\
  $^1$School of Information Science, JAIST,
  Asahidai 1-1, Nomi, Ishikawa 923-1292, Japan\\
  \\
  $^2$State Key Laboratory of Structural Chemistry,
  Fujian Institute of Research on the Structure of Matter,
  Chinese Academy of Sciences, Fuzhou 350002, China \\
  \\
  $^3$National Institute for Materials Science, 1-2-1 Sengen, Tsukuba, Ibaraki 305-0047, Japan\\
  \\
  $^4$University of Tsukuba, 1-1-1 Tennodai, Tsukuba, Ibaraki 305-8577, Japan\\
  \\
  $^5$International School for Advanced Studies (SISSA),
      Via Bonomea 265, 34136, Trieste, Italy\\
  \\
  $^6$Research Center for Advanced Computing
      Infrastructure, JAIST, Asahidai 1-1, Nomi,
      Ishikawa 923-1292, Japan\\
  \\
}

\vspace{10mm}

\date{\today}
\begin{abstract}
The search for hydride compounds that exhibit
high $T_c$ superconductivity has
been extensively studied.
Within the range of binary hydride compounds,
the studies have been developed well including
data-driven searches as a topic of interest.
Toward the search for the ternary systems,
the number of possible combinations grows rapidly,
and hence the power of data-driven search gets
more prominent.
In this study, we constructed various regression models
  to predict $T_c$ for ternary hydride compounds
  and found the extreme gradient boosting (XGBoost) regression
  giving the best performance.
The best performed regression predicts new promising candidates
realizing higher $T_c$, for which
we further identified their possible
crystal structures.
Confirming their lattice and thermodynamical
stabilities, we finally predicted
new ternary hydride superconductors,
YKH$_{12}$ [$C2/m$~(No.12), $T_c$=143.2~K at 240~GPa]
and LaKH$_{12}$ [$R\bar{3}m$~(No.166), $T_c$=99.2~K at 140~GPa]
from first principles.
\end{abstract}
\maketitle

\section{Introduction}
\label{sec.intro}
The compressed polyhydrides are good candidates for
high $T_c$ superconductor due to the high vibration
frequencies provided by the hydrogen atoms,
coupled with the introduction of other elements
to get necessary pre-compression for the entire system
to maintain its metallic and superconducting state
even at lower pressure.
The potential for high $T_c$ has been confirmed
by many theoretical and experimental
studies.~\cite{2004ASH,1968ASH,2008ERE,2015DRO,2015SZC}
The structure searching to get higher $T_c$
for these compounds has been made
mainly within binary compounds,
and some of the synthesis have
reported the achievement of high $T_c$,
{\it e.g.}, LaH$_{10}$ (260~K at 200~GPa),~\cite{2019SOM}
YH$_6$ (224~K at 166~GPa),~\cite{2020TRO}
and
TH$_{10}$ (159~K at 174~GPa)~\cite{2020SEM}
{\it etc}.
Recent theoretical prediction of
Li$_2$MgH$_{16}$ (473~K at 250~GPa)
and the experimental measurement of
carbonaceous sulfur hydride (287~K at $\sim$267~GPa)
indicate
that multi component hydrides could have greater potential
for higher Tc than binary ones.~\cite{2019SUN,2020SNI}
At present, about 10-20 ternary superconducting hydrides
have been proposed, but a small part of them have been
experimentally verified ending up with extremely low
$T_c$.~\cite{2019SUN,2020SNI,2017MA_b,2017MA,2018LI,2019LIA_b,
2017RAH,2019LIA,2020WEI,2019XIE,2017KOK,2019SHA,2019SHA,
2020ZHA_a,2020ZHA_b,2020DIC,2020GUO,2020CUI,2020LV,2020YAN,2015MUR,2019MEN}
According to 'Materials Project (MP) database',~\cite{2013JAI}
the number of ternary compounds amounts
around to five times larger than that of binary compounds
under ambient conditions,
providing us with an exciting field of materials searching.
%
For such problems with a wide search space,
data-driven approaches get to be powerful over
other methods.
For cuprate and iron-based superconductors,
their searchings by using machine learning approaches
have been reported.~\cite{2018STA,2019MAT,2018MER}
For the hydrides superconductor, RbH$_{12}$,
neural networks have been applied.~\cite{2020HUT}
We shall then employ machine learning
techniques to explore ternary polyhydrides
for higher $T_c$.

\vspace{2mm}
The compounds to be targeted were
first narrowed down according to the following policy:
The target ternary system is restricted
within those composed as a combination of
binary hydrides having higher $T_c$ as reported.
From binary hydrides that have been reported
to be
superconducting,~\cite{2008CHE,
2009KIM,2009TSE,2010JIN,2010LI,2010GAO,2010DUA,2011GAO,
2011KIM,2011ABE,2012ZHO_a,2011ZHA,2012ZHO_b,2013GAO,2013ABE, 2013HOO,2013LON,
2013HU,2014XIE,2014YU,2014LI,
2014WAN,2014CHE,2014DUA,
2015YAN,2015HOU,2015DUA,2015ZHA_a,
2015LIU_a,2015SHA,2015LIU_b,2015CHE_a,2015FEN,
2015ZHA_b,2015LIU,2015YU,2015CHE_b,2015ZHA,
2015ERR,2016ISH,2016MAH,2016FU,2016SHA,2016LIU,
2016LI_a,2016ZHO,2016LI_b,2017LI_a,2017ZHU,2017LIU,2017ZEN,2017ISH,
2017LI_b,2017PEN,2017DAV,2017MAJ,2018ZAR,2018SEM,2018YE,2018KVA_a,2018DUR,2018ZHE,
2018ABE,2018ZHU,2018KVA_b,2018WU,2019HEI,
2019LI_b,2019YUA,2019WAN,
2019YAN,2019XI,2019ABE,2020XIE,2020ZHA_a,2020ZHA_b,
2020CUI,2015MA,2020TRO,2019LI_c,
2018KAN}
there can be about 2,800 possible combinations.
Of these, except for those with very low $T_c$,
we can narrow it down to about 1,800 types,
and further, except for those with too high
hydrogen content, to about 1,700 types.
Among these, we limited our search to the Y and La systems
which have tendency to achieve higher $T_c$,
getting about 250 combinations of Y$M$H$_x$ and La$M$H$_x$ ($M =$ Ca, K, and Na).
For these target compounds, a procedure for our virtual
screening via machine-learning and high-throughput {\it ab initio} calculations is as follows:
(1) machine-learning search
for the chemical compositions achieving higher $T_c$,
(2) evolutionary crystal structure search
for the candidate compositions,
(3) stability check for the candiate structures,
(4) {\it ab initio} predictions of $T_c$ for the stable structures.

\begin{figure*}[htbp]
  \begin{center}
    \includegraphics[width=0.85\linewidth]{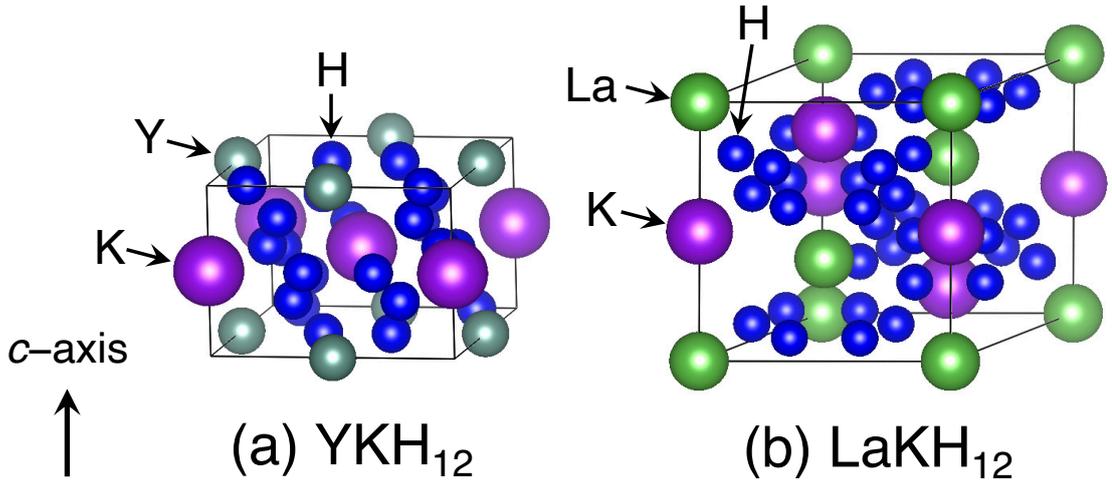}
    \caption{
    Predicted structures for
    (a) YKH$_{12}$ with $C2/m$~(No.12), and
    (b) LaKH$_{12}$ with $R\bar{3}m$~(No.166),
    by the crystal structures search using USPEX.
    }
    \label{fig.structure}
  \end{center}
\end{figure*}

\vspace{2mm}
  (1) For the composition search,
  as described in the ``Method'' section,
  we considered various regression models
  to predict $T_c$ values of the ternary hydrides,
  which were learned with
  theoretically predicted data on $T_c$
  extracted from available literature.~\cite{2008CHE,
2009KIM,2009TSE,2010JIN,2010LI,2010GAO,2010DUA,2011GAO,
2011KIM,2011ABE,2012ZHO_a,2011ZHA,2012ZHO_b,2013GAO,2013ABE, 2013HOO,2013LON,
2013HU,2014XIE,2014YU,2014LI,
2014WAN,2014CHE,2014DUA,
2015YAN,2015HOU,2015DUA,2015ZHA_a,
2015LIU_a,2015SHA,2015LIU_b,2015CHE_a,2015FEN,
2015ZHA_b,2015LIU,2015YU,2015CHE_b,2015ZHA,
2015ERR,2016ISH,2016MAH,2016ZHA,2016FU,2016SHA,2016LIU,
2016LI_a,2016ZHO,2016LI_b,2017RAH,2017MA,
2017MA_b,2017KOK,2017LI_a,2017ZHU,2017LIU,2017ZEN,2017ISH,
2017LI_b,2017PEN,2017DAV,2017MAJ,2018ZAR,2018SEM,
2018LI,2018YE,2018KVA_a,2018DUR,2018ZHE,2018LIU,
2018ABE,2018NAK,2018ZHU,2018KVA_b,2018WU,2019LIA_a,2019SUN,2019HEI,2019GRI,
2019CHE,2019LI_b,2019SHA,2019XIE,2019LIA_b,2019YUA,2019WAN,2019AMS,
2019YAN,2019XI,2019ABE,2019DU,2020GUO,2020LV,2020XIE,2020ZHA_a,2020ZHA_b,2020LI,
2020HAO_a,2020CUI,2015MA,2020TRO,2020DIC,2019LI_c,
2018KAN,2020HAO_b}
  Our descriptors entering the regression as input
  consist of 84 features such as
  chemical compositions, space group, and pressure-dependent
  electronic properties, where
  the composition descriptors were generated
  using the XenonPy software~\cite{2019YAM}
  and the pressure-dependent descriptors
  were computed by the VASP software~\cite{1993KRE,1994KRE,1996KRE_a,1996KRE_b}.
  We checked their prediction performance
  by using the cross validation
  and then the best performed model,
  {\it i.e.}, the extreme gradient boosting algorithm implemented in the XGBoost package,~\cite{2015CHE} was used
  for the high-throughput screening of
  the target ternary compounds.
  Our XGBoost model predicted ternary compositions, YKH$_{12}$ and LaKH$_{12}$,
  to be candidates for achieving higher $T_c$ at 200~GPa.
%
%
(2)
For the predicted chemical compositions,
we further predicted their crystal structures
by using an evolutionary algorithm for crystal structure search implemented
in the USPEX code~\cite{2006GLA}
coupled with {\it ab initio}
geometry optimizations
implemented in the VASP code.~\cite{1993KRE,1994KRE,1996KRE_a,1996KRE_b}
(3)
For the predicted crystal structures,
as shown in Fig.~\ref{fig.structure},
we evaluated their thermodynamical
and structural stabilities by using
convex hull method and
{\it ab initio} phonon calculations.
(4)
Confirming the stabilities,
we finally predicted that
the ternary YKH$_{12}$ and LaKH$_{12}$
are promising candidates to achieve higher $T_c$.
To the best of our knowledge,
this is the first example of
the ternary hydride superconductors
realized by alkali earth metals
($M$=K, +2 valence)
while other preceding studies with
alkali metals ($M$ = Ca, Mg, +1 valence).

\section{Method}
Several ternary superconducting hydrides
predicted theoretically so far,
such as CaYH$_{12}$, YSH$_6$, and
CSH$_7$,~\cite{2019LIA_a,2019LIA_b,2020CUI}
were proposed as a {\it composite} of two
different binary hydrides possibly
to form a compound under high pressure.
Referring 102 published papers
on superconducting hydrides,~\cite{2008CHE,
2009KIM,2009TSE,2010JIN,2010LI,2010GAO,2010DUA,2011GAO,
2011KIM,2011ABE,2012ZHO_a,2011ZHA,2012ZHO_b,2013GAO,2013ABE, 2013HOO,2013LON,
2013HU,2014XIE,2014YU,2014LI,
2014WAN,2014CHE,2014DUA,
2015YAN,2015HOU,2015DUA,2015ZHA_a,
2015LIU_a,2015SHA,2015LIU_b,2015CHE_a,2015FEN,
2015ZHA_b,2015LIU,2015YU,2015CHE_b,2015ZHA,
2015ERR,2016ISH,2016MAH,2016ZHA,2016FU,2016SHA,2016LIU,
2016LI_a,2016ZHO,2016LI_b,2017RAH,2017MA,
2017MA_b,2017KOK,2017LI_a,2017ZHU,2017LIU,2017ZEN,2017ISH,
2017LI_b,2017PEN,2017DAV,2017MAJ,2018ZAR,2018SEM,
2018LI,2018YE,2018KVA_a,2018DUR,2018ZHE,2018LIU,
2018ABE,2018NAK,2018ZHU,2018KVA_b,2018WU,2019LIA_a,2019SUN,2019HEI,2019GRI,
2019CHE,2019LI_b,2019SHA,2019XIE,2019LIA_b,2019YUA,2019WAN,2019AMS,
2019YAN,2019XI,2019ABE,2019DU,2020GUO,2020LV,2020XIE,2020ZHA_a,2020ZHA_b,2020LI,
2020HAO_a,2020CUI,2015MA,2020TRO,2020DIC,2019LI_c,
2018KAN,2020HAO_b}
we obtained 533 superconducting data,
including 181 high-$T_c$ hydrides
($T_c$ higher than liquid nitrogen).
Additionally, we collected 150 kinds of binary compounds.
Excluding those with very low $T_c$ ($T_c <$ 40~K,
McMillan Limit),
81 kinds were left.
They can form 2,867 ternary combinations excluding
the overlap of chemical compositions.
%
Moreover, we excluded those with extremely
higher hydrogen concentration,
making $x\le 16$ for $AM$H$_x$,~\cite{2019SUN}
thereby obtaining 2,366 possible compounds.
%
To make the search space  more compact,
we selected the candidates
only for those with $A$=La and Y
because
La- and Y-based materials have well been verified
as having higher $T_c$
by not only theoretical predictions,
but also experimental observations~\cite{2019SOM,2020TRO}.
Resultant candidates,
Y$M$H$_x$ and La$M$H$_x$,
then amount to 238 compounds, which are input compounds entering the regression.

\vspace{2mm}
It is difficult to obtain all the structural
data from the above pool of published articles.
It is rather practical to use chemical compositions
as the direct descriptor.
In the preceding studies,~\cite{2020FLO,2020SEM_a,2019BOE,2019FLO,2018WAN}
it has been found that
$T_c$ correlates well with (i) space group,~\cite{2019SOM,2019LI_a}
(ii) density of states (DOS) at the Fermi level,
$D\left(E_F\right)$, as a measure of
the applied pressure,~\cite{2020SEM_a,2019HEI,2015BER,2015ERR,2016FU}
as well as (iii) the chemical composition.~\cite{2017KRU,2019BOE}
By the procedures explained below,
we finally set up total 84 descriptors
corresponding to the above three features:
For (i) [space group], we
took the number index for the space group
({\it e.g.,} No.166 for $R\bar{3}m$) as the descriptor.
For (ii) [pressure],
we used the scheme taken in the preceding studies,~\cite{2020MAT,2019MAT}
where the descriptors were composed as
weighted averages over the quantities
for pristine materials composed of each of elements in a compound
(averaging weight is based on the composition ratio).
The quantities were evaluated for
the structure of each pristine material
taken from the Materials Project~\cite{2013JAI}
by using VASP~\cite{1993KRE,1994KRE,1996KRE_a,1996KRE_b}
to get DOS at several values of pressure
(detailed computational conditions were
provided in S.I. (\S\ref{computational}).
For the weighted averaging,
we took the same manner as in
XenonPy.~\cite{2019YAM}
The procedure provides total
56 descriptors for (ii) at this stage.
For (iii) [chemical composition],
we used a XenonPy utility~\cite{2019YAM}
that generates many possible descriptors,
from which we picked up
290 descriptors at the first stage.

\vspace{2mm}
For total 347 descriptors
[290~(iii/Chemical composition) $+$
56~(ii/pressure dependent DOS) $+$
1 (i/space group number)],
we truncated them to avoid overfitting
by excluding those with comparably weaker correlation
with $T_c$.
The truncation can be performed during
the random-forest regressions by monitoring
the correlation using
the scikit-learn library~\cite{2011PED}
(with six trees for this purpose),
finally getting total 84 truncated descriptors
[70~(iii/Chemical composition) $+$
13~(ii/pressure dependent DOS) $+$
1 (space group number)]
as listed in S.I. (\S\ref{descriptors}).

\vspace{2mm}
The above constructed descriptors
were thoroughly incorporated into
four linear Ridge (RD),~\cite{2001HAS}
Bayesian Ridge (RD),~\cite{2006BIS}
LASSO (LS),~\cite{1996TIB,2001HAS,2019YOS,2020YOS}
and Elastic Net (EN)~\cite{2001HAS,2005ZOU})
and three nonlinear Decision-Tree (DT)~\cite{2001HAS},
Random Forest (RF)~\cite{2001HAS,2020CAS,2021YOS},
and Extreme Gradient Boosting (XGBoost)~\cite{2001HAS,2015CHE})
regressors to predict $T_c$ for the target compositions
(see Table~\ref{table.mp});
the (maximum) depths of decision tree
for DT, RF, and XGBoost were set to be 21, 16, and 7, respectively.
To construct the regressiors,
533 data~\cite{2008CHE,
2009KIM,2009TSE,2010JIN,2010LI,2010GAO,2010DUA,2011GAO,
2011KIM,2011ABE,2012ZHO_a,2011ZHA,2012ZHO_b,2013GAO,2013ABE, 2013HOO,2013LON,
2013HU,2014XIE,2014YU,2014LI,
2014WAN,2014CHE,2014DUA,
2015YAN,2015HOU,2015DUA,2015ZHA_a,
2015LIU_a,2015SHA,2015LIU_b,2015CHE_a,2015FEN,
2015ZHA_b,2015LIU,2015YU,2015CHE_b,2015ZHA,
2015ERR,2016ISH,2016MAH,2016ZHA,2016FU,2016SHA,2016LIU,
2016LI_a,2016ZHO,2016LI_b,2017RAH,2017MA,
2017MA_b,2017KOK,2017LI_a,2017ZHU,2017LIU,2017ZEN,2017ISH,
2017LI_b,2017PEN,2017DAV,2017MAJ,2018ZAR,2018SEM,
2018LI,2018YE,2018KVA_a,2018DUR,2018ZHE,2018LIU,
2018ABE,2018NAK,2018ZHU,2018KVA_b,2018WU,2019LIA_a,2019SUN,2019HEI,2019GRI,
2019CHE,2019LI_b,2019SHA,2019XIE,2019LIA_b,2019YUA,2019WAN,2019AMS,
2019YAN,2019XI,2019ABE,2019DU,2020GUO,2020LV,2020XIE,2020ZHA_a,2020ZHA_b,2020LI,
2020HAO_a,2020CUI,2015MA,2020TRO,2020DIC,2019LI_c,
2018KAN,2020HAO_b}
are randomly divided into training and test data
with with the ratio of 80:20.
Hyperparameters in the models were chosen through
the Bayesian optimization technique implemented
in the HyperOpt software package~\cite{2015BER_b}
to minimize the $R^2$ 5-fold cross-validation score.
Model performance was
judged from $R^2$, MAE~(Mean Absolute Error), and
RMSE~(Root Mean Squared Error) as given in
Table~\ref{table.mp}.
Among the above regressors, we found the XGBoost exhibiting
the best performance for the test data, {\it i.e.},
the lowest RMSE, $\Delta T^{\rm RMSE}\sim$ 20~K.
Thus, the XGBoost was chosen as our machine learning
model for $T_c$-prediction used in the successive
high-throughput virtual screening of the ternary compositions.
\begin{table}
 \begin{center}
   \caption{
     Comparison of several regression models
     in terms of $R^2$, MAE~(Mean Absolute Error), and
    RMSE~(Root Mean Squared Error) for the test dataset.
    Each abbreviation means
    'RD'~(Ridge), 'BR'~(Bayesian Ridge),
    'LS'~(Lasso), 'EN'~(Elastic Net),
    'DT'~(Decision Tree),
    'RF'~(Random Forest), and
    'XGB'~(Extreme Gradient Boosting).
     }
     \label{table.mp}
\begin{tabular}{lr r r r r r r r}
& \multicolumn{4}{c}{Linear regression}
& \multicolumn{4}{c}{Nonlinear regression}
\\ \cline{2-5} \cline{7-9}
& RD  & BR & LS  & EN
& & DT & RF & XGB
\\ \hline
$R^2$    & 0.244 & 0.249        & 0.403 & 0.460
& &      0.732       &    0.842      &   0.877
\\ 
MAE  &  34.66    &  37.57     &  34.06      &   32.24
& &   18.58        &      16.34            &    13.53
\\ 
RMSE & 50.34    &  50.17    & 44.75       &  42.56
& &    29.96           &       23.01         &  20.29
\\ \hline
\end{tabular}
 \end{center}
\end{table}

\vspace{2mm}
Once a trained regression is available,
it can immediately predict $T_c$ even for the chemical compositions
with unknown $T_c$ by putting corresponding descriptors
as the input for the regression.
Since we do not know their crystal structures in advance
for the predictions,
we have to assume their space groups
in order to complete input descriptors.
Looking over the existing data,
we found that a space group, $R\bar{3}m$ (No.166),
often gives higher $T_c$ for binary compounds,
so we adopted it as a trial.
The trial setting was proved to be a fair choice
by further verifications
(crystal structural predictions and
{\it ab initio} estimations) in a consistent manner
as explained later.
On the assumption of space group, we predicted $T_c$
for several choices of
$XM$H$_x$ ($X$=Y and La).
For the candidate chemical composition
giving higher $T_c$,
we further predict their crystal structure
by using the USPEX code~\cite{2006GLA}
combined with {\it ab initio} kernel
by VASP.~\cite{1993KRE,1994KRE,1996KRE_a,1996KRE_b}
It randomly generates the 400 structures
among from monomer upto tetramer of
$A$KH$_{12}$ ($A$ = La or Y)
as an 'initial generation' for the generic algorithm.
Each generation evolves 100 structures
according to 40\% heredity, 40\% random,
10\% softmutation, and 10\% transmutation.
A promising candidate structure is
identified when no further evolution
occurs for more than 10 generations.
The candidate is then subject to further
{\it ab initio} geometrical optimizations
by using the Perdew-Burke-Ernzerhof (GGA-PBE)
functional for the exchange-correlation
functional.~\cite{1996PER}
We performed the procedure at the pressure
of 100~GPa, 200~GPa, and 300~GPa,
to get each optimized structure.

\vspace{2mm}
For the predicted crystal structures,
we evaluated the structural stability
by phonon calculations and
thermodynamic stability by
the convex hull method.
For the phonon evaluations,
we used the PhonoPy package~\cite{2015TOG}
combined with {\it ab initio} kernel
by VASP.~\cite{1993KRE,1994KRE,1996KRE_a,1996KRE_b}
Convex hull evaluations were made
by using a utility implemented in USPEX.~\cite{2006GLA}
By the {\it ab initio} phonon calculations,
we finally estimated $T_c$ based on
the Allen-Dynes formalism,~\cite{2017NAK,2016NAK}
to be compared with our data-driven
predictions by the regression
for verification.
Detailed computational conditions
for the phonon calculations
are given in S.I. (\S\ref{computational}).

\section{Results and discussion}
\label{sec.results}
\begin{figure}
  \begin{center}
    \includegraphics[width=\linewidth]{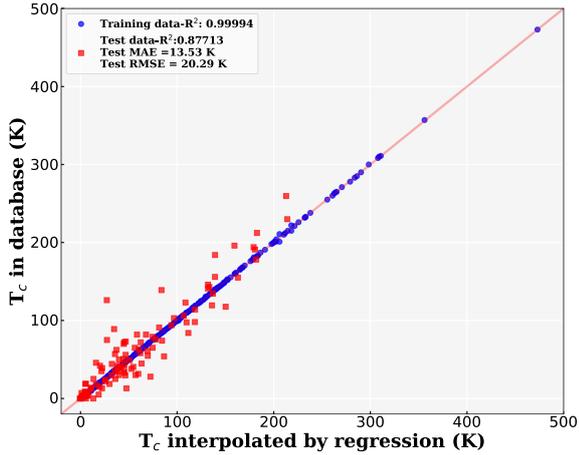}
    \caption{
        Model performance.
        Comparison between $T_c$ in database and that
        interpolated by regression, proving
        the performance of our XGBoost regression model.
        Total 533 data are
        are randomly divided into
        426 training data (80\%)
        and 107 test data (20\%).
    }
    \label{fig.performance}
  \end{center}
\end{figure}

As we mentioned in the previous section (see Table~\ref{table.mp}),
the XGBoost regressor is
the best performed machine learning model.
Fig.~\ref{fig.performance} shows
the performance of our XGBoost model.
Its $R^2$ values are 0.99 and 0.87 for
426 training and 107 test data, respectively,
indicating our regressor is slightly overfitted.
But it exhibits a better performance
than the other models, especially the linear regressions.
Looking at RMSE, the XGBoost value was about 20 K.
This cannot matter for our purpose of the virtual screening
if we keep in mind that out machine-learing prediction
of $T_c$ could deviate from the correspondng {\it ab initio}
prediction by $\pm 20$ K with a 68\% confidence level.

\vspace{2mm}
Getting a good trained regression,
we should be able to perform the
virtual screeing via the machine learning
to search composite candidates with unknown $T_c$ for
high-$T_c$ compounds.
We applied the XGBoost regression
to predict $T_c$ for 238 $AM$H$_x$-type composites,
as we mentioned.
11.7\% of the total composites were found to be
of $AM$H$_{12}$-type ({\it i.e.} $x = 12$) having
higher $T_c$.
We then narrowed down the candidates within
the $AM$H$_{12}$-type candidates to get 28 compounds.
Among the 28 candidates, we found that
the composition YKH$_{12}$ achieves
the highest value of $T_c$, followed by LaKH$_{12}$
as the second highest one, as shown in
Table~\ref{table.Tc}.
For reference, the remaning $T_c$ values are given
in the supporting information (Table~\ref{table.estData}).
\begin{table}
 \begin{center}
   \caption{
   Comparison of the $T_c$ predictions between
   the {\it ab initio} DFT and the XGBoost regression.
     }
     \label{table.Tc}
     \begin{tabular}{lrr}
       \multirow{2}{*}{Composite} & \multicolumn{2}{c}{Predicted $T_c$~[K]} 
       \\ \cline{2-3}
        & Regression  & {\it Ab initio}
       \\
       \hline
       YKH$_{12}$   & 168.9 & 143.2 \\
       LaKH$_{12}$  & 162.8  &  99.2 \\
       \hline
     \end{tabular}
 \end{center}
\end{table}

\vspace{2mm}
For the predicted compositions
to achieve higher $T_c$,
YKH$_{12}$ and LaKH$_{12}$,
we performed further predictions
for their crystal structures
by using USPEX.~\cite{2006GLA}
The predicted structures are shown
in Fig.~\ref{fig.structure}.
The space group $R\bar{3}m$(No.166)
concluded for LaKH$_{12}$ is
consistent with our initial guess
of the structural symmetry as the input
descriptor for our regression.
\begin{figure*}[htbp]
  \begin{center}
    \includegraphics[width=0.38\linewidth]{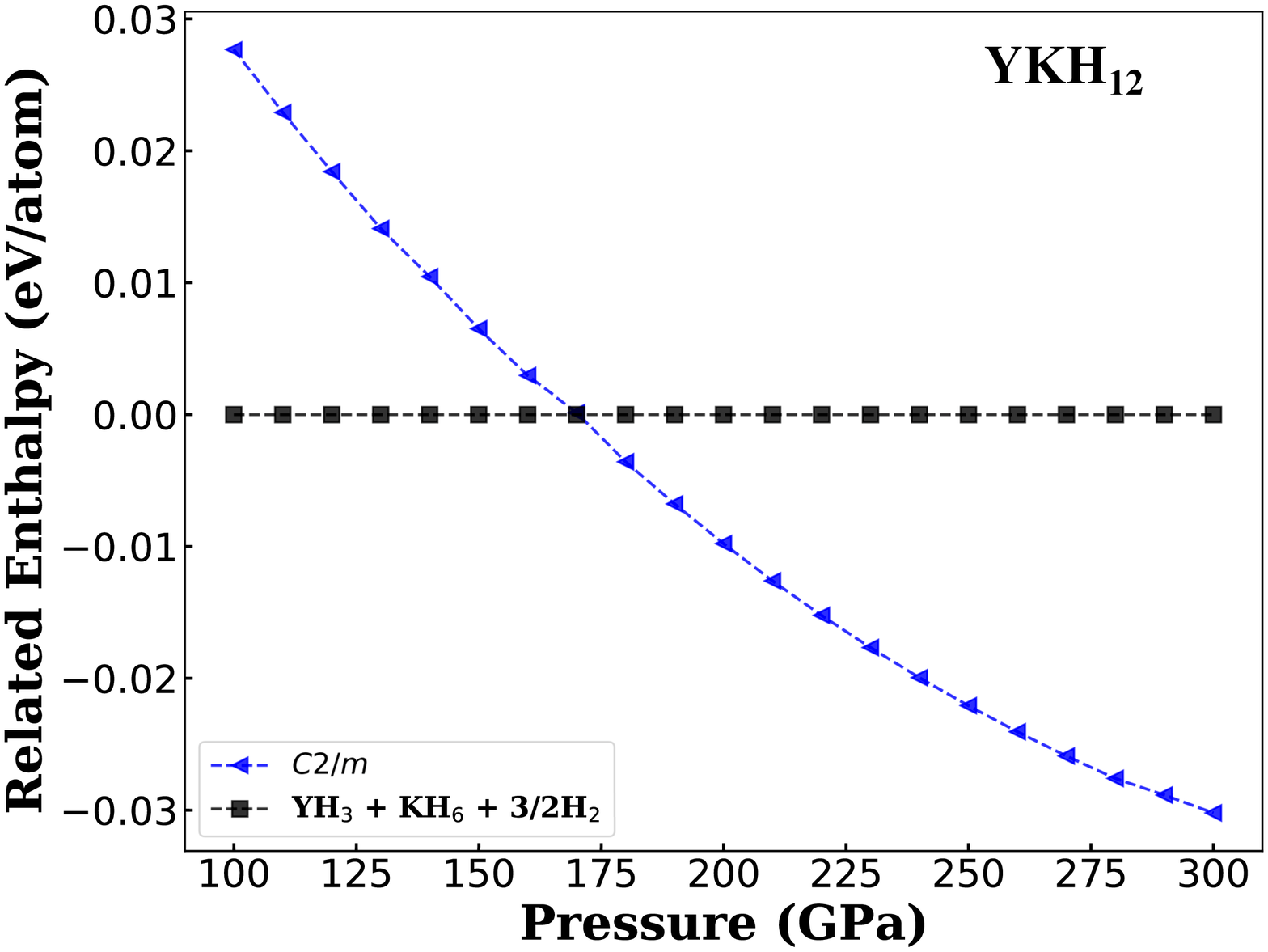}
    \includegraphics[width=0.4\linewidth]{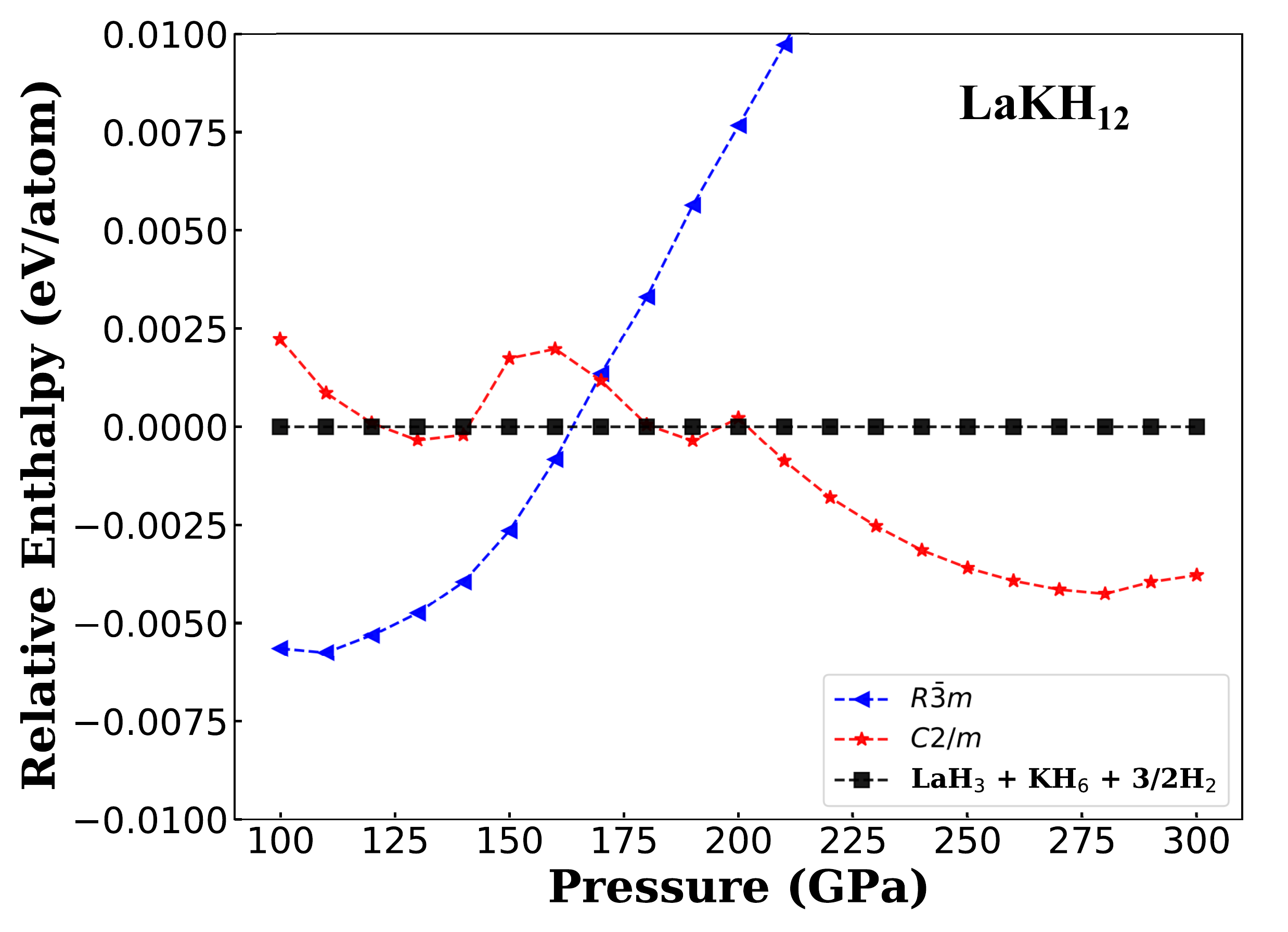}
    \caption{
    Relative enthalpies for
    (a)~YKH$_{12}$ and (b)~LaKH$_{12}$,
    showing the pressure range
    where the compounds are stable.
    }
    \label{fig.stability}
  \end{center}
\end{figure*}

\begin{figure*}[htbp]
  \begin{center}
    \includegraphics[width=8cm]{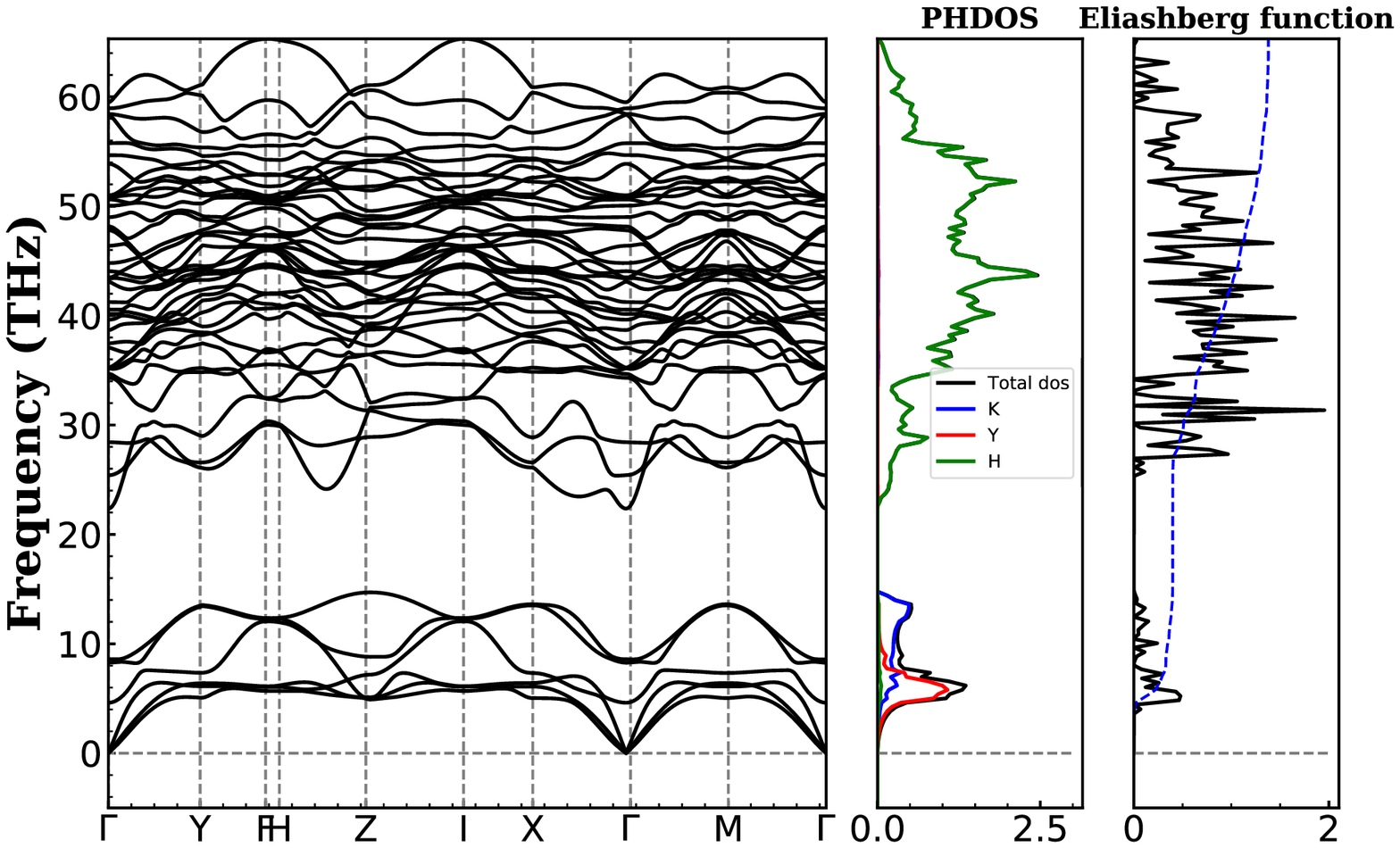}
    \includegraphics[width=8cm]{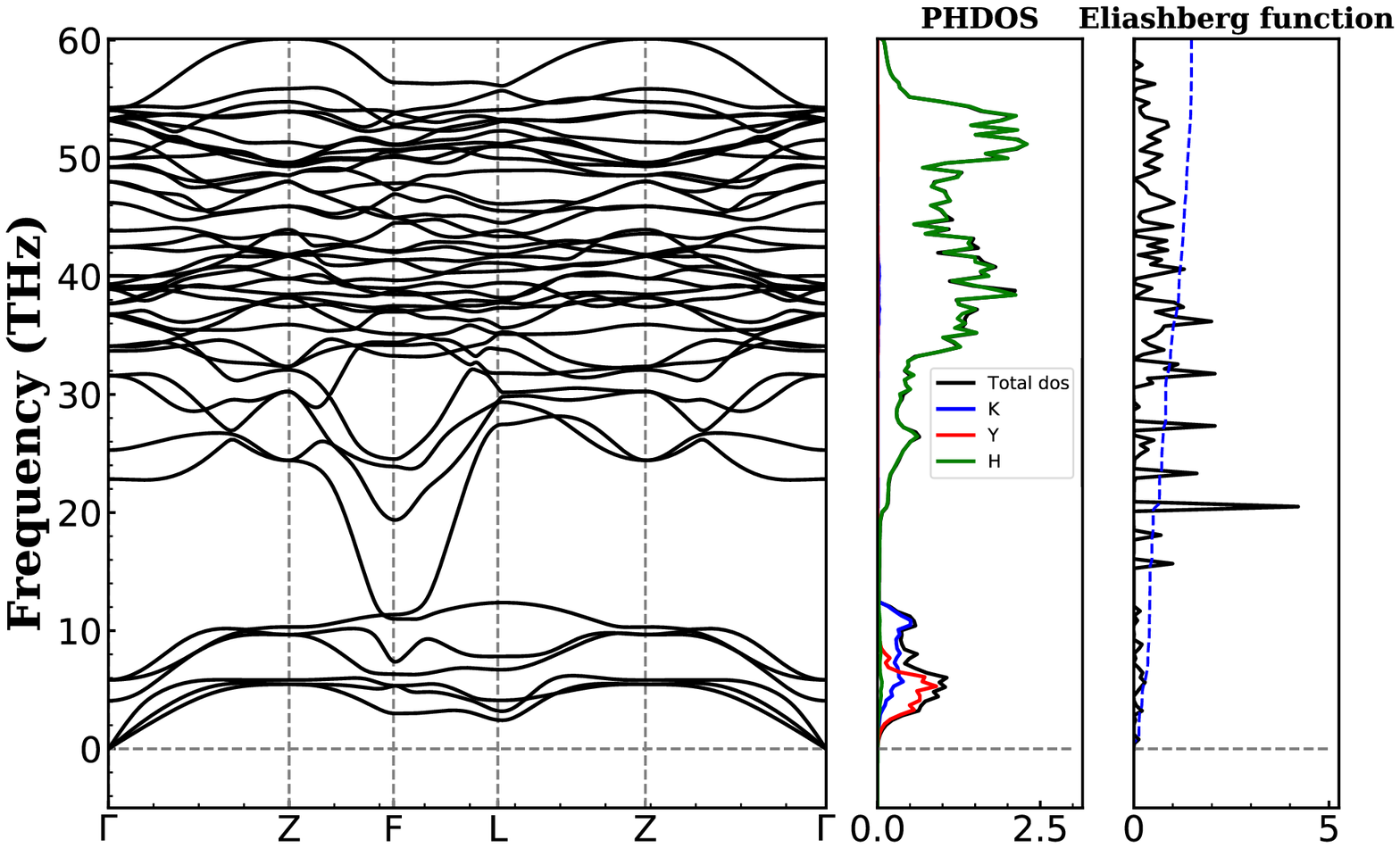}
    \caption{
    Phonon dispersions, phonon DOS, and Eliashberg functions
    for (a)~YKH$_{12}$ at 200~GPa
    and (b)~LaKH$_{12}$ at 160~GPa.
    No imaginary frequency appears
    ensuring the lattice stability for each compound. }
    \label{fig.phononBand}
  \end{center}
\end{figure*}

\vspace{2mm}
To strengthen the reality of predictions,
it is indispensable to estimate the stability of
the predicted structures.
The thermal stability of the structure
is confirmed by comparing a pressure dependence
of relative formation enthalpy within
the USPEX calculations as shown in
Fig.~\ref{fig.stability}.
From the analysis, we can identity
the pressure range where the structures
can stably exist.
Under the pressure range,
we performed {\it ab initio} phonon
calculations to examine the lattice stabilities.
~\cite{2017NAK,2016NAK}
As shown in Fig.~\ref{fig.phononBand},
any imaginary modes do not appear
ensuring the lattice stability for these structures
at the pressure.

\vspace{2mm}
  For YKH$_{12}$ (at 240 GPa) and LaKH$_{12}$ (at 160 GPa),
  we further evaluated their electron-phonon couplings and
  then estimated their $T_c$ values
  based on the Allen-Dynes formalism,~\cite{2017NAK,2016NAK}
  which are listed in Table~\ref{table.Tc}.
  The machine-learning predicted values overestimate
  {\it ab initio} ones by $\sim 20$K and $\sim 60$K for
  YKH$_{12}$ and LaKH$_{12}$, respectively.
  These overestimates lie within almost RMSE and three-times RMSE,
  respectively, which may be thought of as being allowable,
  judging from our XGBoost performance.
\section{Conclusion}
\label{sec.conc}
We performed a data-driven materials searching
for ternary hydrides superconductors
within the range of $AM$H$_{12}$~($A$=La,Y)
composition.
The regression over 533 existing superconductors
was constructed by the random forest method
using 84 descriptors characterizing chemical composition,
space group, and pressure.
Using the regression, we estimated $T_c$
over 239 compositions to get
the prediction of higher $T_c$
achieved by the choice of $M$=K,
YKH$_{12}$ and LaKH$_{12}$.
For the predicted compositions,
we performed evolutionary structure search
to get their crystal structures.
For the structures,
we confirmed their structural stabilities
by using {\it ab initio} phonon calculations
as well as getting $T_c$ estimated by
Allen-Dynes formula.
We finally predicted two new ternary hydrides
superconductors,
YKH$_{12}$ [$C2/m$~(No.12), $T_c$=143.2~K at 240~GPa]
and LaKH$_{12}$ [$R\bar{3}m$~(No.166), $T_c$=99.2~K at 140~GPa].

\section{Acknowledgments}
The computations in this work have been performed
using the facilities of
Research Center for Advanced Computing
Infrastructure (RCACI) at JAIST.
K.H. is grateful for financial support from
the HPCI System Research Project (Project ID: hp190169) and
MEXT-KAKENHI (JP16H06439, JP17K17762, JP19K05029, and JP19H05169).
R.M. is grateful for financial supports from
MEXT-KAKENHI (19H04692 and 16KK0097),
FLAGSHIP2020 (project nos. hp1
90169 and hp190167 at K-computer),
Toyota Motor Corporation, I-O DATA Foundation,
the Air Force Office of Scientific Research
(AFOSR-AOARD/FA2386-17-1-4049;FA2386-19-1-4015),
and JSPS Bilateral Joint Projects (with India DST).

\bibliographystyle{apsrev4-1}
\bibliography{references}

\begin{thebibliography}{155}%
\makeatletter
\providecommand \@ifxundefined [1]{%
 \@ifx{#1\undefined}
}%
\providecommand \@ifnum [1]{%
 \ifnum #1\expandafter \@firstoftwo
 \else \expandafter \@secondoftwo
 \fi
}%
\providecommand \@ifx [1]{%
 \ifx #1\expandafter \@firstoftwo
 \else \expandafter \@secondoftwo
 \fi
}%
\providecommand \natexlab [1]{#1}%
\providecommand \enquote  [1]{``#1''}%
\providecommand \bibnamefont  [1]{#1}%
\providecommand \bibfnamefont [1]{#1}%
\providecommand \citenamefont [1]{#1}%
\providecommand \href@noop [0]{\@secondoftwo}%
\providecommand \href [0]{\begingroup \@sanitize@url \@href}%
\providecommand \@href[1]{\@@startlink{#1}\@@href}%
\providecommand \@@href[1]{\endgroup#1\@@endlink}%
\providecommand \@sanitize@url [0]{\catcode `\\12\catcode `\$12\catcode
  `\&12\catcode `\#12\catcode `\^12\catcode `\_12\catcode `\%12\relax}%
\providecommand \@@startlink[1]{}%
\providecommand \@@endlink[0]{}%
\providecommand \url  [0]{\begingroup\@sanitize@url \@url }%
\providecommand \@url [1]{\endgroup\@href {#1}{\urlprefix }}%
\providecommand \urlprefix  [0]{URL }%
\providecommand \Eprint [0]{\href }%
\providecommand \doibase [0]{http://dx.doi.org/}%
\providecommand \selectlanguage [0]{\@gobble}%
\providecommand \bibinfo  [0]{\@secondoftwo}%
\providecommand \bibfield  [0]{\@secondoftwo}%
\providecommand \translation [1]{[#1]}%
\providecommand \BibitemOpen [0]{}%
\providecommand \bibitemStop [0]{}%
\providecommand \bibitemNoStop [0]{.\EOS\space}%
\providecommand \EOS [0]{\spacefactor3000\relax}%
\providecommand \BibitemShut  [1]{\csname bibitem#1\endcsname}%
\let\auto@bib@innerbib\@empty
\bibitem [{\citenamefont {Ashcroft}(2004)}]{2004ASH}%
  \BibitemOpen
  \bibfield  {author} {\bibinfo {author} {\bibfnamefont {N.}~\bibnamefont
  {Ashcroft}},\ }\href@noop {} {\bibfield  {journal} {\bibinfo  {journal}
  {Physical Review Letters}\ }\textbf {\bibinfo {volume} {92}},\ \bibinfo
  {pages} {187002} (\bibinfo {year} {2004})}\BibitemShut {NoStop}%
\bibitem [{\citenamefont {Ashcroft}(1968)}]{1968ASH}%
  \BibitemOpen
  \bibfield  {author} {\bibinfo {author} {\bibfnamefont {N.~W.}\ \bibnamefont
  {Ashcroft}},\ }\href@noop {} {\bibfield  {journal} {\bibinfo  {journal}
  {Physical Review Letters}\ }\textbf {\bibinfo {volume} {21}},\ \bibinfo
  {pages} {1748} (\bibinfo {year} {1968})}\BibitemShut {NoStop}%
\bibitem [{\citenamefont {Eremets}\ \emph {et~al.}(2008)\citenamefont
  {Eremets}, \citenamefont {Trojan}, \citenamefont {Medvedev}, \citenamefont
  {Tse},\ and\ \citenamefont {Yao}}]{2008ERE}%
  \BibitemOpen
  \bibfield  {author} {\bibinfo {author} {\bibfnamefont {M.}~\bibnamefont
  {Eremets}}, \bibinfo {author} {\bibfnamefont {I.}~\bibnamefont {Trojan}},
  \bibinfo {author} {\bibfnamefont {S.}~\bibnamefont {Medvedev}}, \bibinfo
  {author} {\bibfnamefont {J.}~\bibnamefont {Tse}}, \ and\ \bibinfo {author}
  {\bibfnamefont {Y.}~\bibnamefont {Yao}},\ }\href@noop {} {\bibfield
  {journal} {\bibinfo  {journal} {Science}\ }\textbf {\bibinfo {volume}
  {319}},\ \bibinfo {pages} {1506} (\bibinfo {year} {2008})}\BibitemShut
  {NoStop}%
\bibitem [{\citenamefont {Drozdov}\ \emph {et~al.}(2015)\citenamefont
  {Drozdov}, \citenamefont {Eremets}, \citenamefont {Troyan}, \citenamefont
  {Ksenofontov},\ and\ \citenamefont {Shylin}}]{2015DRO}%
  \BibitemOpen
  \bibfield  {author} {\bibinfo {author} {\bibfnamefont {A.}~\bibnamefont
  {Drozdov}}, \bibinfo {author} {\bibfnamefont {M.}~\bibnamefont {Eremets}},
  \bibinfo {author} {\bibfnamefont {I.}~\bibnamefont {Troyan}}, \bibinfo
  {author} {\bibfnamefont {V.}~\bibnamefont {Ksenofontov}}, \ and\ \bibinfo
  {author} {\bibfnamefont {S.~I.}\ \bibnamefont {Shylin}},\ }\href@noop {}
  {\bibfield  {journal} {\bibinfo  {journal} {Nature}\ }\textbf {\bibinfo
  {volume} {525}},\ \bibinfo {pages} {73} (\bibinfo {year} {2015})}\BibitemShut
  {NoStop}%
\bibitem [{\citenamefont {Szcz{\c{e}}{\'s}niak}\ and\ \citenamefont
  {Zem{\l{}}a}(2015)}]{2015SZC}%
  \BibitemOpen
  \bibfield  {author} {\bibinfo {author} {\bibfnamefont {D.}~\bibnamefont
  {Szcz{\c{e}}{\'s}niak}}\ and\ \bibinfo {author} {\bibfnamefont
  {T.}~\bibnamefont {Zem{\l{}}a}},\ }\href@noop {} {\bibfield  {journal}
  {\bibinfo  {journal} {Superconductor Science and Technology}\ }\textbf
  {\bibinfo {volume} {28}},\ \bibinfo {pages} {085018} (\bibinfo {year}
  {2015})}\BibitemShut {NoStop}%
\bibitem [{\citenamefont {Somayazulu}\ \emph {et~al.}(2019)\citenamefont
  {Somayazulu}, \citenamefont {Ahart}, \citenamefont {Mishra}, \citenamefont
  {Geballe}, \citenamefont {Baldini}, \citenamefont {Meng}, \citenamefont
  {Struzhkin},\ and\ \citenamefont {Hemley}}]{2019SOM}%
  \BibitemOpen
  \bibfield  {author} {\bibinfo {author} {\bibfnamefont {M.}~\bibnamefont
  {Somayazulu}}, \bibinfo {author} {\bibfnamefont {M.}~\bibnamefont {Ahart}},
  \bibinfo {author} {\bibfnamefont {A.~K.}\ \bibnamefont {Mishra}}, \bibinfo
  {author} {\bibfnamefont {Z.~M.}\ \bibnamefont {Geballe}}, \bibinfo {author}
  {\bibfnamefont {M.}~\bibnamefont {Baldini}}, \bibinfo {author} {\bibfnamefont
  {Y.}~\bibnamefont {Meng}}, \bibinfo {author} {\bibfnamefont {V.~V.}\
  \bibnamefont {Struzhkin}}, \ and\ \bibinfo {author} {\bibfnamefont {R.~J.}\
  \bibnamefont {Hemley}},\ }\href@noop {} {\bibfield  {journal} {\bibinfo
  {journal} {Physical review letters}\ }\textbf {\bibinfo {volume} {122}},\
  \bibinfo {pages} {027001} (\bibinfo {year} {2019})}\BibitemShut {NoStop}%
\bibitem [{\citenamefont {Troyan}\ \emph {et~al.}(2020)\citenamefont {Troyan},
  \citenamefont {Semenok}, \citenamefont {Kvashnin}, \citenamefont {Sadakov},
  \citenamefont {Sobolevskiy}, \citenamefont {Pudalov}, \citenamefont
  {Ivanova}, \citenamefont {Prakapenka}, \citenamefont {Greenberg},
  \citenamefont {Gavriliuk} \emph {et~al.}}]{2020TRO}%
  \BibitemOpen
  \bibfield  {author} {\bibinfo {author} {\bibfnamefont {I.~A.}\ \bibnamefont
  {Troyan}}, \bibinfo {author} {\bibfnamefont {D.~V.}\ \bibnamefont {Semenok}},
  \bibinfo {author} {\bibfnamefont {A.~G.}\ \bibnamefont {Kvashnin}}, \bibinfo
  {author} {\bibfnamefont {A.~V.}\ \bibnamefont {Sadakov}}, \bibinfo {author}
  {\bibfnamefont {O.~A.}\ \bibnamefont {Sobolevskiy}}, \bibinfo {author}
  {\bibfnamefont {V.~M.}\ \bibnamefont {Pudalov}}, \bibinfo {author}
  {\bibfnamefont {A.~G.}\ \bibnamefont {Ivanova}}, \bibinfo {author}
  {\bibfnamefont {V.~B.}\ \bibnamefont {Prakapenka}}, \bibinfo {author}
  {\bibfnamefont {E.}~\bibnamefont {Greenberg}}, \bibinfo {author}
  {\bibfnamefont {A.~G.}\ \bibnamefont {Gavriliuk}},  \emph {et~al.},\
  }\href@noop {} {\bibfield  {journal} {\bibinfo  {journal} {arXiv preprint
  arXiv:1908.01534}\ } (\bibinfo {year} {2020})}\BibitemShut {NoStop}%
\bibitem [{\citenamefont {Semenok}\ \emph
  {et~al.}(2020{\natexlab{a}})\citenamefont {Semenok}, \citenamefont {Kruglov},
  \citenamefont {Savkin}, \citenamefont {Kvashnin},\ and\ \citenamefont
  {Oganov}}]{2020SEM}%
  \BibitemOpen
  \bibfield  {author} {\bibinfo {author} {\bibfnamefont {D.~V.}\ \bibnamefont
  {Semenok}}, \bibinfo {author} {\bibfnamefont {I.~A.}\ \bibnamefont
  {Kruglov}}, \bibinfo {author} {\bibfnamefont {I.~A.}\ \bibnamefont {Savkin}},
  \bibinfo {author} {\bibfnamefont {A.~G.}\ \bibnamefont {Kvashnin}}, \ and\
  \bibinfo {author} {\bibfnamefont {A.~R.}\ \bibnamefont {Oganov}},\
  }\href@noop {} {\bibfield  {journal} {\bibinfo  {journal} {Current Opinion in
  Solid State and Materials Science}\ ,\ \bibinfo {pages} {100808}} (\bibinfo
  {year} {2020}{\natexlab{a}})}\BibitemShut {NoStop}%
\bibitem [{\citenamefont {Sun}\ \emph {et~al.}(2019)\citenamefont {Sun},
  \citenamefont {Lv}, \citenamefont {Xie}, \citenamefont {Liu},\ and\
  \citenamefont {Ma}}]{2019SUN}%
  \BibitemOpen
  \bibfield  {author} {\bibinfo {author} {\bibfnamefont {Y.}~\bibnamefont
  {Sun}}, \bibinfo {author} {\bibfnamefont {J.}~\bibnamefont {Lv}}, \bibinfo
  {author} {\bibfnamefont {Y.}~\bibnamefont {Xie}}, \bibinfo {author}
  {\bibfnamefont {H.}~\bibnamefont {Liu}}, \ and\ \bibinfo {author}
  {\bibfnamefont {Y.}~\bibnamefont {Ma}},\ }\href@noop {} {\bibfield  {journal}
  {\bibinfo  {journal} {Physical review letters}\ }\textbf {\bibinfo {volume}
  {123}},\ \bibinfo {pages} {097001} (\bibinfo {year} {2019})}\BibitemShut
  {NoStop}%
\bibitem [{\citenamefont {Snider}\ \emph {et~al.}(2020)\citenamefont {Snider},
  \citenamefont {Dasenbrock-Gammon}, \citenamefont {McBride}, \citenamefont
  {Debessai}, \citenamefont {Vindana}, \citenamefont {Vencatasamy},
  \citenamefont {Lawler}, \citenamefont {Salamat},\ and\ \citenamefont
  {Dias}}]{2020SNI}%
  \BibitemOpen
  \bibfield  {author} {\bibinfo {author} {\bibfnamefont {E.}~\bibnamefont
  {Snider}}, \bibinfo {author} {\bibfnamefont {N.}~\bibnamefont
  {Dasenbrock-Gammon}}, \bibinfo {author} {\bibfnamefont {R.}~\bibnamefont
  {McBride}}, \bibinfo {author} {\bibfnamefont {M.}~\bibnamefont {Debessai}},
  \bibinfo {author} {\bibfnamefont {H.}~\bibnamefont {Vindana}}, \bibinfo
  {author} {\bibfnamefont {K.}~\bibnamefont {Vencatasamy}}, \bibinfo {author}
  {\bibfnamefont {K.~V.}\ \bibnamefont {Lawler}}, \bibinfo {author}
  {\bibfnamefont {A.}~\bibnamefont {Salamat}}, \ and\ \bibinfo {author}
  {\bibfnamefont {R.~P.}\ \bibnamefont {Dias}},\ }\href@noop {} {\bibfield
  {journal} {\bibinfo  {journal} {Nature}\ }\textbf {\bibinfo {volume} {586}},\
  \bibinfo {pages} {373} (\bibinfo {year} {2020})}\BibitemShut {NoStop}%
\bibitem [{\citenamefont {Ma}\ \emph {et~al.}(2017{\natexlab{a}})\citenamefont
  {Ma}, \citenamefont {Duan}, \citenamefont {Shao}, \citenamefont {Yu},
  \citenamefont {Liu}, \citenamefont {Tian}, \citenamefont {Huang},
  \citenamefont {Li}, \citenamefont {Liu},\ and\ \citenamefont
  {Cui}}]{2017MA_b}%
  \BibitemOpen
  \bibfield  {author} {\bibinfo {author} {\bibfnamefont {Y.}~\bibnamefont
  {Ma}}, \bibinfo {author} {\bibfnamefont {D.}~\bibnamefont {Duan}}, \bibinfo
  {author} {\bibfnamefont {Z.}~\bibnamefont {Shao}}, \bibinfo {author}
  {\bibfnamefont {H.}~\bibnamefont {Yu}}, \bibinfo {author} {\bibfnamefont
  {H.}~\bibnamefont {Liu}}, \bibinfo {author} {\bibfnamefont {F.}~\bibnamefont
  {Tian}}, \bibinfo {author} {\bibfnamefont {X.}~\bibnamefont {Huang}},
  \bibinfo {author} {\bibfnamefont {D.}~\bibnamefont {Li}}, \bibinfo {author}
  {\bibfnamefont {B.}~\bibnamefont {Liu}}, \ and\ \bibinfo {author}
  {\bibfnamefont {T.}~\bibnamefont {Cui}},\ }\href@noop {} {\bibfield
  {journal} {\bibinfo  {journal} {Physical Review B}\ }\textbf {\bibinfo
  {volume} {96}},\ \bibinfo {pages} {144518} (\bibinfo {year}
  {2017}{\natexlab{a}})}\BibitemShut {NoStop}%
\bibitem [{\citenamefont {Ma}\ \emph {et~al.}(2017{\natexlab{b}})\citenamefont
  {Ma}, \citenamefont {Duan}, \citenamefont {Shao}, \citenamefont {Li},
  \citenamefont {Wang}, \citenamefont {Yu}, \citenamefont {Tian}, \citenamefont
  {Xie}, \citenamefont {Liu},\ and\ \citenamefont {Cui}}]{2017MA}%
  \BibitemOpen
  \bibfield  {author} {\bibinfo {author} {\bibfnamefont {Y.}~\bibnamefont
  {Ma}}, \bibinfo {author} {\bibfnamefont {D.}~\bibnamefont {Duan}}, \bibinfo
  {author} {\bibfnamefont {Z.}~\bibnamefont {Shao}}, \bibinfo {author}
  {\bibfnamefont {D.}~\bibnamefont {Li}}, \bibinfo {author} {\bibfnamefont
  {L.}~\bibnamefont {Wang}}, \bibinfo {author} {\bibfnamefont {H.}~\bibnamefont
  {Yu}}, \bibinfo {author} {\bibfnamefont {F.}~\bibnamefont {Tian}}, \bibinfo
  {author} {\bibfnamefont {H.}~\bibnamefont {Xie}}, \bibinfo {author}
  {\bibfnamefont {B.}~\bibnamefont {Liu}}, \ and\ \bibinfo {author}
  {\bibfnamefont {T.}~\bibnamefont {Cui}},\ }\href@noop {} {\bibfield
  {journal} {\bibinfo  {journal} {Physical Chemistry Chemical Physics}\
  }\textbf {\bibinfo {volume} {19}},\ \bibinfo {pages} {27406} (\bibinfo {year}
  {2017}{\natexlab{b}})}\BibitemShut {NoStop}%
\bibitem [{\citenamefont {Li}\ \emph {et~al.}(2018)\citenamefont {Li},
  \citenamefont {Liu}, \citenamefont {Tian}, \citenamefont {Wei}, \citenamefont
  {Liu}, \citenamefont {Duan}, \citenamefont {Liu},\ and\ \citenamefont
  {Cui}}]{2018LI}%
  \BibitemOpen
  \bibfield  {author} {\bibinfo {author} {\bibfnamefont {D.}~\bibnamefont
  {Li}}, \bibinfo {author} {\bibfnamefont {Y.}~\bibnamefont {Liu}}, \bibinfo
  {author} {\bibfnamefont {F.-B.}\ \bibnamefont {Tian}}, \bibinfo {author}
  {\bibfnamefont {S.-L.}\ \bibnamefont {Wei}}, \bibinfo {author} {\bibfnamefont
  {Z.}~\bibnamefont {Liu}}, \bibinfo {author} {\bibfnamefont {D.-F.}\
  \bibnamefont {Duan}}, \bibinfo {author} {\bibfnamefont {B.-B.}\ \bibnamefont
  {Liu}}, \ and\ \bibinfo {author} {\bibfnamefont {T.}~\bibnamefont {Cui}},\
  }\href@noop {} {\bibfield  {journal} {\bibinfo  {journal} {Frontiers of
  Physics}\ }\textbf {\bibinfo {volume} {13}},\ \bibinfo {pages} {137107}
  (\bibinfo {year} {2018})}\BibitemShut {NoStop}%
\bibitem [{\citenamefont {Liang}\ \emph
  {et~al.}(2019{\natexlab{a}})\citenamefont {Liang}, \citenamefont {Zhao},
  \citenamefont {Shao}, \citenamefont {Bergara}, \citenamefont {Liu},
  \citenamefont {Wang}, \citenamefont {Sun}, \citenamefont {Zhang},
  \citenamefont {Gao}, \citenamefont {Zhao} \emph {et~al.}}]{2019LIA_b}%
  \BibitemOpen
  \bibfield  {author} {\bibinfo {author} {\bibfnamefont {X.}~\bibnamefont
  {Liang}}, \bibinfo {author} {\bibfnamefont {S.}~\bibnamefont {Zhao}},
  \bibinfo {author} {\bibfnamefont {C.}~\bibnamefont {Shao}}, \bibinfo {author}
  {\bibfnamefont {A.}~\bibnamefont {Bergara}}, \bibinfo {author} {\bibfnamefont
  {H.}~\bibnamefont {Liu}}, \bibinfo {author} {\bibfnamefont {L.}~\bibnamefont
  {Wang}}, \bibinfo {author} {\bibfnamefont {R.}~\bibnamefont {Sun}}, \bibinfo
  {author} {\bibfnamefont {Y.}~\bibnamefont {Zhang}}, \bibinfo {author}
  {\bibfnamefont {Y.}~\bibnamefont {Gao}}, \bibinfo {author} {\bibfnamefont
  {Z.}~\bibnamefont {Zhao}},  \emph {et~al.},\ }\href@noop {} {\bibfield
  {journal} {\bibinfo  {journal} {Physical Review B}\ }\textbf {\bibinfo
  {volume} {100}},\ \bibinfo {pages} {184502} (\bibinfo {year}
  {2019}{\natexlab{a}})}\BibitemShut {NoStop}%
\bibitem [{\citenamefont {Rahm}\ \emph {et~al.}(2017)\citenamefont {Rahm},
  \citenamefont {Hoffmann},\ and\ \citenamefont {Ashcroft}}]{2017RAH}%
  \BibitemOpen
  \bibfield  {author} {\bibinfo {author} {\bibfnamefont {M.}~\bibnamefont
  {Rahm}}, \bibinfo {author} {\bibfnamefont {R.}~\bibnamefont {Hoffmann}}, \
  and\ \bibinfo {author} {\bibfnamefont {N.}~\bibnamefont {Ashcroft}},\
  }\href@noop {} {\bibfield  {journal} {\bibinfo  {journal} {Journal of the
  American Chemical Society}\ }\textbf {\bibinfo {volume} {139}},\ \bibinfo
  {pages} {8740} (\bibinfo {year} {2017})}\BibitemShut {NoStop}%
\bibitem [{\citenamefont {Liang}\ \emph
  {et~al.}(2019{\natexlab{b}})\citenamefont {Liang}, \citenamefont {Bergara},
  \citenamefont {Wang}, \citenamefont {Wen}, \citenamefont {Zhao},
  \citenamefont {Zhou}, \citenamefont {He}, \citenamefont {Gao},\ and\
  \citenamefont {Tian}}]{2019LIA}%
  \BibitemOpen
  \bibfield  {author} {\bibinfo {author} {\bibfnamefont {X.}~\bibnamefont
  {Liang}}, \bibinfo {author} {\bibfnamefont {A.}~\bibnamefont {Bergara}},
  \bibinfo {author} {\bibfnamefont {L.}~\bibnamefont {Wang}}, \bibinfo {author}
  {\bibfnamefont {B.}~\bibnamefont {Wen}}, \bibinfo {author} {\bibfnamefont
  {Z.}~\bibnamefont {Zhao}}, \bibinfo {author} {\bibfnamefont {X.-F.}\
  \bibnamefont {Zhou}}, \bibinfo {author} {\bibfnamefont {J.}~\bibnamefont
  {He}}, \bibinfo {author} {\bibfnamefont {G.}~\bibnamefont {Gao}}, \ and\
  \bibinfo {author} {\bibfnamefont {Y.}~\bibnamefont {Tian}},\ }\href@noop {}
  {\bibfield  {journal} {\bibinfo  {journal} {Physical Review B}\ }\textbf
  {\bibinfo {volume} {99}},\ \bibinfo {pages} {100505} (\bibinfo {year}
  {2019}{\natexlab{b}})}\BibitemShut {NoStop}%
\bibitem [{\citenamefont {Wei}\ \emph {et~al.}(2020)\citenamefont {Wei},
  \citenamefont {Jia}, \citenamefont {Fang}, \citenamefont {Wang},
  \citenamefont {Qian}, \citenamefont {Yuan}, \citenamefont {Selvaraj},
  \citenamefont {Ji},\ and\ \citenamefont {Wei}}]{2020WEI}%
  \BibitemOpen
  \bibfield  {author} {\bibinfo {author} {\bibfnamefont {Y.~K.}\ \bibnamefont
  {Wei}}, \bibinfo {author} {\bibfnamefont {L.~Q.}\ \bibnamefont {Jia}},
  \bibinfo {author} {\bibfnamefont {Y.~Y.}\ \bibnamefont {Fang}}, \bibinfo
  {author} {\bibfnamefont {L.~J.}\ \bibnamefont {Wang}}, \bibinfo {author}
  {\bibfnamefont {Z.~X.}\ \bibnamefont {Qian}}, \bibinfo {author}
  {\bibfnamefont {J.~N.}\ \bibnamefont {Yuan}}, \bibinfo {author}
  {\bibfnamefont {G.}~\bibnamefont {Selvaraj}}, \bibinfo {author}
  {\bibfnamefont {G.~F.}\ \bibnamefont {Ji}}, \ and\ \bibinfo {author}
  {\bibfnamefont {D.~Q.}\ \bibnamefont {Wei}},\ }\href@noop {} {\bibfield
  {journal} {\bibinfo  {journal} {International Journal of Quantum Chemistry}\
  ,\ \bibinfo {pages} {e26459}} (\bibinfo {year} {2020})}\BibitemShut {NoStop}%
\bibitem [{\citenamefont {Xie}\ \emph {et~al.}(2019)\citenamefont {Xie},
  \citenamefont {Duan}, \citenamefont {Shao}, \citenamefont {Song},
  \citenamefont {Wang}, \citenamefont {Xiao}, \citenamefont {Li}, \citenamefont
  {Tian}, \citenamefont {Liu},\ and\ \citenamefont {Cui}}]{2019XIE}%
  \BibitemOpen
  \bibfield  {author} {\bibinfo {author} {\bibfnamefont {H.}~\bibnamefont
  {Xie}}, \bibinfo {author} {\bibfnamefont {D.}~\bibnamefont {Duan}}, \bibinfo
  {author} {\bibfnamefont {Z.}~\bibnamefont {Shao}}, \bibinfo {author}
  {\bibfnamefont {H.}~\bibnamefont {Song}}, \bibinfo {author} {\bibfnamefont
  {Y.}~\bibnamefont {Wang}}, \bibinfo {author} {\bibfnamefont {X.}~\bibnamefont
  {Xiao}}, \bibinfo {author} {\bibfnamefont {D.}~\bibnamefont {Li}}, \bibinfo
  {author} {\bibfnamefont {F.}~\bibnamefont {Tian}}, \bibinfo {author}
  {\bibfnamefont {B.}~\bibnamefont {Liu}}, \ and\ \bibinfo {author}
  {\bibfnamefont {T.}~\bibnamefont {Cui}},\ }\href@noop {} {\bibfield
  {journal} {\bibinfo  {journal} {Journal of Physics: Condensed Matter}\
  }\textbf {\bibinfo {volume} {31}},\ \bibinfo {pages} {245404} (\bibinfo
  {year} {2019})}\BibitemShut {NoStop}%
\bibitem [{\citenamefont {Kokail}\ \emph {et~al.}(2017)\citenamefont {Kokail},
  \citenamefont {von~der Linden},\ and\ \citenamefont {Boeri}}]{2017KOK}%
  \BibitemOpen
  \bibfield  {author} {\bibinfo {author} {\bibfnamefont {C.}~\bibnamefont
  {Kokail}}, \bibinfo {author} {\bibfnamefont {W.}~\bibnamefont {von~der
  Linden}}, \ and\ \bibinfo {author} {\bibfnamefont {L.}~\bibnamefont
  {Boeri}},\ }\href@noop {} {\bibfield  {journal} {\bibinfo  {journal}
  {Physical Review Materials}\ }\textbf {\bibinfo {volume} {1}},\ \bibinfo
  {pages} {074803} (\bibinfo {year} {2017})}\BibitemShut {NoStop}%
\bibitem [{\citenamefont {Shao}\ \emph {et~al.}(2019)\citenamefont {Shao},
  \citenamefont {Duan}, \citenamefont {Ma}, \citenamefont {Yu}, \citenamefont
  {Song}, \citenamefont {Xie}, \citenamefont {Li}, \citenamefont {Tian},
  \citenamefont {Liu},\ and\ \citenamefont {Cui}}]{2019SHA}%
  \BibitemOpen
  \bibfield  {author} {\bibinfo {author} {\bibfnamefont {Z.}~\bibnamefont
  {Shao}}, \bibinfo {author} {\bibfnamefont {D.}~\bibnamefont {Duan}}, \bibinfo
  {author} {\bibfnamefont {Y.}~\bibnamefont {Ma}}, \bibinfo {author}
  {\bibfnamefont {H.}~\bibnamefont {Yu}}, \bibinfo {author} {\bibfnamefont
  {H.}~\bibnamefont {Song}}, \bibinfo {author} {\bibfnamefont {H.}~\bibnamefont
  {Xie}}, \bibinfo {author} {\bibfnamefont {D.}~\bibnamefont {Li}}, \bibinfo
  {author} {\bibfnamefont {F.}~\bibnamefont {Tian}}, \bibinfo {author}
  {\bibfnamefont {B.}~\bibnamefont {Liu}}, \ and\ \bibinfo {author}
  {\bibfnamefont {T.}~\bibnamefont {Cui}},\ }\href@noop {} {\bibfield
  {journal} {\bibinfo  {journal} {npj Computational Materials}\ }\textbf
  {\bibinfo {volume} {5}},\ \bibinfo {pages} {1} (\bibinfo {year}
  {2019})}\BibitemShut {NoStop}%
\bibitem [{\citenamefont {Zhang}\ \emph
  {et~al.}(2020{\natexlab{a}})\citenamefont {Zhang}, \citenamefont {McMahon},
  \citenamefont {Oganov}, \citenamefont {Li}, \citenamefont {Dong},
  \citenamefont {Dong},\ and\ \citenamefont {Wang}}]{2020ZHA_a}%
  \BibitemOpen
  \bibfield  {author} {\bibinfo {author} {\bibfnamefont {J.}~\bibnamefont
  {Zhang}}, \bibinfo {author} {\bibfnamefont {J.~M.}\ \bibnamefont {McMahon}},
  \bibinfo {author} {\bibfnamefont {A.~R.}\ \bibnamefont {Oganov}}, \bibinfo
  {author} {\bibfnamefont {X.}~\bibnamefont {Li}}, \bibinfo {author}
  {\bibfnamefont {X.}~\bibnamefont {Dong}}, \bibinfo {author} {\bibfnamefont
  {H.}~\bibnamefont {Dong}}, \ and\ \bibinfo {author} {\bibfnamefont
  {S.}~\bibnamefont {Wang}},\ }\href@noop {} {\bibfield  {journal} {\bibinfo
  {journal} {Physical Review B}\ }\textbf {\bibinfo {volume} {101}},\ \bibinfo
  {pages} {134108} (\bibinfo {year} {2020}{\natexlab{a}})}\BibitemShut
  {NoStop}%
\bibitem [{\citenamefont {Zhang}\ \emph
  {et~al.}(2020{\natexlab{b}})\citenamefont {Zhang}, \citenamefont {Sun},
  \citenamefont {Li}, \citenamefont {Lv},\ and\ \citenamefont
  {Liu}}]{2020ZHA_b}%
  \BibitemOpen
  \bibfield  {author} {\bibinfo {author} {\bibfnamefont {P.}~\bibnamefont
  {Zhang}}, \bibinfo {author} {\bibfnamefont {Y.}~\bibnamefont {Sun}}, \bibinfo
  {author} {\bibfnamefont {X.}~\bibnamefont {Li}}, \bibinfo {author}
  {\bibfnamefont {J.}~\bibnamefont {Lv}}, \ and\ \bibinfo {author}
  {\bibfnamefont {H.}~\bibnamefont {Liu}},\ }\href@noop {} {\bibfield
  {journal} {\bibinfo  {journal} {Physical Review B}\ }\textbf {\bibinfo
  {volume} {102}},\ \bibinfo {pages} {184103} (\bibinfo {year}
  {2020}{\natexlab{b}})}\BibitemShut {NoStop}%
\bibitem [{\citenamefont {Di~Cataldo}\ \emph {et~al.}(2020)\citenamefont
  {Di~Cataldo}, \citenamefont {von~der Linden},\ and\ \citenamefont
  {Boeri}}]{2020DIC}%
  \BibitemOpen
  \bibfield  {author} {\bibinfo {author} {\bibfnamefont {S.}~\bibnamefont
  {Di~Cataldo}}, \bibinfo {author} {\bibfnamefont {W.}~\bibnamefont {von~der
  Linden}}, \ and\ \bibinfo {author} {\bibfnamefont {L.}~\bibnamefont
  {Boeri}},\ }\href@noop {} {\bibfield  {journal} {\bibinfo  {journal}
  {Physical Review B}\ }\textbf {\bibinfo {volume} {102}},\ \bibinfo {pages}
  {014516} (\bibinfo {year} {2020})}\BibitemShut {NoStop}%
\bibitem [{\citenamefont {Guo}\ \emph {et~al.}(2020)\citenamefont {Guo},
  \citenamefont {Wang}, \citenamefont {Chen}, \citenamefont {Lu}, \citenamefont
  {Ho},\ and\ \citenamefont {Wang}}]{2020GUO}%
  \BibitemOpen
  \bibfield  {author} {\bibinfo {author} {\bibfnamefont {X.}~\bibnamefont
  {Guo}}, \bibinfo {author} {\bibfnamefont {R.-L.}\ \bibnamefont {Wang}},
  \bibinfo {author} {\bibfnamefont {H.-L.}\ \bibnamefont {Chen}}, \bibinfo
  {author} {\bibfnamefont {W.-C.}\ \bibnamefont {Lu}}, \bibinfo {author}
  {\bibfnamefont {K.}~\bibnamefont {Ho}}, \ and\ \bibinfo {author}
  {\bibfnamefont {C.}~\bibnamefont {Wang}},\ }\href@noop {} {\bibfield
  {journal} {\bibinfo  {journal} {Physics Letters A}\ }\textbf {\bibinfo
  {volume} {384}},\ \bibinfo {pages} {126189} (\bibinfo {year}
  {2020})}\BibitemShut {NoStop}%
\bibitem [{\citenamefont {Cui}\ \emph {et~al.}(2020)\citenamefont {Cui},
  \citenamefont {Bi}, \citenamefont {Shi}, \citenamefont {Li}, \citenamefont
  {Liu}, \citenamefont {Zurek},\ and\ \citenamefont {Hemley}}]{2020CUI}%
  \BibitemOpen
  \bibfield  {author} {\bibinfo {author} {\bibfnamefont {W.}~\bibnamefont
  {Cui}}, \bibinfo {author} {\bibfnamefont {T.}~\bibnamefont {Bi}}, \bibinfo
  {author} {\bibfnamefont {J.}~\bibnamefont {Shi}}, \bibinfo {author}
  {\bibfnamefont {Y.}~\bibnamefont {Li}}, \bibinfo {author} {\bibfnamefont
  {H.}~\bibnamefont {Liu}}, \bibinfo {author} {\bibfnamefont {E.}~\bibnamefont
  {Zurek}}, \ and\ \bibinfo {author} {\bibfnamefont {R.~J.}\ \bibnamefont
  {Hemley}},\ }\href@noop {} {\bibfield  {journal} {\bibinfo  {journal}
  {Physical Review B}\ }\textbf {\bibinfo {volume} {101}},\ \bibinfo {pages}
  {134504} (\bibinfo {year} {2020})}\BibitemShut {NoStop}%
\bibitem [{\citenamefont {Lv}\ \emph {et~al.}(2020)\citenamefont {Lv},
  \citenamefont {Zhang}, \citenamefont {Li}, \citenamefont {Hai}, \citenamefont
  {Lu}, \citenamefont {Li},\ and\ \citenamefont {Zhong}}]{2020LV}%
  \BibitemOpen
  \bibfield  {author} {\bibinfo {author} {\bibfnamefont {H.-Y.}\ \bibnamefont
  {Lv}}, \bibinfo {author} {\bibfnamefont {S.-Y.}\ \bibnamefont {Zhang}},
  \bibinfo {author} {\bibfnamefont {M.-H.}\ \bibnamefont {Li}}, \bibinfo
  {author} {\bibfnamefont {Y.-L.}\ \bibnamefont {Hai}}, \bibinfo {author}
  {\bibfnamefont {N.}~\bibnamefont {Lu}}, \bibinfo {author} {\bibfnamefont
  {W.-J.}\ \bibnamefont {Li}}, \ and\ \bibinfo {author} {\bibfnamefont {G.-H.}\
  \bibnamefont {Zhong}},\ }\href@noop {} {\bibfield  {journal} {\bibinfo
  {journal} {Physical Chemistry Chemical Physics}\ }\textbf {\bibinfo {volume}
  {22}},\ \bibinfo {pages} {1069} (\bibinfo {year} {2020})}\BibitemShut
  {NoStop}%
\bibitem [{\citenamefont {Yan}\ \emph {et~al.}(2020)\citenamefont {Yan},
  \citenamefont {Bi}, \citenamefont {Geng}, \citenamefont {Wang},\ and\
  \citenamefont {Zurek}}]{2020YAN}%
  \BibitemOpen
  \bibfield  {author} {\bibinfo {author} {\bibfnamefont {Y.}~\bibnamefont
  {Yan}}, \bibinfo {author} {\bibfnamefont {T.}~\bibnamefont {Bi}}, \bibinfo
  {author} {\bibfnamefont {N.}~\bibnamefont {Geng}}, \bibinfo {author}
  {\bibfnamefont {X.}~\bibnamefont {Wang}}, \ and\ \bibinfo {author}
  {\bibfnamefont {E.}~\bibnamefont {Zurek}},\ }\href@noop {} {\bibfield
  {journal} {\bibinfo  {journal} {The Journal of Physical Chemistry Letters}\
  }\textbf {\bibinfo {volume} {11}},\ \bibinfo {pages} {9629} (\bibinfo {year}
  {2020})}\BibitemShut {NoStop}%
\bibitem [{\citenamefont {Muramatsu}\ \emph {et~al.}(2015)\citenamefont
  {Muramatsu}, \citenamefont {Wanene}, \citenamefont {Somayazulu},
  \citenamefont {Vinitsky}, \citenamefont {Chandra}, \citenamefont {Strobel},
  \citenamefont {Struzhkin},\ and\ \citenamefont {Hemley}}]{2015MUR}%
  \BibitemOpen
  \bibfield  {author} {\bibinfo {author} {\bibfnamefont {T.}~\bibnamefont
  {Muramatsu}}, \bibinfo {author} {\bibfnamefont {W.~K.}\ \bibnamefont
  {Wanene}}, \bibinfo {author} {\bibfnamefont {M.}~\bibnamefont {Somayazulu}},
  \bibinfo {author} {\bibfnamefont {E.}~\bibnamefont {Vinitsky}}, \bibinfo
  {author} {\bibfnamefont {D.}~\bibnamefont {Chandra}}, \bibinfo {author}
  {\bibfnamefont {T.~A.}\ \bibnamefont {Strobel}}, \bibinfo {author}
  {\bibfnamefont {V.~V.}\ \bibnamefont {Struzhkin}}, \ and\ \bibinfo {author}
  {\bibfnamefont {R.~J.}\ \bibnamefont {Hemley}},\ }\href@noop {} {\bibfield
  {journal} {\bibinfo  {journal} {The Journal of Physical Chemistry C}\
  }\textbf {\bibinfo {volume} {119}},\ \bibinfo {pages} {18007} (\bibinfo
  {year} {2015})}\BibitemShut {NoStop}%
\bibitem [{\citenamefont {Meng}\ \emph {et~al.}(2019)\citenamefont {Meng},
  \citenamefont {Sakata}, \citenamefont {Shimizu}, \citenamefont {Iijima},
  \citenamefont {Saitoh}, \citenamefont {Sato}, \citenamefont {Takagi},\ and\
  \citenamefont {Orimo}}]{2019MEN}%
  \BibitemOpen
  \bibfield  {author} {\bibinfo {author} {\bibfnamefont {D.}~\bibnamefont
  {Meng}}, \bibinfo {author} {\bibfnamefont {M.}~\bibnamefont {Sakata}},
  \bibinfo {author} {\bibfnamefont {K.}~\bibnamefont {Shimizu}}, \bibinfo
  {author} {\bibfnamefont {Y.}~\bibnamefont {Iijima}}, \bibinfo {author}
  {\bibfnamefont {H.}~\bibnamefont {Saitoh}}, \bibinfo {author} {\bibfnamefont
  {T.}~\bibnamefont {Sato}}, \bibinfo {author} {\bibfnamefont {S.}~\bibnamefont
  {Takagi}}, \ and\ \bibinfo {author} {\bibfnamefont {S.-i.}\ \bibnamefont
  {Orimo}},\ }\href@noop {} {\bibfield  {journal} {\bibinfo  {journal}
  {Physical Review B}\ }\textbf {\bibinfo {volume} {99}},\ \bibinfo {pages}
  {024508} (\bibinfo {year} {2019})}\BibitemShut {NoStop}%
\bibitem [{\citenamefont {Jain}\ \emph {et~al.}(2013)\citenamefont {Jain},
  \citenamefont {Ong}, \citenamefont {Hautier}, \citenamefont {Chen},
  \citenamefont {Richards}, \citenamefont {Dacek}, \citenamefont {Cholia},
  \citenamefont {Gunter}, \citenamefont {Skinner}, \citenamefont {Ceder} \emph
  {et~al.}}]{2013JAI}%
  \BibitemOpen
  \bibfield  {author} {\bibinfo {author} {\bibfnamefont {A.}~\bibnamefont
  {Jain}}, \bibinfo {author} {\bibfnamefont {S.~P.}\ \bibnamefont {Ong}},
  \bibinfo {author} {\bibfnamefont {G.}~\bibnamefont {Hautier}}, \bibinfo
  {author} {\bibfnamefont {W.}~\bibnamefont {Chen}}, \bibinfo {author}
  {\bibfnamefont {W.~D.}\ \bibnamefont {Richards}}, \bibinfo {author}
  {\bibfnamefont {S.}~\bibnamefont {Dacek}}, \bibinfo {author} {\bibfnamefont
  {S.}~\bibnamefont {Cholia}}, \bibinfo {author} {\bibfnamefont
  {D.}~\bibnamefont {Gunter}}, \bibinfo {author} {\bibfnamefont
  {D.}~\bibnamefont {Skinner}}, \bibinfo {author} {\bibfnamefont
  {G.}~\bibnamefont {Ceder}},  \emph {et~al.},\ }\href@noop {} {\bibfield
  {journal} {\bibinfo  {journal} {Apl Materials}\ }\textbf {\bibinfo {volume}
  {1}},\ \bibinfo {pages} {011002} (\bibinfo {year} {2013})}\BibitemShut
  {NoStop}%
\bibitem [{\citenamefont {Stanev}\ \emph {et~al.}(2018)\citenamefont {Stanev},
  \citenamefont {Oses}, \citenamefont {Kusne}, \citenamefont {Rodriguez},
  \citenamefont {Paglione}, \citenamefont {Curtarolo},\ and\ \citenamefont
  {Takeuchi}}]{2018STA}%
  \BibitemOpen
  \bibfield  {author} {\bibinfo {author} {\bibfnamefont {V.}~\bibnamefont
  {Stanev}}, \bibinfo {author} {\bibfnamefont {C.}~\bibnamefont {Oses}},
  \bibinfo {author} {\bibfnamefont {A.~G.}\ \bibnamefont {Kusne}}, \bibinfo
  {author} {\bibfnamefont {E.}~\bibnamefont {Rodriguez}}, \bibinfo {author}
  {\bibfnamefont {J.}~\bibnamefont {Paglione}}, \bibinfo {author}
  {\bibfnamefont {S.}~\bibnamefont {Curtarolo}}, \ and\ \bibinfo {author}
  {\bibfnamefont {I.}~\bibnamefont {Takeuchi}},\ }\href@noop {} {\bibfield
  {journal} {\bibinfo  {journal} {npj Computational Materials}\ }\textbf
  {\bibinfo {volume} {4}},\ \bibinfo {pages} {1} (\bibinfo {year}
  {2018})}\BibitemShut {NoStop}%
\bibitem [{\citenamefont {Matsumoto}\ and\ \citenamefont
  {Horide}(2019)}]{2019MAT}%
  \BibitemOpen
  \bibfield  {author} {\bibinfo {author} {\bibfnamefont {K.}~\bibnamefont
  {Matsumoto}}\ and\ \bibinfo {author} {\bibfnamefont {T.}~\bibnamefont
  {Horide}},\ }\href@noop {} {\bibfield  {journal} {\bibinfo  {journal}
  {Applied Physics Express}\ }\textbf {\bibinfo {volume} {12}},\ \bibinfo
  {pages} {073003} (\bibinfo {year} {2019})}\BibitemShut {NoStop}%
\bibitem [{\citenamefont {Meredig}\ \emph {et~al.}(2018)\citenamefont
  {Meredig}, \citenamefont {Antono}, \citenamefont {Church}, \citenamefont
  {Hutchinson}, \citenamefont {Ling}, \citenamefont {Paradiso}, \citenamefont
  {Blaiszik}, \citenamefont {Foster}, \citenamefont {Gibbons}, \citenamefont
  {Hattrick-Simpers} \emph {et~al.}}]{2018MER}%
  \BibitemOpen
  \bibfield  {author} {\bibinfo {author} {\bibfnamefont {B.}~\bibnamefont
  {Meredig}}, \bibinfo {author} {\bibfnamefont {E.}~\bibnamefont {Antono}},
  \bibinfo {author} {\bibfnamefont {C.}~\bibnamefont {Church}}, \bibinfo
  {author} {\bibfnamefont {M.}~\bibnamefont {Hutchinson}}, \bibinfo {author}
  {\bibfnamefont {J.}~\bibnamefont {Ling}}, \bibinfo {author} {\bibfnamefont
  {S.}~\bibnamefont {Paradiso}}, \bibinfo {author} {\bibfnamefont
  {B.}~\bibnamefont {Blaiszik}}, \bibinfo {author} {\bibfnamefont
  {I.}~\bibnamefont {Foster}}, \bibinfo {author} {\bibfnamefont
  {B.}~\bibnamefont {Gibbons}}, \bibinfo {author} {\bibfnamefont
  {J.}~\bibnamefont {Hattrick-Simpers}},  \emph {et~al.},\ }\href@noop {}
  {\bibfield  {journal} {\bibinfo  {journal} {Molecular Systems Design \&
  Engineering}\ }\textbf {\bibinfo {volume} {3}},\ \bibinfo {pages} {819}
  (\bibinfo {year} {2018})}\BibitemShut {NoStop}%
\bibitem [{\citenamefont {Hutcheon}\ \emph {et~al.}(2020)\citenamefont
  {Hutcheon}, \citenamefont {Shipley},\ and\ \citenamefont {Needs}}]{2020HUT}%
  \BibitemOpen
  \bibfield  {author} {\bibinfo {author} {\bibfnamefont {M.~J.}\ \bibnamefont
  {Hutcheon}}, \bibinfo {author} {\bibfnamefont {A.~M.}\ \bibnamefont
  {Shipley}}, \ and\ \bibinfo {author} {\bibfnamefont {R.~J.}\ \bibnamefont
  {Needs}},\ }\href@noop {} {\bibfield  {journal} {\bibinfo  {journal}
  {Physical Review B}\ }\textbf {\bibinfo {volume} {101}},\ \bibinfo {pages}
  {144505} (\bibinfo {year} {2020})}\BibitemShut {NoStop}%
\bibitem [{\citenamefont {Chen}\ \emph {et~al.}(2008)\citenamefont {Chen},
  \citenamefont {Wang}, \citenamefont {Struzhkin}, \citenamefont {Mao},
  \citenamefont {Hemley},\ and\ \citenamefont {Lin}}]{2008CHE}%
  \BibitemOpen
  \bibfield  {author} {\bibinfo {author} {\bibfnamefont {X.-J.}\ \bibnamefont
  {Chen}}, \bibinfo {author} {\bibfnamefont {J.-L.}\ \bibnamefont {Wang}},
  \bibinfo {author} {\bibfnamefont {V.~V.}\ \bibnamefont {Struzhkin}}, \bibinfo
  {author} {\bibfnamefont {H.-k.}\ \bibnamefont {Mao}}, \bibinfo {author}
  {\bibfnamefont {R.~J.}\ \bibnamefont {Hemley}}, \ and\ \bibinfo {author}
  {\bibfnamefont {H.-Q.}\ \bibnamefont {Lin}},\ }\href@noop {} {\bibfield
  {journal} {\bibinfo  {journal} {Physical review letters}\ }\textbf {\bibinfo
  {volume} {101}},\ \bibinfo {pages} {077002} (\bibinfo {year}
  {2008})}\BibitemShut {NoStop}%
\bibitem [{\citenamefont {Kim}\ \emph {et~al.}(2009)\citenamefont {Kim},
  \citenamefont {Scheicher},\ and\ \citenamefont {Ahuja}}]{2009KIM}%
  \BibitemOpen
  \bibfield  {author} {\bibinfo {author} {\bibfnamefont {D.~Y.}\ \bibnamefont
  {Kim}}, \bibinfo {author} {\bibfnamefont {R.~H.}\ \bibnamefont {Scheicher}},
  \ and\ \bibinfo {author} {\bibfnamefont {R.}~\bibnamefont {Ahuja}},\
  }\href@noop {} {\bibfield  {journal} {\bibinfo  {journal} {Physical review
  letters}\ }\textbf {\bibinfo {volume} {103}},\ \bibinfo {pages} {077002}
  (\bibinfo {year} {2009})}\BibitemShut {NoStop}%
\bibitem [{\citenamefont {John}\ \emph {et~al.}(2009)\citenamefont {John},
  \citenamefont {Song}, \citenamefont {Yao}, \citenamefont {Smith},
  \citenamefont {Desgreniers},\ and\ \citenamefont {Klug}}]{2009TSE}%
  \BibitemOpen
  \bibfield  {author} {\bibinfo {author} {\bibfnamefont {S.~T.}\ \bibnamefont
  {John}}, \bibinfo {author} {\bibfnamefont {Z.}~\bibnamefont {Song}}, \bibinfo
  {author} {\bibfnamefont {Y.}~\bibnamefont {Yao}}, \bibinfo {author}
  {\bibfnamefont {J.~S.}\ \bibnamefont {Smith}}, \bibinfo {author}
  {\bibfnamefont {S.}~\bibnamefont {Desgreniers}}, \ and\ \bibinfo {author}
  {\bibfnamefont {D.~D.}\ \bibnamefont {Klug}},\ }\href@noop {} {\bibfield
  {journal} {\bibinfo  {journal} {Solid state communications}\ }\textbf
  {\bibinfo {volume} {149}},\ \bibinfo {pages} {1944} (\bibinfo {year}
  {2009})}\BibitemShut {NoStop}%
\bibitem [{\citenamefont {Jin}\ \emph {et~al.}(2010)\citenamefont {Jin},
  \citenamefont {Meng}, \citenamefont {He}, \citenamefont {Ma}, \citenamefont
  {Liu}, \citenamefont {Cui}, \citenamefont {Zou},\ and\ \citenamefont
  {Mao}}]{2010JIN}%
  \BibitemOpen
  \bibfield  {author} {\bibinfo {author} {\bibfnamefont {X.}~\bibnamefont
  {Jin}}, \bibinfo {author} {\bibfnamefont {X.}~\bibnamefont {Meng}}, \bibinfo
  {author} {\bibfnamefont {Z.}~\bibnamefont {He}}, \bibinfo {author}
  {\bibfnamefont {Y.}~\bibnamefont {Ma}}, \bibinfo {author} {\bibfnamefont
  {B.}~\bibnamefont {Liu}}, \bibinfo {author} {\bibfnamefont {T.}~\bibnamefont
  {Cui}}, \bibinfo {author} {\bibfnamefont {G.}~\bibnamefont {Zou}}, \ and\
  \bibinfo {author} {\bibfnamefont {H.-k.}\ \bibnamefont {Mao}},\ }\href@noop
  {} {\bibfield  {journal} {\bibinfo  {journal} {Proceedings of the National
  Academy of Sciences}\ }\textbf {\bibinfo {volume} {107}},\ \bibinfo {pages}
  {9969} (\bibinfo {year} {2010})}\BibitemShut {NoStop}%
\bibitem [{\citenamefont {Li}\ \emph {et~al.}(2010)\citenamefont {Li},
  \citenamefont {Gao}, \citenamefont {Xie}, \citenamefont {Ma}, \citenamefont
  {Cui},\ and\ \citenamefont {Zou}}]{2010LI}%
  \BibitemOpen
  \bibfield  {author} {\bibinfo {author} {\bibfnamefont {Y.}~\bibnamefont
  {Li}}, \bibinfo {author} {\bibfnamefont {G.}~\bibnamefont {Gao}}, \bibinfo
  {author} {\bibfnamefont {Y.}~\bibnamefont {Xie}}, \bibinfo {author}
  {\bibfnamefont {Y.}~\bibnamefont {Ma}}, \bibinfo {author} {\bibfnamefont
  {T.}~\bibnamefont {Cui}}, \ and\ \bibinfo {author} {\bibfnamefont
  {G.}~\bibnamefont {Zou}},\ }\href@noop {} {\bibfield  {journal} {\bibinfo
  {journal} {Proceedings of the National Academy of Sciences}\ }\textbf
  {\bibinfo {volume} {107}},\ \bibinfo {pages} {15708} (\bibinfo {year}
  {2010})}\BibitemShut {NoStop}%
\bibitem [{\citenamefont {Gao}\ \emph {et~al.}(2010)\citenamefont {Gao},
  \citenamefont {Oganov}, \citenamefont {Li}, \citenamefont {Li}, \citenamefont
  {Wang}, \citenamefont {Cui}, \citenamefont {Ma}, \citenamefont {Bergara},
  \citenamefont {Lyakhov}, \citenamefont {Iitaka} \emph {et~al.}}]{2010GAO}%
  \BibitemOpen
  \bibfield  {author} {\bibinfo {author} {\bibfnamefont {G.}~\bibnamefont
  {Gao}}, \bibinfo {author} {\bibfnamefont {A.~R.}\ \bibnamefont {Oganov}},
  \bibinfo {author} {\bibfnamefont {P.}~\bibnamefont {Li}}, \bibinfo {author}
  {\bibfnamefont {Z.}~\bibnamefont {Li}}, \bibinfo {author} {\bibfnamefont
  {H.}~\bibnamefont {Wang}}, \bibinfo {author} {\bibfnamefont {T.}~\bibnamefont
  {Cui}}, \bibinfo {author} {\bibfnamefont {Y.}~\bibnamefont {Ma}}, \bibinfo
  {author} {\bibfnamefont {A.}~\bibnamefont {Bergara}}, \bibinfo {author}
  {\bibfnamefont {A.~O.}\ \bibnamefont {Lyakhov}}, \bibinfo {author}
  {\bibfnamefont {T.}~\bibnamefont {Iitaka}},  \emph {et~al.},\ }\href@noop {}
  {\bibfield  {journal} {\bibinfo  {journal} {Proceedings of the National
  Academy of Sciences}\ }\textbf {\bibinfo {volume} {107}},\ \bibinfo {pages}
  {1317} (\bibinfo {year} {2010})}\BibitemShut {NoStop}%
\bibitem [{\citenamefont {Duan}\ \emph {et~al.}(2010)\citenamefont {Duan},
  \citenamefont {Tian}, \citenamefont {He}, \citenamefont {Meng}, \citenamefont
  {Wang}, \citenamefont {Chen}, \citenamefont {Zhao}, \citenamefont {Liu},\
  and\ \citenamefont {Cui}}]{2010DUA}%
  \BibitemOpen
  \bibfield  {author} {\bibinfo {author} {\bibfnamefont {D.}~\bibnamefont
  {Duan}}, \bibinfo {author} {\bibfnamefont {F.}~\bibnamefont {Tian}}, \bibinfo
  {author} {\bibfnamefont {Z.}~\bibnamefont {He}}, \bibinfo {author}
  {\bibfnamefont {X.}~\bibnamefont {Meng}}, \bibinfo {author} {\bibfnamefont
  {L.}~\bibnamefont {Wang}}, \bibinfo {author} {\bibfnamefont {C.}~\bibnamefont
  {Chen}}, \bibinfo {author} {\bibfnamefont {X.}~\bibnamefont {Zhao}}, \bibinfo
  {author} {\bibfnamefont {B.}~\bibnamefont {Liu}}, \ and\ \bibinfo {author}
  {\bibfnamefont {T.}~\bibnamefont {Cui}},\ }\href@noop {} {\bibfield
  {journal} {\bibinfo  {journal} {The Journal of chemical physics}\ }\textbf
  {\bibinfo {volume} {133}},\ \bibinfo {pages} {074509} (\bibinfo {year}
  {2010})}\BibitemShut {NoStop}%
\bibitem [{\citenamefont {Gao}\ \emph {et~al.}(2011)\citenamefont {Gao},
  \citenamefont {Wang}, \citenamefont {Bergara}, \citenamefont {Li},
  \citenamefont {Liu},\ and\ \citenamefont {Ma}}]{2011GAO}%
  \BibitemOpen
  \bibfield  {author} {\bibinfo {author} {\bibfnamefont {G.}~\bibnamefont
  {Gao}}, \bibinfo {author} {\bibfnamefont {H.}~\bibnamefont {Wang}}, \bibinfo
  {author} {\bibfnamefont {A.}~\bibnamefont {Bergara}}, \bibinfo {author}
  {\bibfnamefont {Y.}~\bibnamefont {Li}}, \bibinfo {author} {\bibfnamefont
  {G.}~\bibnamefont {Liu}}, \ and\ \bibinfo {author} {\bibfnamefont
  {Y.}~\bibnamefont {Ma}},\ }\href@noop {} {\bibfield  {journal} {\bibinfo
  {journal} {Physical Review B}\ }\textbf {\bibinfo {volume} {84}},\ \bibinfo
  {pages} {064118} (\bibinfo {year} {2011})}\BibitemShut {NoStop}%
\bibitem [{\citenamefont {Kim}\ \emph {et~al.}(2011)\citenamefont {Kim},
  \citenamefont {Scheicher}, \citenamefont {Pickard}, \citenamefont {Needs},\
  and\ \citenamefont {Ahuja}}]{2011KIM}%
  \BibitemOpen
  \bibfield  {author} {\bibinfo {author} {\bibfnamefont {D.~Y.}\ \bibnamefont
  {Kim}}, \bibinfo {author} {\bibfnamefont {R.~H.}\ \bibnamefont {Scheicher}},
  \bibinfo {author} {\bibfnamefont {C.~J.}\ \bibnamefont {Pickard}}, \bibinfo
  {author} {\bibfnamefont {R.}~\bibnamefont {Needs}}, \ and\ \bibinfo {author}
  {\bibfnamefont {R.}~\bibnamefont {Ahuja}},\ }\href@noop {} {\bibfield
  {journal} {\bibinfo  {journal} {Physical review letters}\ }\textbf {\bibinfo
  {volume} {107}},\ \bibinfo {pages} {117002} (\bibinfo {year}
  {2011})}\BibitemShut {NoStop}%
\bibitem [{\citenamefont {Abe}\ and\ \citenamefont {Ashcroft}(2011)}]{2011ABE}%
  \BibitemOpen
  \bibfield  {author} {\bibinfo {author} {\bibfnamefont {K.}~\bibnamefont
  {Abe}}\ and\ \bibinfo {author} {\bibfnamefont {N.}~\bibnamefont {Ashcroft}},\
  }\href@noop {} {\bibfield  {journal} {\bibinfo  {journal} {Physical Review
  B}\ }\textbf {\bibinfo {volume} {84}},\ \bibinfo {pages} {104118} (\bibinfo
  {year} {2011})}\BibitemShut {NoStop}%
\bibitem [{\citenamefont {Zhong}\ \emph {et~al.}(2012)\citenamefont {Zhong},
  \citenamefont {Zhang}, \citenamefont {Chen}, \citenamefont {Li},
  \citenamefont {Zhang},\ and\ \citenamefont {Lin}}]{2012ZHO_a}%
  \BibitemOpen
  \bibfield  {author} {\bibinfo {author} {\bibfnamefont {G.}~\bibnamefont
  {Zhong}}, \bibinfo {author} {\bibfnamefont {C.}~\bibnamefont {Zhang}},
  \bibinfo {author} {\bibfnamefont {X.}~\bibnamefont {Chen}}, \bibinfo {author}
  {\bibfnamefont {Y.}~\bibnamefont {Li}}, \bibinfo {author} {\bibfnamefont
  {R.}~\bibnamefont {Zhang}}, \ and\ \bibinfo {author} {\bibfnamefont
  {H.}~\bibnamefont {Lin}},\ }\href@noop {} {\bibfield  {journal} {\bibinfo
  {journal} {The Journal of Physical Chemistry C}\ }\textbf {\bibinfo {volume}
  {116}},\ \bibinfo {pages} {5225} (\bibinfo {year} {2012})}\BibitemShut
  {NoStop}%
\bibitem [{\citenamefont {Zhang}\ \emph {et~al.}(2011)\citenamefont {Zhang},
  \citenamefont {Chen},\ and\ \citenamefont {Lin}}]{2011ZHA}%
  \BibitemOpen
  \bibfield  {author} {\bibinfo {author} {\bibfnamefont {C.}~\bibnamefont
  {Zhang}}, \bibinfo {author} {\bibfnamefont {X.-J.}\ \bibnamefont {Chen}}, \
  and\ \bibinfo {author} {\bibfnamefont {H.-Q.}\ \bibnamefont {Lin}},\
  }\href@noop {} {\bibfield  {journal} {\bibinfo  {journal} {Journal of
  Physics: Condensed Matter}\ }\textbf {\bibinfo {volume} {24}},\ \bibinfo
  {pages} {035701} (\bibinfo {year} {2011})}\BibitemShut {NoStop}%
\bibitem [{\citenamefont {Zhou}\ \emph {et~al.}(2012)\citenamefont {Zhou},
  \citenamefont {Jin}, \citenamefont {Meng}, \citenamefont {Bao}, \citenamefont
  {Ma}, \citenamefont {Liu},\ and\ \citenamefont {Cui}}]{2012ZHO_b}%
  \BibitemOpen
  \bibfield  {author} {\bibinfo {author} {\bibfnamefont {D.}~\bibnamefont
  {Zhou}}, \bibinfo {author} {\bibfnamefont {X.}~\bibnamefont {Jin}}, \bibinfo
  {author} {\bibfnamefont {X.}~\bibnamefont {Meng}}, \bibinfo {author}
  {\bibfnamefont {G.}~\bibnamefont {Bao}}, \bibinfo {author} {\bibfnamefont
  {Y.}~\bibnamefont {Ma}}, \bibinfo {author} {\bibfnamefont {B.}~\bibnamefont
  {Liu}}, \ and\ \bibinfo {author} {\bibfnamefont {T.}~\bibnamefont {Cui}},\
  }\href@noop {} {\bibfield  {journal} {\bibinfo  {journal} {Physical Review
  B}\ }\textbf {\bibinfo {volume} {86}},\ \bibinfo {pages} {014118} (\bibinfo
  {year} {2012})}\BibitemShut {NoStop}%
\bibitem [{\citenamefont {Gao}\ \emph {et~al.}(2013)\citenamefont {Gao},
  \citenamefont {Hoffmann}, \citenamefont {Ashcroft}, \citenamefont {Liu},
  \citenamefont {Bergara},\ and\ \citenamefont {Ma}}]{2013GAO}%
  \BibitemOpen
  \bibfield  {author} {\bibinfo {author} {\bibfnamefont {G.}~\bibnamefont
  {Gao}}, \bibinfo {author} {\bibfnamefont {R.}~\bibnamefont {Hoffmann}},
  \bibinfo {author} {\bibfnamefont {N.~W.}\ \bibnamefont {Ashcroft}}, \bibinfo
  {author} {\bibfnamefont {H.}~\bibnamefont {Liu}}, \bibinfo {author}
  {\bibfnamefont {A.}~\bibnamefont {Bergara}}, \ and\ \bibinfo {author}
  {\bibfnamefont {Y.}~\bibnamefont {Ma}},\ }\href@noop {} {\bibfield  {journal}
  {\bibinfo  {journal} {Physical Review B}\ }\textbf {\bibinfo {volume} {88}},\
  \bibinfo {pages} {184104} (\bibinfo {year} {2013})}\BibitemShut {NoStop}%
\bibitem [{\citenamefont {Abe}\ and\ \citenamefont {Ashcroft}(2013)}]{2013ABE}%
  \BibitemOpen
  \bibfield  {author} {\bibinfo {author} {\bibfnamefont {K.}~\bibnamefont
  {Abe}}\ and\ \bibinfo {author} {\bibfnamefont {N.}~\bibnamefont {Ashcroft}},\
  }\href@noop {} {\bibfield  {journal} {\bibinfo  {journal} {Physical Review
  B}\ }\textbf {\bibinfo {volume} {88}},\ \bibinfo {pages} {174110} (\bibinfo
  {year} {2013})}\BibitemShut {NoStop}%
\bibitem [{\citenamefont {Hooper}\ \emph {et~al.}(2013)\citenamefont {Hooper},
  \citenamefont {Altintas}, \citenamefont {Shamp},\ and\ \citenamefont
  {Zurek}}]{2013HOO}%
  \BibitemOpen
  \bibfield  {author} {\bibinfo {author} {\bibfnamefont {J.}~\bibnamefont
  {Hooper}}, \bibinfo {author} {\bibfnamefont {B.}~\bibnamefont {Altintas}},
  \bibinfo {author} {\bibfnamefont {A.}~\bibnamefont {Shamp}}, \ and\ \bibinfo
  {author} {\bibfnamefont {E.}~\bibnamefont {Zurek}},\ }\href@noop {}
  {\bibfield  {journal} {\bibinfo  {journal} {The Journal of Physical Chemistry
  C}\ }\textbf {\bibinfo {volume} {117}},\ \bibinfo {pages} {2982} (\bibinfo
  {year} {2013})}\BibitemShut {NoStop}%
\bibitem [{\citenamefont {Lonie}\ \emph {et~al.}(2013)\citenamefont {Lonie},
  \citenamefont {Hooper}, \citenamefont {Altintas},\ and\ \citenamefont
  {Zurek}}]{2013LON}%
  \BibitemOpen
  \bibfield  {author} {\bibinfo {author} {\bibfnamefont {D.~C.}\ \bibnamefont
  {Lonie}}, \bibinfo {author} {\bibfnamefont {J.}~\bibnamefont {Hooper}},
  \bibinfo {author} {\bibfnamefont {B.}~\bibnamefont {Altintas}}, \ and\
  \bibinfo {author} {\bibfnamefont {E.}~\bibnamefont {Zurek}},\ }\href@noop {}
  {\bibfield  {journal} {\bibinfo  {journal} {Physical Review B}\ }\textbf
  {\bibinfo {volume} {87}},\ \bibinfo {pages} {054107} (\bibinfo {year}
  {2013})}\BibitemShut {NoStop}%
\bibitem [{\citenamefont {Hu}\ \emph {et~al.}(2013)\citenamefont {Hu},
  \citenamefont {Oganov}, \citenamefont {Zhu}, \citenamefont {Qian},
  \citenamefont {Frapper}, \citenamefont {Lyakhov},\ and\ \citenamefont
  {Zhou}}]{2013HU}%
  \BibitemOpen
  \bibfield  {author} {\bibinfo {author} {\bibfnamefont {C.-H.}\ \bibnamefont
  {Hu}}, \bibinfo {author} {\bibfnamefont {A.~R.}\ \bibnamefont {Oganov}},
  \bibinfo {author} {\bibfnamefont {Q.}~\bibnamefont {Zhu}}, \bibinfo {author}
  {\bibfnamefont {G.-R.}\ \bibnamefont {Qian}}, \bibinfo {author}
  {\bibfnamefont {G.}~\bibnamefont {Frapper}}, \bibinfo {author} {\bibfnamefont
  {A.~O.}\ \bibnamefont {Lyakhov}}, \ and\ \bibinfo {author} {\bibfnamefont
  {H.-Y.}\ \bibnamefont {Zhou}},\ }\href@noop {} {\bibfield  {journal}
  {\bibinfo  {journal} {Physical review letters}\ }\textbf {\bibinfo {volume}
  {110}},\ \bibinfo {pages} {165504} (\bibinfo {year} {2013})}\BibitemShut
  {NoStop}%
\bibitem [{\citenamefont {Xie}\ \emph {et~al.}(2014)\citenamefont {Xie},
  \citenamefont {Li}, \citenamefont {Oganov},\ and\ \citenamefont
  {Wang}}]{2014XIE}%
  \BibitemOpen
  \bibfield  {author} {\bibinfo {author} {\bibfnamefont {Y.}~\bibnamefont
  {Xie}}, \bibinfo {author} {\bibfnamefont {Q.}~\bibnamefont {Li}}, \bibinfo
  {author} {\bibfnamefont {A.~R.}\ \bibnamefont {Oganov}}, \ and\ \bibinfo
  {author} {\bibfnamefont {H.}~\bibnamefont {Wang}},\ }\href@noop {} {\bibfield
   {journal} {\bibinfo  {journal} {Acta Crystallographica Section C: Structural
  Chemistry}\ }\textbf {\bibinfo {volume} {70}},\ \bibinfo {pages} {104}
  (\bibinfo {year} {2014})}\BibitemShut {NoStop}%
\bibitem [{\citenamefont {Yu}\ \emph {et~al.}(2014)\citenamefont {Yu},
  \citenamefont {Zeng}, \citenamefont {Oganov}, \citenamefont {Hu},
  \citenamefont {Frapper},\ and\ \citenamefont {Zhang}}]{2014YU}%
  \BibitemOpen
  \bibfield  {author} {\bibinfo {author} {\bibfnamefont {S.}~\bibnamefont
  {Yu}}, \bibinfo {author} {\bibfnamefont {Q.}~\bibnamefont {Zeng}}, \bibinfo
  {author} {\bibfnamefont {A.~R.}\ \bibnamefont {Oganov}}, \bibinfo {author}
  {\bibfnamefont {C.}~\bibnamefont {Hu}}, \bibinfo {author} {\bibfnamefont
  {G.}~\bibnamefont {Frapper}}, \ and\ \bibinfo {author} {\bibfnamefont
  {L.}~\bibnamefont {Zhang}},\ }\href@noop {} {\bibfield  {journal} {\bibinfo
  {journal} {AIP Advances}\ }\textbf {\bibinfo {volume} {4}},\ \bibinfo {pages}
  {107118} (\bibinfo {year} {2014})}\BibitemShut {NoStop}%
\bibitem [{\citenamefont {Li}\ \emph {et~al.}(2014)\citenamefont {Li},
  \citenamefont {Hao}, \citenamefont {Liu}, \citenamefont {Li},\ and\
  \citenamefont {Ma}}]{2014LI}%
  \BibitemOpen
  \bibfield  {author} {\bibinfo {author} {\bibfnamefont {Y.}~\bibnamefont
  {Li}}, \bibinfo {author} {\bibfnamefont {J.}~\bibnamefont {Hao}}, \bibinfo
  {author} {\bibfnamefont {H.}~\bibnamefont {Liu}}, \bibinfo {author}
  {\bibfnamefont {Y.}~\bibnamefont {Li}}, \ and\ \bibinfo {author}
  {\bibfnamefont {Y.}~\bibnamefont {Ma}},\ }\href@noop {} {\bibfield  {journal}
  {\bibinfo  {journal} {The Journal of chemical physics}\ }\textbf {\bibinfo
  {volume} {140}},\ \bibinfo {pages} {174712} (\bibinfo {year}
  {2014})}\BibitemShut {NoStop}%
\bibitem [{\citenamefont {Wang}\ \emph {et~al.}(2014)\citenamefont {Wang},
  \citenamefont {Yao}, \citenamefont {Zhu}, \citenamefont {Liu}, \citenamefont
  {Iitaka}, \citenamefont {Wang},\ and\ \citenamefont {Ma}}]{2014WAN}%
  \BibitemOpen
  \bibfield  {author} {\bibinfo {author} {\bibfnamefont {Z.}~\bibnamefont
  {Wang}}, \bibinfo {author} {\bibfnamefont {Y.}~\bibnamefont {Yao}}, \bibinfo
  {author} {\bibfnamefont {L.}~\bibnamefont {Zhu}}, \bibinfo {author}
  {\bibfnamefont {H.}~\bibnamefont {Liu}}, \bibinfo {author} {\bibfnamefont
  {T.}~\bibnamefont {Iitaka}}, \bibinfo {author} {\bibfnamefont
  {H.}~\bibnamefont {Wang}}, \ and\ \bibinfo {author} {\bibfnamefont
  {Y.}~\bibnamefont {Ma}},\ }\href@noop {} {\bibfield  {journal} {\bibinfo
  {journal} {The Journal of chemical physics}\ }\textbf {\bibinfo {volume}
  {140}},\ \bibinfo {pages} {124707} (\bibinfo {year} {2014})}\BibitemShut
  {NoStop}%
\bibitem [{\citenamefont {Chen}\ \emph {et~al.}(2014)\citenamefont {Chen},
  \citenamefont {Tian}, \citenamefont {Duan}, \citenamefont {Bao},
  \citenamefont {Jin}, \citenamefont {Liu},\ and\ \citenamefont
  {Cui}}]{2014CHE}%
  \BibitemOpen
  \bibfield  {author} {\bibinfo {author} {\bibfnamefont {C.}~\bibnamefont
  {Chen}}, \bibinfo {author} {\bibfnamefont {F.}~\bibnamefont {Tian}}, \bibinfo
  {author} {\bibfnamefont {D.}~\bibnamefont {Duan}}, \bibinfo {author}
  {\bibfnamefont {K.}~\bibnamefont {Bao}}, \bibinfo {author} {\bibfnamefont
  {X.}~\bibnamefont {Jin}}, \bibinfo {author} {\bibfnamefont {B.}~\bibnamefont
  {Liu}}, \ and\ \bibinfo {author} {\bibfnamefont {T.}~\bibnamefont {Cui}},\
  }\href@noop {} {\bibfield  {journal} {\bibinfo  {journal} {The Journal of
  Chemical Physics}\ }\textbf {\bibinfo {volume} {140}},\ \bibinfo {pages}
  {114703} (\bibinfo {year} {2014})}\BibitemShut {NoStop}%
\bibitem [{\citenamefont {Duan}\ \emph {et~al.}(2014)\citenamefont {Duan},
  \citenamefont {Liu}, \citenamefont {Tian}, \citenamefont {Li}, \citenamefont
  {Huang}, \citenamefont {Zhao}, \citenamefont {Yu}, \citenamefont {Liu},
  \citenamefont {Tian},\ and\ \citenamefont {Cui}}]{2014DUA}%
  \BibitemOpen
  \bibfield  {author} {\bibinfo {author} {\bibfnamefont {D.}~\bibnamefont
  {Duan}}, \bibinfo {author} {\bibfnamefont {Y.}~\bibnamefont {Liu}}, \bibinfo
  {author} {\bibfnamefont {F.}~\bibnamefont {Tian}}, \bibinfo {author}
  {\bibfnamefont {D.}~\bibnamefont {Li}}, \bibinfo {author} {\bibfnamefont
  {X.}~\bibnamefont {Huang}}, \bibinfo {author} {\bibfnamefont
  {Z.}~\bibnamefont {Zhao}}, \bibinfo {author} {\bibfnamefont {H.}~\bibnamefont
  {Yu}}, \bibinfo {author} {\bibfnamefont {B.}~\bibnamefont {Liu}}, \bibinfo
  {author} {\bibfnamefont {W.}~\bibnamefont {Tian}}, \ and\ \bibinfo {author}
  {\bibfnamefont {T.}~\bibnamefont {Cui}},\ }\href@noop {} {\bibfield
  {journal} {\bibinfo  {journal} {Scientific reports}\ }\textbf {\bibinfo
  {volume} {4}},\ \bibinfo {pages} {6968} (\bibinfo {year} {2014})}\BibitemShut
  {NoStop}%
\bibitem [{\citenamefont {Yan}\ \emph {et~al.}(2015)\citenamefont {Yan},
  \citenamefont {Chen}, \citenamefont {Kuang},\ and\ \citenamefont
  {Xiang}}]{2015YAN}%
  \BibitemOpen
  \bibfield  {author} {\bibinfo {author} {\bibfnamefont {X.}~\bibnamefont
  {Yan}}, \bibinfo {author} {\bibfnamefont {Y.}~\bibnamefont {Chen}}, \bibinfo
  {author} {\bibfnamefont {X.}~\bibnamefont {Kuang}}, \ and\ \bibinfo {author}
  {\bibfnamefont {S.}~\bibnamefont {Xiang}},\ }\href@noop {} {\bibfield
  {journal} {\bibinfo  {journal} {The Journal of chemical physics}\ }\textbf
  {\bibinfo {volume} {143}},\ \bibinfo {pages} {124310} (\bibinfo {year}
  {2015})}\BibitemShut {NoStop}%
\bibitem [{\citenamefont {Hou}\ \emph {et~al.}(2015)\citenamefont {Hou},
  \citenamefont {Zhao}, \citenamefont {Tian}, \citenamefont {Li}, \citenamefont
  {Duan}, \citenamefont {Zhao}, \citenamefont {Chu}, \citenamefont {Liu},\ and\
  \citenamefont {Cui}}]{2015HOU}%
  \BibitemOpen
  \bibfield  {author} {\bibinfo {author} {\bibfnamefont {P.}~\bibnamefont
  {Hou}}, \bibinfo {author} {\bibfnamefont {X.}~\bibnamefont {Zhao}}, \bibinfo
  {author} {\bibfnamefont {F.}~\bibnamefont {Tian}}, \bibinfo {author}
  {\bibfnamefont {D.}~\bibnamefont {Li}}, \bibinfo {author} {\bibfnamefont
  {D.}~\bibnamefont {Duan}}, \bibinfo {author} {\bibfnamefont {Z.}~\bibnamefont
  {Zhao}}, \bibinfo {author} {\bibfnamefont {B.}~\bibnamefont {Chu}}, \bibinfo
  {author} {\bibfnamefont {B.}~\bibnamefont {Liu}}, \ and\ \bibinfo {author}
  {\bibfnamefont {T.}~\bibnamefont {Cui}},\ }\href@noop {} {\bibfield
  {journal} {\bibinfo  {journal} {RSC advances}\ }\textbf {\bibinfo {volume}
  {5}},\ \bibinfo {pages} {5096} (\bibinfo {year} {2015})}\BibitemShut
  {NoStop}%
\bibitem [{\citenamefont {Duan}\ \emph {et~al.}(2015)\citenamefont {Duan},
  \citenamefont {Tian}, \citenamefont {Liu}, \citenamefont {Huang},
  \citenamefont {Li}, \citenamefont {Yu}, \citenamefont {Ma}, \citenamefont
  {Liu},\ and\ \citenamefont {Cui}}]{2015DUA}%
  \BibitemOpen
  \bibfield  {author} {\bibinfo {author} {\bibfnamefont {D.}~\bibnamefont
  {Duan}}, \bibinfo {author} {\bibfnamefont {F.}~\bibnamefont {Tian}}, \bibinfo
  {author} {\bibfnamefont {Y.}~\bibnamefont {Liu}}, \bibinfo {author}
  {\bibfnamefont {X.}~\bibnamefont {Huang}}, \bibinfo {author} {\bibfnamefont
  {D.}~\bibnamefont {Li}}, \bibinfo {author} {\bibfnamefont {H.}~\bibnamefont
  {Yu}}, \bibinfo {author} {\bibfnamefont {Y.}~\bibnamefont {Ma}}, \bibinfo
  {author} {\bibfnamefont {B.}~\bibnamefont {Liu}}, \ and\ \bibinfo {author}
  {\bibfnamefont {T.}~\bibnamefont {Cui}},\ }\href@noop {} {\bibfield
  {journal} {\bibinfo  {journal} {Physical Chemistry Chemical Physics}\
  }\textbf {\bibinfo {volume} {17}},\ \bibinfo {pages} {32335} (\bibinfo {year}
  {2015})}\BibitemShut {NoStop}%
\bibitem [{\citenamefont {Zhang}\ \emph
  {et~al.}(2015{\natexlab{a}})\citenamefont {Zhang}, \citenamefont {Jin},
  \citenamefont {Lv}, \citenamefont {Zhuang}, \citenamefont {Liu},
  \citenamefont {Lv}, \citenamefont {Bao}, \citenamefont {Li}, \citenamefont
  {Liu},\ and\ \citenamefont {Cui}}]{2015ZHA_a}%
  \BibitemOpen
  \bibfield  {author} {\bibinfo {author} {\bibfnamefont {H.}~\bibnamefont
  {Zhang}}, \bibinfo {author} {\bibfnamefont {X.}~\bibnamefont {Jin}}, \bibinfo
  {author} {\bibfnamefont {Y.}~\bibnamefont {Lv}}, \bibinfo {author}
  {\bibfnamefont {Q.}~\bibnamefont {Zhuang}}, \bibinfo {author} {\bibfnamefont
  {Y.}~\bibnamefont {Liu}}, \bibinfo {author} {\bibfnamefont {Q.}~\bibnamefont
  {Lv}}, \bibinfo {author} {\bibfnamefont {K.}~\bibnamefont {Bao}}, \bibinfo
  {author} {\bibfnamefont {D.}~\bibnamefont {Li}}, \bibinfo {author}
  {\bibfnamefont {B.}~\bibnamefont {Liu}}, \ and\ \bibinfo {author}
  {\bibfnamefont {T.}~\bibnamefont {Cui}},\ }\href@noop {} {\bibfield
  {journal} {\bibinfo  {journal} {Scientific reports}\ }\textbf {\bibinfo
  {volume} {5}},\ \bibinfo {pages} {8845} (\bibinfo {year}
  {2015}{\natexlab{a}})}\BibitemShut {NoStop}%
\bibitem [{\citenamefont {Liu}\ \emph {et~al.}(2015{\natexlab{a}})\citenamefont
  {Liu}, \citenamefont {Huang}, \citenamefont {Duan}, \citenamefont {Tian},
  \citenamefont {Liu}, \citenamefont {Li}, \citenamefont {Zhao}, \citenamefont
  {Sha}, \citenamefont {Yu}, \citenamefont {Zhang} \emph {et~al.}}]{2015LIU_a}%
  \BibitemOpen
  \bibfield  {author} {\bibinfo {author} {\bibfnamefont {Y.}~\bibnamefont
  {Liu}}, \bibinfo {author} {\bibfnamefont {X.}~\bibnamefont {Huang}}, \bibinfo
  {author} {\bibfnamefont {D.}~\bibnamefont {Duan}}, \bibinfo {author}
  {\bibfnamefont {F.}~\bibnamefont {Tian}}, \bibinfo {author} {\bibfnamefont
  {H.}~\bibnamefont {Liu}}, \bibinfo {author} {\bibfnamefont {D.}~\bibnamefont
  {Li}}, \bibinfo {author} {\bibfnamefont {Z.}~\bibnamefont {Zhao}}, \bibinfo
  {author} {\bibfnamefont {X.}~\bibnamefont {Sha}}, \bibinfo {author}
  {\bibfnamefont {H.}~\bibnamefont {Yu}}, \bibinfo {author} {\bibfnamefont
  {H.}~\bibnamefont {Zhang}},  \emph {et~al.},\ }\href@noop {} {\bibfield
  {journal} {\bibinfo  {journal} {Scientific reports}\ }\textbf {\bibinfo
  {volume} {5}},\ \bibinfo {pages} {11381} (\bibinfo {year}
  {2015}{\natexlab{a}})}\BibitemShut {NoStop}%
\bibitem [{\citenamefont {Shamp}\ and\ \citenamefont {Zurek}(2015)}]{2015SHA}%
  \BibitemOpen
  \bibfield  {author} {\bibinfo {author} {\bibfnamefont {A.}~\bibnamefont
  {Shamp}}\ and\ \bibinfo {author} {\bibfnamefont {E.}~\bibnamefont {Zurek}},\
  }\href@noop {} {\bibfield  {journal} {\bibinfo  {journal} {The Journal of
  Physical Chemistry Letters}\ }\textbf {\bibinfo {volume} {6}},\ \bibinfo
  {pages} {4067} (\bibinfo {year} {2015})}\BibitemShut {NoStop}%
\bibitem [{\citenamefont {Liu}\ \emph {et~al.}(2015{\natexlab{b}})\citenamefont
  {Liu}, \citenamefont {Duan}, \citenamefont {Huang}, \citenamefont {Tian},
  \citenamefont {Li}, \citenamefont {Sha}, \citenamefont {Wang}, \citenamefont
  {Zhang}, \citenamefont {Yang}, \citenamefont {Liu} \emph
  {et~al.}}]{2015LIU_b}%
  \BibitemOpen
  \bibfield  {author} {\bibinfo {author} {\bibfnamefont {Y.}~\bibnamefont
  {Liu}}, \bibinfo {author} {\bibfnamefont {D.}~\bibnamefont {Duan}}, \bibinfo
  {author} {\bibfnamefont {X.}~\bibnamefont {Huang}}, \bibinfo {author}
  {\bibfnamefont {F.}~\bibnamefont {Tian}}, \bibinfo {author} {\bibfnamefont
  {D.}~\bibnamefont {Li}}, \bibinfo {author} {\bibfnamefont {X.}~\bibnamefont
  {Sha}}, \bibinfo {author} {\bibfnamefont {C.}~\bibnamefont {Wang}}, \bibinfo
  {author} {\bibfnamefont {H.}~\bibnamefont {Zhang}}, \bibinfo {author}
  {\bibfnamefont {T.}~\bibnamefont {Yang}}, \bibinfo {author} {\bibfnamefont
  {B.}~\bibnamefont {Liu}},  \emph {et~al.},\ }\href@noop {} {\bibfield
  {journal} {\bibinfo  {journal} {The Journal of Physical Chemistry C}\
  }\textbf {\bibinfo {volume} {119}},\ \bibinfo {pages} {15905} (\bibinfo
  {year} {2015}{\natexlab{b}})}\BibitemShut {NoStop}%
\bibitem [{\citenamefont {Chen}\ \emph
  {et~al.}(2015{\natexlab{a}})\citenamefont {Chen}, \citenamefont {Xu},
  \citenamefont {Sun},\ and\ \citenamefont {Wang}}]{2015CHE_a}%
  \BibitemOpen
  \bibfield  {author} {\bibinfo {author} {\bibfnamefont {C.}~\bibnamefont
  {Chen}}, \bibinfo {author} {\bibfnamefont {Y.}~\bibnamefont {Xu}}, \bibinfo
  {author} {\bibfnamefont {X.}~\bibnamefont {Sun}}, \ and\ \bibinfo {author}
  {\bibfnamefont {S.}~\bibnamefont {Wang}},\ }\href@noop {} {\bibfield
  {journal} {\bibinfo  {journal} {The Journal of Physical Chemistry C}\
  }\textbf {\bibinfo {volume} {119}},\ \bibinfo {pages} {17039} (\bibinfo
  {year} {2015}{\natexlab{a}})}\BibitemShut {NoStop}%
\bibitem [{\citenamefont {Feng}\ \emph {et~al.}(2015)\citenamefont {Feng},
  \citenamefont {Zhang}, \citenamefont {Gao}, \citenamefont {Liu},\ and\
  \citenamefont {Wang}}]{2015FEN}%
  \BibitemOpen
  \bibfield  {author} {\bibinfo {author} {\bibfnamefont {X.}~\bibnamefont
  {Feng}}, \bibinfo {author} {\bibfnamefont {J.}~\bibnamefont {Zhang}},
  \bibinfo {author} {\bibfnamefont {G.}~\bibnamefont {Gao}}, \bibinfo {author}
  {\bibfnamefont {H.}~\bibnamefont {Liu}}, \ and\ \bibinfo {author}
  {\bibfnamefont {H.}~\bibnamefont {Wang}},\ }\href@noop {} {\bibfield
  {journal} {\bibinfo  {journal} {RSC advances}\ }\textbf {\bibinfo {volume}
  {5}},\ \bibinfo {pages} {59292} (\bibinfo {year} {2015})}\BibitemShut
  {NoStop}%
\bibitem [{\citenamefont {Zhang}\ \emph
  {et~al.}(2015{\natexlab{b}})\citenamefont {Zhang}, \citenamefont {Jin},
  \citenamefont {Lv}, \citenamefont {Zhuang}, \citenamefont {Lv}, \citenamefont
  {Liu}, \citenamefont {Bao}, \citenamefont {Li}, \citenamefont {Liu},\ and\
  \citenamefont {Cui}}]{2015ZHA_b}%
  \BibitemOpen
  \bibfield  {author} {\bibinfo {author} {\bibfnamefont {H.}~\bibnamefont
  {Zhang}}, \bibinfo {author} {\bibfnamefont {X.}~\bibnamefont {Jin}}, \bibinfo
  {author} {\bibfnamefont {Y.}~\bibnamefont {Lv}}, \bibinfo {author}
  {\bibfnamefont {Q.}~\bibnamefont {Zhuang}}, \bibinfo {author} {\bibfnamefont
  {Q.}~\bibnamefont {Lv}}, \bibinfo {author} {\bibfnamefont {Y.}~\bibnamefont
  {Liu}}, \bibinfo {author} {\bibfnamefont {K.}~\bibnamefont {Bao}}, \bibinfo
  {author} {\bibfnamefont {D.}~\bibnamefont {Li}}, \bibinfo {author}
  {\bibfnamefont {B.}~\bibnamefont {Liu}}, \ and\ \bibinfo {author}
  {\bibfnamefont {T.}~\bibnamefont {Cui}},\ }\href@noop {} {\bibfield
  {journal} {\bibinfo  {journal} {Physical Chemistry Chemical Physics}\
  }\textbf {\bibinfo {volume} {17}},\ \bibinfo {pages} {27630} (\bibinfo {year}
  {2015}{\natexlab{b}})}\BibitemShut {NoStop}%
\bibitem [{\citenamefont {Liu}\ \emph {et~al.}(2015{\natexlab{c}})\citenamefont
  {Liu}, \citenamefont {Duan}, \citenamefont {Tian}, \citenamefont {Liu},
  \citenamefont {Wang}, \citenamefont {Huang}, \citenamefont {Li},
  \citenamefont {Ma}, \citenamefont {Liu},\ and\ \citenamefont
  {Cui}}]{2015LIU}%
  \BibitemOpen
  \bibfield  {author} {\bibinfo {author} {\bibfnamefont {Y.}~\bibnamefont
  {Liu}}, \bibinfo {author} {\bibfnamefont {D.}~\bibnamefont {Duan}}, \bibinfo
  {author} {\bibfnamefont {F.}~\bibnamefont {Tian}}, \bibinfo {author}
  {\bibfnamefont {H.}~\bibnamefont {Liu}}, \bibinfo {author} {\bibfnamefont
  {C.}~\bibnamefont {Wang}}, \bibinfo {author} {\bibfnamefont {X.}~\bibnamefont
  {Huang}}, \bibinfo {author} {\bibfnamefont {D.}~\bibnamefont {Li}}, \bibinfo
  {author} {\bibfnamefont {Y.}~\bibnamefont {Ma}}, \bibinfo {author}
  {\bibfnamefont {B.}~\bibnamefont {Liu}}, \ and\ \bibinfo {author}
  {\bibfnamefont {T.}~\bibnamefont {Cui}},\ }\href@noop {} {\bibfield
  {journal} {\bibinfo  {journal} {Inorganic chemistry}\ }\textbf {\bibinfo
  {volume} {54}},\ \bibinfo {pages} {9924} (\bibinfo {year}
  {2015}{\natexlab{c}})}\BibitemShut {NoStop}%
\bibitem [{\citenamefont {Yu}\ \emph {et~al.}(2015)\citenamefont {Yu},
  \citenamefont {Jia}, \citenamefont {Frapper}, \citenamefont {Li},
  \citenamefont {Oganov}, \citenamefont {Zeng},\ and\ \citenamefont
  {Zhang}}]{2015YU}%
  \BibitemOpen
  \bibfield  {author} {\bibinfo {author} {\bibfnamefont {S.}~\bibnamefont
  {Yu}}, \bibinfo {author} {\bibfnamefont {X.}~\bibnamefont {Jia}}, \bibinfo
  {author} {\bibfnamefont {G.}~\bibnamefont {Frapper}}, \bibinfo {author}
  {\bibfnamefont {D.}~\bibnamefont {Li}}, \bibinfo {author} {\bibfnamefont
  {A.~R.}\ \bibnamefont {Oganov}}, \bibinfo {author} {\bibfnamefont
  {Q.}~\bibnamefont {Zeng}}, \ and\ \bibinfo {author} {\bibfnamefont
  {L.}~\bibnamefont {Zhang}},\ }\href@noop {} {\bibfield  {journal} {\bibinfo
  {journal} {Scientific reports}\ }\textbf {\bibinfo {volume} {5}},\ \bibinfo
  {pages} {17764} (\bibinfo {year} {2015})}\BibitemShut {NoStop}%
\bibitem [{\citenamefont {Cheng}\ \emph {et~al.}(2015)\citenamefont {Cheng},
  \citenamefont {Zhang}, \citenamefont {Wang}, \citenamefont {Zhong},
  \citenamefont {Yang}, \citenamefont {Chen},\ and\ \citenamefont
  {Lin}}]{2015CHE_b}%
  \BibitemOpen
  \bibfield  {author} {\bibinfo {author} {\bibfnamefont {Y.}~\bibnamefont
  {Cheng}}, \bibinfo {author} {\bibfnamefont {C.}~\bibnamefont {Zhang}},
  \bibinfo {author} {\bibfnamefont {T.}~\bibnamefont {Wang}}, \bibinfo {author}
  {\bibfnamefont {G.}~\bibnamefont {Zhong}}, \bibinfo {author} {\bibfnamefont
  {C.}~\bibnamefont {Yang}}, \bibinfo {author} {\bibfnamefont {X.-J.}\
  \bibnamefont {Chen}}, \ and\ \bibinfo {author} {\bibfnamefont {H.-Q.}\
  \bibnamefont {Lin}},\ }\href@noop {} {\bibfield  {journal} {\bibinfo
  {journal} {Scientific reports}\ }\textbf {\bibinfo {volume} {5}},\ \bibinfo
  {pages} {16475} (\bibinfo {year} {2015})}\BibitemShut {NoStop}%
\bibitem [{\citenamefont {Zhang}\ \emph
  {et~al.}(2015{\natexlab{c}})\citenamefont {Zhang}, \citenamefont {Wang},
  \citenamefont {Zhang}, \citenamefont {Liu}, \citenamefont {Zhong},
  \citenamefont {Song}, \citenamefont {Yang}, \citenamefont {Zhang},\ and\
  \citenamefont {Ma}}]{2015ZHA}%
  \BibitemOpen
  \bibfield  {author} {\bibinfo {author} {\bibfnamefont {S.}~\bibnamefont
  {Zhang}}, \bibinfo {author} {\bibfnamefont {Y.}~\bibnamefont {Wang}},
  \bibinfo {author} {\bibfnamefont {J.}~\bibnamefont {Zhang}}, \bibinfo
  {author} {\bibfnamefont {H.}~\bibnamefont {Liu}}, \bibinfo {author}
  {\bibfnamefont {X.}~\bibnamefont {Zhong}}, \bibinfo {author} {\bibfnamefont
  {H.-F.}\ \bibnamefont {Song}}, \bibinfo {author} {\bibfnamefont
  {G.}~\bibnamefont {Yang}}, \bibinfo {author} {\bibfnamefont {L.}~\bibnamefont
  {Zhang}}, \ and\ \bibinfo {author} {\bibfnamefont {Y.}~\bibnamefont {Ma}},\
  }\href@noop {} {\bibfield  {journal} {\bibinfo  {journal} {Scientific
  reports}\ }\textbf {\bibinfo {volume} {5}},\ \bibinfo {pages} {1} (\bibinfo
  {year} {2015}{\natexlab{c}})}\BibitemShut {NoStop}%
\bibitem [{\citenamefont {Errea}\ \emph {et~al.}(2015)\citenamefont {Errea},
  \citenamefont {Calandra}, \citenamefont {Pickard}, \citenamefont {Nelson},
  \citenamefont {Needs}, \citenamefont {Li}, \citenamefont {Liu}, \citenamefont
  {Zhang}, \citenamefont {Ma},\ and\ \citenamefont {Mauri}}]{2015ERR}%
  \BibitemOpen
  \bibfield  {author} {\bibinfo {author} {\bibfnamefont {I.}~\bibnamefont
  {Errea}}, \bibinfo {author} {\bibfnamefont {M.}~\bibnamefont {Calandra}},
  \bibinfo {author} {\bibfnamefont {C.~J.}\ \bibnamefont {Pickard}}, \bibinfo
  {author} {\bibfnamefont {J.}~\bibnamefont {Nelson}}, \bibinfo {author}
  {\bibfnamefont {R.~J.}\ \bibnamefont {Needs}}, \bibinfo {author}
  {\bibfnamefont {Y.}~\bibnamefont {Li}}, \bibinfo {author} {\bibfnamefont
  {H.}~\bibnamefont {Liu}}, \bibinfo {author} {\bibfnamefont {Y.}~\bibnamefont
  {Zhang}}, \bibinfo {author} {\bibfnamefont {Y.}~\bibnamefont {Ma}}, \ and\
  \bibinfo {author} {\bibfnamefont {F.}~\bibnamefont {Mauri}},\ }\href@noop {}
  {\bibfield  {journal} {\bibinfo  {journal} {Physical Review Letters}\
  }\textbf {\bibinfo {volume} {114}},\ \bibinfo {pages} {157004} (\bibinfo
  {year} {2015})}\BibitemShut {NoStop}%
\bibitem [{\citenamefont {Ishikawa}\ \emph {et~al.}(2016)\citenamefont
  {Ishikawa}, \citenamefont {Nakanishi}, \citenamefont {Shimizu}, \citenamefont
  {Katayama-Yoshida}, \citenamefont {Oda},\ and\ \citenamefont
  {Suzuki}}]{2016ISH}%
  \BibitemOpen
  \bibfield  {author} {\bibinfo {author} {\bibfnamefont {T.}~\bibnamefont
  {Ishikawa}}, \bibinfo {author} {\bibfnamefont {A.}~\bibnamefont {Nakanishi}},
  \bibinfo {author} {\bibfnamefont {K.}~\bibnamefont {Shimizu}}, \bibinfo
  {author} {\bibfnamefont {H.}~\bibnamefont {Katayama-Yoshida}}, \bibinfo
  {author} {\bibfnamefont {T.}~\bibnamefont {Oda}}, \ and\ \bibinfo {author}
  {\bibfnamefont {N.}~\bibnamefont {Suzuki}},\ }\href@noop {} {\bibfield
  {journal} {\bibinfo  {journal} {Scientific reports}\ }\textbf {\bibinfo
  {volume} {6}},\ \bibinfo {pages} {23160} (\bibinfo {year}
  {2016})}\BibitemShut {NoStop}%
\bibitem [{\citenamefont {Esfahani}\ \emph {et~al.}(2016)\citenamefont
  {Esfahani}, \citenamefont {Wang}, \citenamefont {Oganov}, \citenamefont
  {Dong}, \citenamefont {Zhu}, \citenamefont {Wang}, \citenamefont {Rakitin},\
  and\ \citenamefont {Zhou}}]{2016MAH}%
  \BibitemOpen
  \bibfield  {author} {\bibinfo {author} {\bibfnamefont {M.~M.~D.}\
  \bibnamefont {Esfahani}}, \bibinfo {author} {\bibfnamefont {Z.}~\bibnamefont
  {Wang}}, \bibinfo {author} {\bibfnamefont {A.~R.}\ \bibnamefont {Oganov}},
  \bibinfo {author} {\bibfnamefont {H.}~\bibnamefont {Dong}}, \bibinfo {author}
  {\bibfnamefont {Q.}~\bibnamefont {Zhu}}, \bibinfo {author} {\bibfnamefont
  {S.}~\bibnamefont {Wang}}, \bibinfo {author} {\bibfnamefont {M.~S.}\
  \bibnamefont {Rakitin}}, \ and\ \bibinfo {author} {\bibfnamefont {X.-F.}\
  \bibnamefont {Zhou}},\ }\href@noop {} {\bibfield  {journal} {\bibinfo
  {journal} {Scientific reports}\ }\textbf {\bibinfo {volume} {6}},\ \bibinfo
  {pages} {22873} (\bibinfo {year} {2016})}\BibitemShut {NoStop}%
\bibitem [{\citenamefont {Fu}\ \emph {et~al.}(2016)\citenamefont {Fu},
  \citenamefont {Du}, \citenamefont {Zhang}, \citenamefont {Peng},
  \citenamefont {Zhang}, \citenamefont {Pickard}, \citenamefont {Needs},
  \citenamefont {Singh}, \citenamefont {Zheng},\ and\ \citenamefont
  {Ma}}]{2016FU}%
  \BibitemOpen
  \bibfield  {author} {\bibinfo {author} {\bibfnamefont {Y.}~\bibnamefont
  {Fu}}, \bibinfo {author} {\bibfnamefont {X.}~\bibnamefont {Du}}, \bibinfo
  {author} {\bibfnamefont {L.}~\bibnamefont {Zhang}}, \bibinfo {author}
  {\bibfnamefont {F.}~\bibnamefont {Peng}}, \bibinfo {author} {\bibfnamefont
  {M.}~\bibnamefont {Zhang}}, \bibinfo {author} {\bibfnamefont {C.~J.}\
  \bibnamefont {Pickard}}, \bibinfo {author} {\bibfnamefont {R.~J.}\
  \bibnamefont {Needs}}, \bibinfo {author} {\bibfnamefont {D.~J.}\ \bibnamefont
  {Singh}}, \bibinfo {author} {\bibfnamefont {W.}~\bibnamefont {Zheng}}, \ and\
  \bibinfo {author} {\bibfnamefont {Y.}~\bibnamefont {Ma}},\ }\href@noop {}
  {\bibfield  {journal} {\bibinfo  {journal} {Chemistry of Materials}\ }\textbf
  {\bibinfo {volume} {28}},\ \bibinfo {pages} {1746} (\bibinfo {year}
  {2016})}\BibitemShut {NoStop}%
\bibitem [{\citenamefont {Shamp}\ \emph {et~al.}(2016)\citenamefont {Shamp},
  \citenamefont {Terpstra}, \citenamefont {Bi}, \citenamefont {Falls},
  \citenamefont {Avery},\ and\ \citenamefont {Zurek}}]{2016SHA}%
  \BibitemOpen
  \bibfield  {author} {\bibinfo {author} {\bibfnamefont {A.}~\bibnamefont
  {Shamp}}, \bibinfo {author} {\bibfnamefont {T.}~\bibnamefont {Terpstra}},
  \bibinfo {author} {\bibfnamefont {T.}~\bibnamefont {Bi}}, \bibinfo {author}
  {\bibfnamefont {Z.}~\bibnamefont {Falls}}, \bibinfo {author} {\bibfnamefont
  {P.}~\bibnamefont {Avery}}, \ and\ \bibinfo {author} {\bibfnamefont
  {E.}~\bibnamefont {Zurek}},\ }\href@noop {} {\bibfield  {journal} {\bibinfo
  {journal} {Journal of the American Chemical Society}\ }\textbf {\bibinfo
  {volume} {138}},\ \bibinfo {pages} {1884} (\bibinfo {year}
  {2016})}\BibitemShut {NoStop}%
\bibitem [{\citenamefont {Liu}\ \emph {et~al.}(2016)\citenamefont {Liu},
  \citenamefont {Duan}, \citenamefont {Tian}, \citenamefont {Wang},
  \citenamefont {Ma}, \citenamefont {Li}, \citenamefont {Huang}, \citenamefont
  {Liu},\ and\ \citenamefont {Cui}}]{2016LIU}%
  \BibitemOpen
  \bibfield  {author} {\bibinfo {author} {\bibfnamefont {Y.}~\bibnamefont
  {Liu}}, \bibinfo {author} {\bibfnamefont {D.}~\bibnamefont {Duan}}, \bibinfo
  {author} {\bibfnamefont {F.}~\bibnamefont {Tian}}, \bibinfo {author}
  {\bibfnamefont {C.}~\bibnamefont {Wang}}, \bibinfo {author} {\bibfnamefont
  {Y.}~\bibnamefont {Ma}}, \bibinfo {author} {\bibfnamefont {D.}~\bibnamefont
  {Li}}, \bibinfo {author} {\bibfnamefont {X.}~\bibnamefont {Huang}}, \bibinfo
  {author} {\bibfnamefont {B.}~\bibnamefont {Liu}}, \ and\ \bibinfo {author}
  {\bibfnamefont {T.}~\bibnamefont {Cui}},\ }\href@noop {} {\bibfield
  {journal} {\bibinfo  {journal} {Physical Chemistry Chemical Physics}\
  }\textbf {\bibinfo {volume} {18}},\ \bibinfo {pages} {1516} (\bibinfo {year}
  {2016})}\BibitemShut {NoStop}%
\bibitem [{\citenamefont {Li}\ \emph {et~al.}(2016{\natexlab{a}})\citenamefont
  {Li}, \citenamefont {Liu},\ and\ \citenamefont {Peng}}]{2016LI_a}%
  \BibitemOpen
  \bibfield  {author} {\bibinfo {author} {\bibfnamefont {X.}~\bibnamefont
  {Li}}, \bibinfo {author} {\bibfnamefont {H.}~\bibnamefont {Liu}}, \ and\
  \bibinfo {author} {\bibfnamefont {F.}~\bibnamefont {Peng}},\ }\href@noop {}
  {\bibfield  {journal} {\bibinfo  {journal} {Physical Chemistry Chemical
  Physics}\ }\textbf {\bibinfo {volume} {18}},\ \bibinfo {pages} {28791}
  (\bibinfo {year} {2016}{\natexlab{a}})}\BibitemShut {NoStop}%
\bibitem [{\citenamefont {Zhong}\ \emph {et~al.}(2016)\citenamefont {Zhong},
  \citenamefont {Wang}, \citenamefont {Zhang}, \citenamefont {Liu},
  \citenamefont {Zhang}, \citenamefont {Song}, \citenamefont {Yang},
  \citenamefont {Zhang},\ and\ \citenamefont {Ma}}]{2016ZHO}%
  \BibitemOpen
  \bibfield  {author} {\bibinfo {author} {\bibfnamefont {X.}~\bibnamefont
  {Zhong}}, \bibinfo {author} {\bibfnamefont {H.}~\bibnamefont {Wang}},
  \bibinfo {author} {\bibfnamefont {J.}~\bibnamefont {Zhang}}, \bibinfo
  {author} {\bibfnamefont {H.}~\bibnamefont {Liu}}, \bibinfo {author}
  {\bibfnamefont {S.}~\bibnamefont {Zhang}}, \bibinfo {author} {\bibfnamefont
  {H.-F.}\ \bibnamefont {Song}}, \bibinfo {author} {\bibfnamefont
  {G.}~\bibnamefont {Yang}}, \bibinfo {author} {\bibfnamefont {L.}~\bibnamefont
  {Zhang}}, \ and\ \bibinfo {author} {\bibfnamefont {Y.}~\bibnamefont {Ma}},\
  }\href@noop {} {\bibfield  {journal} {\bibinfo  {journal} {Physical Review
  Letters}\ }\textbf {\bibinfo {volume} {116}},\ \bibinfo {pages} {057002}
  (\bibinfo {year} {2016})}\BibitemShut {NoStop}%
\bibitem [{\citenamefont {Li}\ \emph {et~al.}(2016{\natexlab{b}})\citenamefont
  {Li}, \citenamefont {Wang}, \citenamefont {Liu}, \citenamefont {Zhang},
  \citenamefont {Hao}, \citenamefont {Pickard}, \citenamefont {Nelson},
  \citenamefont {Needs}, \citenamefont {Li}, \citenamefont {Huang} \emph
  {et~al.}}]{2016LI_b}%
  \BibitemOpen
  \bibfield  {author} {\bibinfo {author} {\bibfnamefont {Y.}~\bibnamefont
  {Li}}, \bibinfo {author} {\bibfnamefont {L.}~\bibnamefont {Wang}}, \bibinfo
  {author} {\bibfnamefont {H.}~\bibnamefont {Liu}}, \bibinfo {author}
  {\bibfnamefont {Y.}~\bibnamefont {Zhang}}, \bibinfo {author} {\bibfnamefont
  {J.}~\bibnamefont {Hao}}, \bibinfo {author} {\bibfnamefont {C.~J.}\
  \bibnamefont {Pickard}}, \bibinfo {author} {\bibfnamefont {J.~R.}\
  \bibnamefont {Nelson}}, \bibinfo {author} {\bibfnamefont {R.~J.}\
  \bibnamefont {Needs}}, \bibinfo {author} {\bibfnamefont {W.}~\bibnamefont
  {Li}}, \bibinfo {author} {\bibfnamefont {Y.}~\bibnamefont {Huang}},  \emph
  {et~al.},\ }\href@noop {} {\bibfield  {journal} {\bibinfo  {journal}
  {Physical Review B}\ }\textbf {\bibinfo {volume} {93}},\ \bibinfo {pages}
  {020103} (\bibinfo {year} {2016}{\natexlab{b}})}\BibitemShut {NoStop}%
\bibitem [{\citenamefont {Li}\ and\ \citenamefont {Peng}(2017)}]{2017LI_a}%
  \BibitemOpen
  \bibfield  {author} {\bibinfo {author} {\bibfnamefont {X.}~\bibnamefont
  {Li}}\ and\ \bibinfo {author} {\bibfnamefont {F.}~\bibnamefont {Peng}},\
  }\href@noop {} {\bibfield  {journal} {\bibinfo  {journal} {Inorganic
  chemistry}\ }\textbf {\bibinfo {volume} {56}},\ \bibinfo {pages} {13759}
  (\bibinfo {year} {2017})}\BibitemShut {NoStop}%
\bibitem [{\citenamefont {Zhuang}\ \emph {et~al.}(2017)\citenamefont {Zhuang},
  \citenamefont {Jin}, \citenamefont {Cui}, \citenamefont {Ma}, \citenamefont
  {Lv}, \citenamefont {Li}, \citenamefont {Zhang}, \citenamefont {Meng},\ and\
  \citenamefont {Bao}}]{2017ZHU}%
  \BibitemOpen
  \bibfield  {author} {\bibinfo {author} {\bibfnamefont {Q.}~\bibnamefont
  {Zhuang}}, \bibinfo {author} {\bibfnamefont {X.}~\bibnamefont {Jin}},
  \bibinfo {author} {\bibfnamefont {T.}~\bibnamefont {Cui}}, \bibinfo {author}
  {\bibfnamefont {Y.}~\bibnamefont {Ma}}, \bibinfo {author} {\bibfnamefont
  {Q.}~\bibnamefont {Lv}}, \bibinfo {author} {\bibfnamefont {Y.}~\bibnamefont
  {Li}}, \bibinfo {author} {\bibfnamefont {H.}~\bibnamefont {Zhang}}, \bibinfo
  {author} {\bibfnamefont {X.}~\bibnamefont {Meng}}, \ and\ \bibinfo {author}
  {\bibfnamefont {K.}~\bibnamefont {Bao}},\ }\href@noop {} {\bibfield
  {journal} {\bibinfo  {journal} {Inorganic chemistry}\ }\textbf {\bibinfo
  {volume} {56}},\ \bibinfo {pages} {3901} (\bibinfo {year}
  {2017})}\BibitemShut {NoStop}%
\bibitem [{\citenamefont {Liu}\ \emph {et~al.}(2017)\citenamefont {Liu},
  \citenamefont {Naumov}, \citenamefont {Hoffmann}, \citenamefont {Ashcroft},\
  and\ \citenamefont {Hemley}}]{2017LIU}%
  \BibitemOpen
  \bibfield  {author} {\bibinfo {author} {\bibfnamefont {H.}~\bibnamefont
  {Liu}}, \bibinfo {author} {\bibfnamefont {I.~I.}\ \bibnamefont {Naumov}},
  \bibinfo {author} {\bibfnamefont {R.}~\bibnamefont {Hoffmann}}, \bibinfo
  {author} {\bibfnamefont {N.}~\bibnamefont {Ashcroft}}, \ and\ \bibinfo
  {author} {\bibfnamefont {R.~J.}\ \bibnamefont {Hemley}},\ }\href@noop {}
  {\bibfield  {journal} {\bibinfo  {journal} {Proceedings of the National
  Academy of Sciences}\ }\textbf {\bibinfo {volume} {114}},\ \bibinfo {pages}
  {6990} (\bibinfo {year} {2017})}\BibitemShut {NoStop}%
\bibitem [{\citenamefont {Zeng}\ \emph {et~al.}(2017)\citenamefont {Zeng},
  \citenamefont {Yu}, \citenamefont {Li}, \citenamefont {Oganov},\ and\
  \citenamefont {Frapper}}]{2017ZEN}%
  \BibitemOpen
  \bibfield  {author} {\bibinfo {author} {\bibfnamefont {Q.}~\bibnamefont
  {Zeng}}, \bibinfo {author} {\bibfnamefont {S.}~\bibnamefont {Yu}}, \bibinfo
  {author} {\bibfnamefont {D.}~\bibnamefont {Li}}, \bibinfo {author}
  {\bibfnamefont {A.~R.}\ \bibnamefont {Oganov}}, \ and\ \bibinfo {author}
  {\bibfnamefont {G.}~\bibnamefont {Frapper}},\ }\href@noop {} {\bibfield
  {journal} {\bibinfo  {journal} {Physical Chemistry Chemical Physics}\
  }\textbf {\bibinfo {volume} {19}},\ \bibinfo {pages} {8236} (\bibinfo {year}
  {2017})}\BibitemShut {NoStop}%
\bibitem [{\citenamefont {Ishikawa}\ \emph {et~al.}(2017)\citenamefont
  {Ishikawa}, \citenamefont {Nakanishi}, \citenamefont {Shimizu},\ and\
  \citenamefont {Oda}}]{2017ISH}%
  \BibitemOpen
  \bibfield  {author} {\bibinfo {author} {\bibfnamefont {T.}~\bibnamefont
  {Ishikawa}}, \bibinfo {author} {\bibfnamefont {A.}~\bibnamefont {Nakanishi}},
  \bibinfo {author} {\bibfnamefont {K.}~\bibnamefont {Shimizu}}, \ and\
  \bibinfo {author} {\bibfnamefont {T.}~\bibnamefont {Oda}},\ }\href@noop {}
  {\bibfield  {journal} {\bibinfo  {journal} {Journal of the Physical Society
  of Japan}\ }\textbf {\bibinfo {volume} {86}},\ \bibinfo {pages} {124711}
  (\bibinfo {year} {2017})}\BibitemShut {NoStop}%
\bibitem [{\citenamefont {Li}\ \emph {et~al.}(2017)\citenamefont {Li},
  \citenamefont {Hu},\ and\ \citenamefont {Huang}}]{2017LI_b}%
  \BibitemOpen
  \bibfield  {author} {\bibinfo {author} {\bibfnamefont {X.-F.}\ \bibnamefont
  {Li}}, \bibinfo {author} {\bibfnamefont {Z.-Y.}\ \bibnamefont {Hu}}, \ and\
  \bibinfo {author} {\bibfnamefont {B.}~\bibnamefont {Huang}},\ }\href@noop {}
  {\bibfield  {journal} {\bibinfo  {journal} {Physical Chemistry Chemical
  Physics}\ }\textbf {\bibinfo {volume} {19}},\ \bibinfo {pages} {3538}
  (\bibinfo {year} {2017})}\BibitemShut {NoStop}%
\bibitem [{\citenamefont {Peng}\ \emph {et~al.}(2017)\citenamefont {Peng},
  \citenamefont {Sun}, \citenamefont {Pickard}, \citenamefont {Needs},
  \citenamefont {Wu},\ and\ \citenamefont {Ma}}]{2017PEN}%
  \BibitemOpen
  \bibfield  {author} {\bibinfo {author} {\bibfnamefont {F.}~\bibnamefont
  {Peng}}, \bibinfo {author} {\bibfnamefont {Y.}~\bibnamefont {Sun}}, \bibinfo
  {author} {\bibfnamefont {C.~J.}\ \bibnamefont {Pickard}}, \bibinfo {author}
  {\bibfnamefont {R.~J.}\ \bibnamefont {Needs}}, \bibinfo {author}
  {\bibfnamefont {Q.}~\bibnamefont {Wu}}, \ and\ \bibinfo {author}
  {\bibfnamefont {Y.}~\bibnamefont {Ma}},\ }\href@noop {} {\bibfield  {journal}
  {\bibinfo  {journal} {Physical review letters}\ }\textbf {\bibinfo {volume}
  {119}},\ \bibinfo {pages} {107001} (\bibinfo {year} {2017})}\BibitemShut
  {NoStop}%
\bibitem [{\citenamefont {Esfahani}\ \emph {et~al.}(2017)\citenamefont
  {Esfahani}, \citenamefont {Oganov}, \citenamefont {Niu},\ and\ \citenamefont
  {Zhang}}]{2017DAV}%
  \BibitemOpen
  \bibfield  {author} {\bibinfo {author} {\bibfnamefont {M.~M.~D.}\
  \bibnamefont {Esfahani}}, \bibinfo {author} {\bibfnamefont {A.~R.}\
  \bibnamefont {Oganov}}, \bibinfo {author} {\bibfnamefont {H.}~\bibnamefont
  {Niu}}, \ and\ \bibinfo {author} {\bibfnamefont {J.}~\bibnamefont {Zhang}},\
  }\href@noop {} {\bibfield  {journal} {\bibinfo  {journal} {Physical Review
  B}\ }\textbf {\bibinfo {volume} {95}},\ \bibinfo {pages} {134506} (\bibinfo
  {year} {2017})}\BibitemShut {NoStop}%
\bibitem [{\citenamefont {Majumdar}\ \emph {et~al.}(2017)\citenamefont
  {Majumdar}, \citenamefont {John}, \citenamefont {Wu},\ and\ \citenamefont
  {Yao}}]{2017MAJ}%
  \BibitemOpen
  \bibfield  {author} {\bibinfo {author} {\bibfnamefont {A.}~\bibnamefont
  {Majumdar}}, \bibinfo {author} {\bibfnamefont {S.~T.}\ \bibnamefont {John}},
  \bibinfo {author} {\bibfnamefont {M.}~\bibnamefont {Wu}}, \ and\ \bibinfo
  {author} {\bibfnamefont {Y.}~\bibnamefont {Yao}},\ }\href@noop {} {\bibfield
  {journal} {\bibinfo  {journal} {Physical Review B}\ }\textbf {\bibinfo
  {volume} {96}},\ \bibinfo {pages} {201107} (\bibinfo {year}
  {2017})}\BibitemShut {NoStop}%
\bibitem [{\citenamefont {Zarifi}\ \emph {et~al.}(2018)\citenamefont {Zarifi},
  \citenamefont {Bi}, \citenamefont {Liu},\ and\ \citenamefont
  {Zurek}}]{2018ZAR}%
  \BibitemOpen
  \bibfield  {author} {\bibinfo {author} {\bibfnamefont {N.}~\bibnamefont
  {Zarifi}}, \bibinfo {author} {\bibfnamefont {T.}~\bibnamefont {Bi}}, \bibinfo
  {author} {\bibfnamefont {H.}~\bibnamefont {Liu}}, \ and\ \bibinfo {author}
  {\bibfnamefont {E.}~\bibnamefont {Zurek}},\ }\href@noop {} {\bibfield
  {journal} {\bibinfo  {journal} {The Journal of Physical Chemistry C}\
  }\textbf {\bibinfo {volume} {122}},\ \bibinfo {pages} {24262} (\bibinfo
  {year} {2018})}\BibitemShut {NoStop}%
\bibitem [{\citenamefont {Semenok}\ \emph {et~al.}(2018)\citenamefont
  {Semenok}, \citenamefont {Kvashnin}, \citenamefont {Kruglov},\ and\
  \citenamefont {Oganov}}]{2018SEM}%
  \BibitemOpen
  \bibfield  {author} {\bibinfo {author} {\bibfnamefont {D.~V.}\ \bibnamefont
  {Semenok}}, \bibinfo {author} {\bibfnamefont {A.~G.}\ \bibnamefont
  {Kvashnin}}, \bibinfo {author} {\bibfnamefont {I.~A.}\ \bibnamefont
  {Kruglov}}, \ and\ \bibinfo {author} {\bibfnamefont {A.~R.}\ \bibnamefont
  {Oganov}},\ }\href@noop {} {\bibfield  {journal} {\bibinfo  {journal} {The
  journal of physical chemistry letters}\ }\textbf {\bibinfo {volume} {9}},\
  \bibinfo {pages} {1920} (\bibinfo {year} {2018})}\BibitemShut {NoStop}%
\bibitem [{\citenamefont {Ye}\ \emph {et~al.}(2018)\citenamefont {Ye},
  \citenamefont {Zarifi}, \citenamefont {Zurek}, \citenamefont {Hoffmann},\
  and\ \citenamefont {Ashcroft}}]{2018YE}%
  \BibitemOpen
  \bibfield  {author} {\bibinfo {author} {\bibfnamefont {X.}~\bibnamefont
  {Ye}}, \bibinfo {author} {\bibfnamefont {N.}~\bibnamefont {Zarifi}}, \bibinfo
  {author} {\bibfnamefont {E.}~\bibnamefont {Zurek}}, \bibinfo {author}
  {\bibfnamefont {R.}~\bibnamefont {Hoffmann}}, \ and\ \bibinfo {author}
  {\bibfnamefont {N.}~\bibnamefont {Ashcroft}},\ }\href@noop {} {\bibfield
  {journal} {\bibinfo  {journal} {The Journal of Physical Chemistry C}\
  }\textbf {\bibinfo {volume} {122}},\ \bibinfo {pages} {6298} (\bibinfo {year}
  {2018})}\BibitemShut {NoStop}%
\bibitem [{\citenamefont {Kvashnin}\ \emph
  {et~al.}(2018{\natexlab{a}})\citenamefont {Kvashnin}, \citenamefont
  {Semenok}, \citenamefont {Kruglov}, \citenamefont {Wrona},\ and\
  \citenamefont {Oganov}}]{2018KVA_a}%
  \BibitemOpen
  \bibfield  {author} {\bibinfo {author} {\bibfnamefont {A.~G.}\ \bibnamefont
  {Kvashnin}}, \bibinfo {author} {\bibfnamefont {D.~V.}\ \bibnamefont
  {Semenok}}, \bibinfo {author} {\bibfnamefont {I.~A.}\ \bibnamefont
  {Kruglov}}, \bibinfo {author} {\bibfnamefont {I.~A.}\ \bibnamefont {Wrona}},
  \ and\ \bibinfo {author} {\bibfnamefont {A.~R.}\ \bibnamefont {Oganov}},\
  }\href@noop {} {\bibfield  {journal} {\bibinfo  {journal} {ACS applied
  materials \& interfaces}\ }\textbf {\bibinfo {volume} {10}},\ \bibinfo
  {pages} {43809} (\bibinfo {year} {2018}{\natexlab{a}})}\BibitemShut {NoStop}%
\bibitem [{\citenamefont {Durajski}\ and\ \citenamefont
  {Szcz{\c{e}}{\'s}niak}(2018)}]{2018DUR}%
  \BibitemOpen
  \bibfield  {author} {\bibinfo {author} {\bibfnamefont {A.~P.}\ \bibnamefont
  {Durajski}}\ and\ \bibinfo {author} {\bibfnamefont {R.}~\bibnamefont
  {Szcz{\c{e}}{\'s}niak}},\ }\href@noop {} {\bibfield  {journal} {\bibinfo
  {journal} {The Journal of chemical physics}\ }\textbf {\bibinfo {volume}
  {149}},\ \bibinfo {pages} {074101} (\bibinfo {year} {2018})}\BibitemShut
  {NoStop}%
\bibitem [{\citenamefont {Zheng}\ \emph {et~al.}(2018)\citenamefont {Zheng},
  \citenamefont {Zhang}, \citenamefont {Sun}, \citenamefont {Zhang},
  \citenamefont {Lin}, \citenamefont {Yang},\ and\ \citenamefont
  {Bergara}}]{2018ZHE}%
  \BibitemOpen
  \bibfield  {author} {\bibinfo {author} {\bibfnamefont {S.}~\bibnamefont
  {Zheng}}, \bibinfo {author} {\bibfnamefont {S.}~\bibnamefont {Zhang}},
  \bibinfo {author} {\bibfnamefont {Y.}~\bibnamefont {Sun}}, \bibinfo {author}
  {\bibfnamefont {J.}~\bibnamefont {Zhang}}, \bibinfo {author} {\bibfnamefont
  {J.}~\bibnamefont {Lin}}, \bibinfo {author} {\bibfnamefont {G.}~\bibnamefont
  {Yang}}, \ and\ \bibinfo {author} {\bibfnamefont {A.}~\bibnamefont
  {Bergara}},\ }\href@noop {} {\bibfield  {journal} {\bibinfo  {journal}
  {Frontiers in Physics}\ }\textbf {\bibinfo {volume} {6}},\ \bibinfo {pages}
  {101} (\bibinfo {year} {2018})}\BibitemShut {NoStop}%
\bibitem [{\citenamefont {Abe}(2018)}]{2018ABE}%
  \BibitemOpen
  \bibfield  {author} {\bibinfo {author} {\bibfnamefont {K.}~\bibnamefont
  {Abe}},\ }\href@noop {} {\bibfield  {journal} {\bibinfo  {journal} {Physical
  Review B}\ }\textbf {\bibinfo {volume} {98}},\ \bibinfo {pages} {134103}
  (\bibinfo {year} {2018})}\BibitemShut {NoStop}%
\bibitem [{\citenamefont {Zhuang}\ \emph {et~al.}(2018)\citenamefont {Zhuang},
  \citenamefont {Jin}, \citenamefont {Cui}, \citenamefont {Zhang},
  \citenamefont {Li}, \citenamefont {Li}, \citenamefont {Bao},\ and\
  \citenamefont {Liu}}]{2018ZHU}%
  \BibitemOpen
  \bibfield  {author} {\bibinfo {author} {\bibfnamefont {Q.}~\bibnamefont
  {Zhuang}}, \bibinfo {author} {\bibfnamefont {X.}~\bibnamefont {Jin}},
  \bibinfo {author} {\bibfnamefont {T.}~\bibnamefont {Cui}}, \bibinfo {author}
  {\bibfnamefont {D.}~\bibnamefont {Zhang}}, \bibinfo {author} {\bibfnamefont
  {Y.}~\bibnamefont {Li}}, \bibinfo {author} {\bibfnamefont {X.}~\bibnamefont
  {Li}}, \bibinfo {author} {\bibfnamefont {K.}~\bibnamefont {Bao}}, \ and\
  \bibinfo {author} {\bibfnamefont {B.}~\bibnamefont {Liu}},\ }\href@noop {}
  {\bibfield  {journal} {\bibinfo  {journal} {Physical Review B}\ }\textbf
  {\bibinfo {volume} {98}},\ \bibinfo {pages} {024514} (\bibinfo {year}
  {2018})}\BibitemShut {NoStop}%
\bibitem [{\citenamefont {Kvashnin}\ \emph
  {et~al.}(2018{\natexlab{b}})\citenamefont {Kvashnin}, \citenamefont
  {Kruglov}, \citenamefont {Semenok},\ and\ \citenamefont
  {Oganov}}]{2018KVA_b}%
  \BibitemOpen
  \bibfield  {author} {\bibinfo {author} {\bibfnamefont {A.~G.}\ \bibnamefont
  {Kvashnin}}, \bibinfo {author} {\bibfnamefont {I.~A.}\ \bibnamefont
  {Kruglov}}, \bibinfo {author} {\bibfnamefont {D.~V.}\ \bibnamefont
  {Semenok}}, \ and\ \bibinfo {author} {\bibfnamefont {A.~R.}\ \bibnamefont
  {Oganov}},\ }\href@noop {} {\bibfield  {journal} {\bibinfo  {journal} {The
  Journal of Physical Chemistry C}\ }\textbf {\bibinfo {volume} {122}},\
  \bibinfo {pages} {4731} (\bibinfo {year} {2018}{\natexlab{b}})}\BibitemShut
  {NoStop}%
\bibitem [{\citenamefont {Wu}\ \emph {et~al.}(2018)\citenamefont {Wu},
  \citenamefont {Zhao}, \citenamefont {Chen}, \citenamefont {Wang},
  \citenamefont {Chen}, \citenamefont {Guo}, \citenamefont {Zang},\ and\
  \citenamefont {Lu}}]{2018WU}%
  \BibitemOpen
  \bibfield  {author} {\bibinfo {author} {\bibfnamefont {J.}~\bibnamefont
  {Wu}}, \bibinfo {author} {\bibfnamefont {L.-Z.}\ \bibnamefont {Zhao}},
  \bibinfo {author} {\bibfnamefont {H.-L.}\ \bibnamefont {Chen}}, \bibinfo
  {author} {\bibfnamefont {D.}~\bibnamefont {Wang}}, \bibinfo {author}
  {\bibfnamefont {J.-Y.}\ \bibnamefont {Chen}}, \bibinfo {author}
  {\bibfnamefont {X.}~\bibnamefont {Guo}}, \bibinfo {author} {\bibfnamefont
  {Q.-J.}\ \bibnamefont {Zang}}, \ and\ \bibinfo {author} {\bibfnamefont
  {W.-C.}\ \bibnamefont {Lu}},\ }\href@noop {} {\bibfield  {journal} {\bibinfo
  {journal} {physica status solidi (b)}\ }\textbf {\bibinfo {volume} {255}},\
  \bibinfo {pages} {1800224} (\bibinfo {year} {2018})}\BibitemShut {NoStop}%
\bibitem [{\citenamefont {Heil}\ \emph {et~al.}(2019)\citenamefont {Heil},
  \citenamefont {Di~Cataldo}, \citenamefont {Bachelet},\ and\ \citenamefont
  {Boeri}}]{2019HEI}%
  \BibitemOpen
  \bibfield  {author} {\bibinfo {author} {\bibfnamefont {C.}~\bibnamefont
  {Heil}}, \bibinfo {author} {\bibfnamefont {S.}~\bibnamefont {Di~Cataldo}},
  \bibinfo {author} {\bibfnamefont {G.~B.}\ \bibnamefont {Bachelet}}, \ and\
  \bibinfo {author} {\bibfnamefont {L.}~\bibnamefont {Boeri}},\ }\href@noop {}
  {\bibfield  {journal} {\bibinfo  {journal} {Physical Review B}\ }\textbf
  {\bibinfo {volume} {99}},\ \bibinfo {pages} {220502} (\bibinfo {year}
  {2019})}\BibitemShut {NoStop}%
\bibitem [{\citenamefont {Li}\ \emph {et~al.}(2019{\natexlab{a}})\citenamefont
  {Li}, \citenamefont {Huang}, \citenamefont {Duan}, \citenamefont {Pickard},
  \citenamefont {Zhou}, \citenamefont {Xie}, \citenamefont {Zhuang},
  \citenamefont {Huang}, \citenamefont {Zhou}, \citenamefont {Liu} \emph
  {et~al.}}]{2019LI_b}%
  \BibitemOpen
  \bibfield  {author} {\bibinfo {author} {\bibfnamefont {X.}~\bibnamefont
  {Li}}, \bibinfo {author} {\bibfnamefont {X.}~\bibnamefont {Huang}}, \bibinfo
  {author} {\bibfnamefont {D.}~\bibnamefont {Duan}}, \bibinfo {author}
  {\bibfnamefont {C.~J.}\ \bibnamefont {Pickard}}, \bibinfo {author}
  {\bibfnamefont {D.}~\bibnamefont {Zhou}}, \bibinfo {author} {\bibfnamefont
  {H.}~\bibnamefont {Xie}}, \bibinfo {author} {\bibfnamefont {Q.}~\bibnamefont
  {Zhuang}}, \bibinfo {author} {\bibfnamefont {Y.}~\bibnamefont {Huang}},
  \bibinfo {author} {\bibfnamefont {Q.}~\bibnamefont {Zhou}}, \bibinfo {author}
  {\bibfnamefont {B.}~\bibnamefont {Liu}},  \emph {et~al.},\ }\href@noop {}
  {\bibfield  {journal} {\bibinfo  {journal} {Nature communications}\ }\textbf
  {\bibinfo {volume} {10}},\ \bibinfo {pages} {1} (\bibinfo {year}
  {2019}{\natexlab{a}})}\BibitemShut {NoStop}%
\bibitem [{\citenamefont {Yuan}\ \emph {et~al.}(2019)\citenamefont {Yuan},
  \citenamefont {Li}, \citenamefont {Fang}, \citenamefont {Liu}, \citenamefont
  {Pei}, \citenamefont {Li}, \citenamefont {Zheng}, \citenamefont {Yang},\ and\
  \citenamefont {Wang}}]{2019YUA}%
  \BibitemOpen
  \bibfield  {author} {\bibinfo {author} {\bibfnamefont {Y.}~\bibnamefont
  {Yuan}}, \bibinfo {author} {\bibfnamefont {Y.}~\bibnamefont {Li}}, \bibinfo
  {author} {\bibfnamefont {G.}~\bibnamefont {Fang}}, \bibinfo {author}
  {\bibfnamefont {G.}~\bibnamefont {Liu}}, \bibinfo {author} {\bibfnamefont
  {C.}~\bibnamefont {Pei}}, \bibinfo {author} {\bibfnamefont {X.}~\bibnamefont
  {Li}}, \bibinfo {author} {\bibfnamefont {H.}~\bibnamefont {Zheng}}, \bibinfo
  {author} {\bibfnamefont {K.}~\bibnamefont {Yang}}, \ and\ \bibinfo {author}
  {\bibfnamefont {L.}~\bibnamefont {Wang}},\ }\href@noop {} {\bibfield
  {journal} {\bibinfo  {journal} {National Science Review}\ }\textbf {\bibinfo
  {volume} {6}},\ \bibinfo {pages} {524} (\bibinfo {year} {2019})}\BibitemShut
  {NoStop}%
\bibitem [{\citenamefont {Wang}\ \emph {et~al.}(2019)\citenamefont {Wang},
  \citenamefont {Zhang}, \citenamefont {Chen}, \citenamefont {Wu},
  \citenamefont {Zang},\ and\ \citenamefont {Lu}}]{2019WAN}%
  \BibitemOpen
  \bibfield  {author} {\bibinfo {author} {\bibfnamefont {D.}~\bibnamefont
  {Wang}}, \bibinfo {author} {\bibfnamefont {H.}~\bibnamefont {Zhang}},
  \bibinfo {author} {\bibfnamefont {H.-L.}\ \bibnamefont {Chen}}, \bibinfo
  {author} {\bibfnamefont {J.}~\bibnamefont {Wu}}, \bibinfo {author}
  {\bibfnamefont {Q.-J.}\ \bibnamefont {Zang}}, \ and\ \bibinfo {author}
  {\bibfnamefont {W.-C.}\ \bibnamefont {Lu}},\ }\href@noop {} {\bibfield
  {journal} {\bibinfo  {journal} {Physics Letters A}\ }\textbf {\bibinfo
  {volume} {383}},\ \bibinfo {pages} {774} (\bibinfo {year}
  {2019})}\BibitemShut {NoStop}%
\bibitem [{\citenamefont {Yang}\ \emph {et~al.}(2019)\citenamefont {Yang},
  \citenamefont {Lu}, \citenamefont {Li}, \citenamefont {Xue}, \citenamefont
  {Zang}, \citenamefont {Ho},\ and\ \citenamefont {Wang}}]{2019YAN}%
  \BibitemOpen
  \bibfield  {author} {\bibinfo {author} {\bibfnamefont {W.-H.}\ \bibnamefont
  {Yang}}, \bibinfo {author} {\bibfnamefont {W.-C.}\ \bibnamefont {Lu}},
  \bibinfo {author} {\bibfnamefont {S.-D.}\ \bibnamefont {Li}}, \bibinfo
  {author} {\bibfnamefont {X.-Y.}\ \bibnamefont {Xue}}, \bibinfo {author}
  {\bibfnamefont {Q.-J.}\ \bibnamefont {Zang}}, \bibinfo {author}
  {\bibfnamefont {K.-M.}\ \bibnamefont {Ho}}, \ and\ \bibinfo {author}
  {\bibfnamefont {C.-Z.}\ \bibnamefont {Wang}},\ }\href@noop {} {\bibfield
  {journal} {\bibinfo  {journal} {Physical Chemistry Chemical Physics}\
  }\textbf {\bibinfo {volume} {21}},\ \bibinfo {pages} {5466} (\bibinfo {year}
  {2019})}\BibitemShut {NoStop}%
\bibitem [{\citenamefont {Xi}\ \emph {et~al.}(2019)\citenamefont {Xi},
  \citenamefont {Jing}, \citenamefont {Li}, \citenamefont {Deng}, \citenamefont
  {Cao},\ and\ \citenamefont {Yang}}]{2019XI}%
  \BibitemOpen
  \bibfield  {author} {\bibinfo {author} {\bibfnamefont {R.}~\bibnamefont
  {Xi}}, \bibinfo {author} {\bibfnamefont {Y.}~\bibnamefont {Jing}}, \bibinfo
  {author} {\bibfnamefont {J.}~\bibnamefont {Li}}, \bibinfo {author}
  {\bibfnamefont {Y.}~\bibnamefont {Deng}}, \bibinfo {author} {\bibfnamefont
  {X.}~\bibnamefont {Cao}}, \ and\ \bibinfo {author} {\bibfnamefont
  {G.}~\bibnamefont {Yang}},\ }\href@noop {} {\bibfield  {journal} {\bibinfo
  {journal} {The Journal of Physical Chemistry C}\ }\textbf {\bibinfo {volume}
  {123}},\ \bibinfo {pages} {24243} (\bibinfo {year} {2019})}\BibitemShut
  {NoStop}%
\bibitem [{\citenamefont {Abe}(2019)}]{2019ABE}%
  \BibitemOpen
  \bibfield  {author} {\bibinfo {author} {\bibfnamefont {K.}~\bibnamefont
  {Abe}},\ }\href@noop {} {\bibfield  {journal} {\bibinfo  {journal} {Physical
  Review B}\ }\textbf {\bibinfo {volume} {100}},\ \bibinfo {pages} {174105}
  (\bibinfo {year} {2019})}\BibitemShut {NoStop}%
\bibitem [{\citenamefont {Xie}\ \emph {et~al.}(2020)\citenamefont {Xie},
  \citenamefont {Zhang}, \citenamefont {Duan}, \citenamefont {Huang},
  \citenamefont {Huang}, \citenamefont {Song}, \citenamefont {Feng},
  \citenamefont {Yao}, \citenamefont {Pickard},\ and\ \citenamefont
  {Cui}}]{2020XIE}%
  \BibitemOpen
  \bibfield  {author} {\bibinfo {author} {\bibfnamefont {H.}~\bibnamefont
  {Xie}}, \bibinfo {author} {\bibfnamefont {W.}~\bibnamefont {Zhang}}, \bibinfo
  {author} {\bibfnamefont {D.}~\bibnamefont {Duan}}, \bibinfo {author}
  {\bibfnamefont {X.}~\bibnamefont {Huang}}, \bibinfo {author} {\bibfnamefont
  {Y.}~\bibnamefont {Huang}}, \bibinfo {author} {\bibfnamefont
  {H.}~\bibnamefont {Song}}, \bibinfo {author} {\bibfnamefont {X.}~\bibnamefont
  {Feng}}, \bibinfo {author} {\bibfnamefont {Y.}~\bibnamefont {Yao}}, \bibinfo
  {author} {\bibfnamefont {C.~J.}\ \bibnamefont {Pickard}}, \ and\ \bibinfo
  {author} {\bibfnamefont {T.}~\bibnamefont {Cui}},\ }\href@noop {} {\bibfield
  {journal} {\bibinfo  {journal} {The Journal of Physical Chemistry Letters}\
  }\textbf {\bibinfo {volume} {11}},\ \bibinfo {pages} {646} (\bibinfo {year}
  {2020})}\BibitemShut {NoStop}%
\bibitem [{\citenamefont {Ma}\ \emph {et~al.}(2015)\citenamefont {Ma},
  \citenamefont {Duan}, \citenamefont {Li}, \citenamefont {Liu}, \citenamefont
  {Tian}, \citenamefont {Yu}, \citenamefont {Xu}, \citenamefont {Shao},
  \citenamefont {Liu},\ and\ \citenamefont {Cui}}]{2015MA}%
  \BibitemOpen
  \bibfield  {author} {\bibinfo {author} {\bibfnamefont {Y.}~\bibnamefont
  {Ma}}, \bibinfo {author} {\bibfnamefont {D.}~\bibnamefont {Duan}}, \bibinfo
  {author} {\bibfnamefont {D.}~\bibnamefont {Li}}, \bibinfo {author}
  {\bibfnamefont {Y.}~\bibnamefont {Liu}}, \bibinfo {author} {\bibfnamefont
  {F.}~\bibnamefont {Tian}}, \bibinfo {author} {\bibfnamefont {H.}~\bibnamefont
  {Yu}}, \bibinfo {author} {\bibfnamefont {C.}~\bibnamefont {Xu}}, \bibinfo
  {author} {\bibfnamefont {Z.}~\bibnamefont {Shao}}, \bibinfo {author}
  {\bibfnamefont {B.}~\bibnamefont {Liu}}, \ and\ \bibinfo {author}
  {\bibfnamefont {T.}~\bibnamefont {Cui}},\ }\href@noop {} {\bibfield
  {journal} {\bibinfo  {journal} {arXiv preprint arXiv:1511.05291}\ } (\bibinfo
  {year} {2015})}\BibitemShut {NoStop}%
\bibitem [{\citenamefont {Li}\ \emph {et~al.}(2019{\natexlab{b}})\citenamefont
  {Li}, \citenamefont {Li}, \citenamefont {Wang}, \citenamefont {Liu},
  \citenamefont {Li},\ and\ \citenamefont {Liu}}]{2019LI_c}%
  \BibitemOpen
  \bibfield  {author} {\bibinfo {author} {\bibfnamefont {H.}~\bibnamefont
  {Li}}, \bibinfo {author} {\bibfnamefont {X.}~\bibnamefont {Li}}, \bibinfo
  {author} {\bibfnamefont {H.}~\bibnamefont {Wang}}, \bibinfo {author}
  {\bibfnamefont {G.}~\bibnamefont {Liu}}, \bibinfo {author} {\bibfnamefont
  {Y.}~\bibnamefont {Li}}, \ and\ \bibinfo {author} {\bibfnamefont
  {H.}~\bibnamefont {Liu}},\ }\href@noop {} {\bibfield  {journal} {\bibinfo
  {journal} {New Journal of Physics}\ }\textbf {\bibinfo {volume} {21}},\
  \bibinfo {pages} {123009} (\bibinfo {year} {2019}{\natexlab{b}})}\BibitemShut
  {NoStop}%
\bibitem [{\citenamefont {Kanagaprabha}\ and\ \citenamefont
  {Rajeswarapalanichamy}(2018)}]{2018KAN}%
  \BibitemOpen
  \bibfield  {author} {\bibinfo {author} {\bibfnamefont {S.}~\bibnamefont
  {Kanagaprabha}}\ and\ \bibinfo {author} {\bibfnamefont {R.}~\bibnamefont
  {Rajeswarapalanichamy}},\ }\href@noop {} {\bibfield  {journal} {\bibinfo
  {journal} {Journal of Materials Science Research and Reviews}\ ,\ \bibinfo
  {pages} {1}} (\bibinfo {year} {2018})}\BibitemShut {NoStop}%
\bibitem [{\citenamefont {Zhang}\ \emph {et~al.}(2016)\citenamefont {Zhang},
  \citenamefont {Zhu}, \citenamefont {Liu},\ and\ \citenamefont
  {Yang}}]{2016ZHA}%
  \BibitemOpen
  \bibfield  {author} {\bibinfo {author} {\bibfnamefont {S.}~\bibnamefont
  {Zhang}}, \bibinfo {author} {\bibfnamefont {L.}~\bibnamefont {Zhu}}, \bibinfo
  {author} {\bibfnamefont {H.}~\bibnamefont {Liu}}, \ and\ \bibinfo {author}
  {\bibfnamefont {G.}~\bibnamefont {Yang}},\ }\href@noop {} {\bibfield
  {journal} {\bibinfo  {journal} {Inorganic Chemistry}\ }\textbf {\bibinfo
  {volume} {55}},\ \bibinfo {pages} {11434} (\bibinfo {year}
  {2016})}\BibitemShut {NoStop}%
\bibitem [{\citenamefont {Liu}\ \emph {et~al.}(2018)\citenamefont {Liu},
  \citenamefont {Cui}, \citenamefont {Shi}, \citenamefont {Zhu}, \citenamefont
  {Chen}, \citenamefont {Lin}, \citenamefont {Su}, \citenamefont {Ma},
  \citenamefont {Yang}, \citenamefont {Xu} \emph {et~al.}}]{2018LIU}%
  \BibitemOpen
  \bibfield  {author} {\bibinfo {author} {\bibfnamefont {B.}~\bibnamefont
  {Liu}}, \bibinfo {author} {\bibfnamefont {W.}~\bibnamefont {Cui}}, \bibinfo
  {author} {\bibfnamefont {J.}~\bibnamefont {Shi}}, \bibinfo {author}
  {\bibfnamefont {L.}~\bibnamefont {Zhu}}, \bibinfo {author} {\bibfnamefont
  {J.}~\bibnamefont {Chen}}, \bibinfo {author} {\bibfnamefont {S.}~\bibnamefont
  {Lin}}, \bibinfo {author} {\bibfnamefont {R.}~\bibnamefont {Su}}, \bibinfo
  {author} {\bibfnamefont {J.}~\bibnamefont {Ma}}, \bibinfo {author}
  {\bibfnamefont {K.}~\bibnamefont {Yang}}, \bibinfo {author} {\bibfnamefont
  {M.}~\bibnamefont {Xu}},  \emph {et~al.},\ }\href@noop {} {\bibfield
  {journal} {\bibinfo  {journal} {Physical Review B}\ }\textbf {\bibinfo
  {volume} {98}},\ \bibinfo {pages} {174101} (\bibinfo {year}
  {2018})}\BibitemShut {NoStop}%
\bibitem [{\citenamefont {Nakanishi}\ \emph {et~al.}(2018)\citenamefont
  {Nakanishi}, \citenamefont {Ishikawa},\ and\ \citenamefont
  {Shimizu}}]{2018NAK}%
  \BibitemOpen
  \bibfield  {author} {\bibinfo {author} {\bibfnamefont {A.}~\bibnamefont
  {Nakanishi}}, \bibinfo {author} {\bibfnamefont {T.}~\bibnamefont {Ishikawa}},
  \ and\ \bibinfo {author} {\bibfnamefont {K.}~\bibnamefont {Shimizu}},\
  }\href@noop {} {\bibfield  {journal} {\bibinfo  {journal} {Journal of the
  Physical Society of Japan}\ }\textbf {\bibinfo {volume} {87}},\ \bibinfo
  {pages} {124711} (\bibinfo {year} {2018})}\BibitemShut {NoStop}%
\bibitem [{\citenamefont {Liang}\ \emph
  {et~al.}(2019{\natexlab{c}})\citenamefont {Liang}, \citenamefont {Bergara},
  \citenamefont {Wang}, \citenamefont {Wen}, \citenamefont {Zhao},
  \citenamefont {Zhou}, \citenamefont {He}, \citenamefont {Gao},\ and\
  \citenamefont {Tian}}]{2019LIA_a}%
  \BibitemOpen
  \bibfield  {author} {\bibinfo {author} {\bibfnamefont {X.}~\bibnamefont
  {Liang}}, \bibinfo {author} {\bibfnamefont {A.}~\bibnamefont {Bergara}},
  \bibinfo {author} {\bibfnamefont {L.}~\bibnamefont {Wang}}, \bibinfo {author}
  {\bibfnamefont {B.}~\bibnamefont {Wen}}, \bibinfo {author} {\bibfnamefont
  {Z.}~\bibnamefont {Zhao}}, \bibinfo {author} {\bibfnamefont {X.-F.}\
  \bibnamefont {Zhou}}, \bibinfo {author} {\bibfnamefont {J.}~\bibnamefont
  {He}}, \bibinfo {author} {\bibfnamefont {G.}~\bibnamefont {Gao}}, \ and\
  \bibinfo {author} {\bibfnamefont {Y.}~\bibnamefont {Tian}},\ }\href@noop {}
  {\bibfield  {journal} {\bibinfo  {journal} {Physical Review B}\ }\textbf
  {\bibinfo {volume} {99}},\ \bibinfo {pages} {100505} (\bibinfo {year}
  {2019}{\natexlab{c}})}\BibitemShut {NoStop}%
\bibitem [{\citenamefont {Grishakov}\ \emph {et~al.}(2019)\citenamefont
  {Grishakov}, \citenamefont {Degtyarenko},\ and\ \citenamefont
  {Mazur}}]{2019GRI}%
  \BibitemOpen
  \bibfield  {author} {\bibinfo {author} {\bibfnamefont {K.}~\bibnamefont
  {Grishakov}}, \bibinfo {author} {\bibfnamefont {N.}~\bibnamefont
  {Degtyarenko}}, \ and\ \bibinfo {author} {\bibfnamefont {E.}~\bibnamefont
  {Mazur}},\ }\href@noop {} {\bibfield  {journal} {\bibinfo  {journal} {Journal
  of Experimental and Theoretical Physics}\ }\textbf {\bibinfo {volume}
  {128}},\ \bibinfo {pages} {105} (\bibinfo {year} {2019})}\BibitemShut
  {NoStop}%
\bibitem [{\citenamefont {Chen}\ \emph {et~al.}(2019)\citenamefont {Chen},
  \citenamefont {Xue}, \citenamefont {Chen}, \citenamefont {Liu}, \citenamefont
  {Zang},\ and\ \citenamefont {Lu}}]{2019CHE}%
  \BibitemOpen
  \bibfield  {author} {\bibinfo {author} {\bibfnamefont {J.}~\bibnamefont
  {Chen}}, \bibinfo {author} {\bibfnamefont {X.}~\bibnamefont {Xue}}, \bibinfo
  {author} {\bibfnamefont {H.}~\bibnamefont {Chen}}, \bibinfo {author}
  {\bibfnamefont {H.}~\bibnamefont {Liu}}, \bibinfo {author} {\bibfnamefont
  {Q.}~\bibnamefont {Zang}}, \ and\ \bibinfo {author} {\bibfnamefont
  {W.}~\bibnamefont {Lu}},\ }\href@noop {} {\bibfield  {journal} {\bibinfo
  {journal} {The Journal of Physical Chemistry C}\ }\textbf {\bibinfo {volume}
  {123}},\ \bibinfo {pages} {28008} (\bibinfo {year} {2019})}\BibitemShut
  {NoStop}%
\bibitem [{\citenamefont {Amsler}(2019)}]{2019AMS}%
  \BibitemOpen
  \bibfield  {author} {\bibinfo {author} {\bibfnamefont {M.}~\bibnamefont
  {Amsler}},\ }\href@noop {} {\bibfield  {journal} {\bibinfo  {journal}
  {Physical Review B}\ }\textbf {\bibinfo {volume} {99}},\ \bibinfo {pages}
  {060102} (\bibinfo {year} {2019})}\BibitemShut {NoStop}%
\bibitem [{\citenamefont {Du}\ \emph {et~al.}(2019)\citenamefont {Du},
  \citenamefont {Zhang}, \citenamefont {Lin}, \citenamefont {Zhang},
  \citenamefont {Bergara},\ and\ \citenamefont {Yang}}]{2019DU}%
  \BibitemOpen
  \bibfield  {author} {\bibinfo {author} {\bibfnamefont {X.}~\bibnamefont
  {Du}}, \bibinfo {author} {\bibfnamefont {S.}~\bibnamefont {Zhang}}, \bibinfo
  {author} {\bibfnamefont {J.}~\bibnamefont {Lin}}, \bibinfo {author}
  {\bibfnamefont {X.}~\bibnamefont {Zhang}}, \bibinfo {author} {\bibfnamefont
  {A.}~\bibnamefont {Bergara}}, \ and\ \bibinfo {author} {\bibfnamefont
  {G.}~\bibnamefont {Yang}},\ }\href@noop {} {\bibfield  {journal} {\bibinfo
  {journal} {Physical Review B}\ }\textbf {\bibinfo {volume} {100}},\ \bibinfo
  {pages} {134110} (\bibinfo {year} {2019})}\BibitemShut {NoStop}%
\bibitem [{\citenamefont {Li}\ \emph {et~al.}(2020)\citenamefont {Li},
  \citenamefont {Xie}, \citenamefont {Sun}, \citenamefont {Huang},
  \citenamefont {Liu}, \citenamefont {Chen},\ and\ \citenamefont
  {Ma}}]{2020LI}%
  \BibitemOpen
  \bibfield  {author} {\bibinfo {author} {\bibfnamefont {X.}~\bibnamefont
  {Li}}, \bibinfo {author} {\bibfnamefont {Y.}~\bibnamefont {Xie}}, \bibinfo
  {author} {\bibfnamefont {Y.}~\bibnamefont {Sun}}, \bibinfo {author}
  {\bibfnamefont {P.}~\bibnamefont {Huang}}, \bibinfo {author} {\bibfnamefont
  {H.}~\bibnamefont {Liu}}, \bibinfo {author} {\bibfnamefont {C.}~\bibnamefont
  {Chen}}, \ and\ \bibinfo {author} {\bibfnamefont {Y.}~\bibnamefont {Ma}},\
  }\href@noop {} {\bibfield  {journal} {\bibinfo  {journal} {The Journal of
  Physical Chemistry Letters}\ }\textbf {\bibinfo {volume} {11}},\ \bibinfo
  {pages} {935} (\bibinfo {year} {2020})}\BibitemShut {NoStop}%
\bibitem [{\citenamefont {Hao}\ \emph {et~al.}(2020{\natexlab{a}})\citenamefont
  {Hao}, \citenamefont {Yuan}, \citenamefont {Guo}, \citenamefont {Zhang},
  \citenamefont {Luo}, \citenamefont {Gao}, \citenamefont {Ling}, \citenamefont
  {Chen}, \citenamefont {Zhao},\ and\ \citenamefont {Yu}}]{2020HAO_a}%
  \BibitemOpen
  \bibfield  {author} {\bibinfo {author} {\bibfnamefont {L.}~\bibnamefont
  {Hao}}, \bibinfo {author} {\bibfnamefont {Z.}~\bibnamefont {Yuan}}, \bibinfo
  {author} {\bibfnamefont {X.}~\bibnamefont {Guo}}, \bibinfo {author}
  {\bibfnamefont {Y.}~\bibnamefont {Zhang}}, \bibinfo {author} {\bibfnamefont
  {K.}~\bibnamefont {Luo}}, \bibinfo {author} {\bibfnamefont {Y.}~\bibnamefont
  {Gao}}, \bibinfo {author} {\bibfnamefont {F.}~\bibnamefont {Ling}}, \bibinfo
  {author} {\bibfnamefont {X.}~\bibnamefont {Chen}}, \bibinfo {author}
  {\bibfnamefont {Z.}~\bibnamefont {Zhao}}, \ and\ \bibinfo {author}
  {\bibfnamefont {D.}~\bibnamefont {Yu}},\ }\href@noop {} {\bibfield  {journal}
  {\bibinfo  {journal} {Physics Letters A}\ ,\ \bibinfo {pages} {126525}}
  (\bibinfo {year} {2020}{\natexlab{a}})}\BibitemShut {NoStop}%
\bibitem [{\citenamefont {Hao}\ \emph {et~al.}(2020{\natexlab{b}})\citenamefont
  {Hao}, \citenamefont {Yuan}, \citenamefont {Guo}, \citenamefont {Zhang},
  \citenamefont {Luo}, \citenamefont {Gao}, \citenamefont {Ling}, \citenamefont
  {Chen}, \citenamefont {Zhao},\ and\ \citenamefont {Yu}}]{2020HAO_b}%
  \BibitemOpen
  \bibfield  {author} {\bibinfo {author} {\bibfnamefont {L.}~\bibnamefont
  {Hao}}, \bibinfo {author} {\bibfnamefont {Z.}~\bibnamefont {Yuan}}, \bibinfo
  {author} {\bibfnamefont {X.}~\bibnamefont {Guo}}, \bibinfo {author}
  {\bibfnamefont {Y.}~\bibnamefont {Zhang}}, \bibinfo {author} {\bibfnamefont
  {K.}~\bibnamefont {Luo}}, \bibinfo {author} {\bibfnamefont {Y.}~\bibnamefont
  {Gao}}, \bibinfo {author} {\bibfnamefont {F.}~\bibnamefont {Ling}}, \bibinfo
  {author} {\bibfnamefont {X.}~\bibnamefont {Chen}}, \bibinfo {author}
  {\bibfnamefont {Z.}~\bibnamefont {Zhao}}, \ and\ \bibinfo {author}
  {\bibfnamefont {D.}~\bibnamefont {Yu}},\ }\href@noop {} {\bibfield  {journal}
  {\bibinfo  {journal} {Physics Letters A}\ ,\ \bibinfo {pages} {126525}}
  (\bibinfo {year} {2020}{\natexlab{b}})}\BibitemShut {NoStop}%
\bibitem [{\citenamefont {Yamada}\ \emph {et~al.}(2019)\citenamefont {Yamada},
  \citenamefont {Liu}, \citenamefont {Wu}, \citenamefont {Koyama},
  \citenamefont {Ju}, \citenamefont {Shiomi}, \citenamefont {Morikawa},\ and\
  \citenamefont {Yoshida}}]{2019YAM}%
  \BibitemOpen
  \bibfield  {author} {\bibinfo {author} {\bibfnamefont {H.}~\bibnamefont
  {Yamada}}, \bibinfo {author} {\bibfnamefont {C.}~\bibnamefont {Liu}},
  \bibinfo {author} {\bibfnamefont {S.}~\bibnamefont {Wu}}, \bibinfo {author}
  {\bibfnamefont {Y.}~\bibnamefont {Koyama}}, \bibinfo {author} {\bibfnamefont
  {S.}~\bibnamefont {Ju}}, \bibinfo {author} {\bibfnamefont {J.}~\bibnamefont
  {Shiomi}}, \bibinfo {author} {\bibfnamefont {J.}~\bibnamefont {Morikawa}}, \
  and\ \bibinfo {author} {\bibfnamefont {R.}~\bibnamefont {Yoshida}},\
  }\href@noop {} {\bibfield  {journal} {\bibinfo  {journal} {ACS central
  science}\ }\textbf {\bibinfo {volume} {5}},\ \bibinfo {pages} {1717}
  (\bibinfo {year} {2019})}\BibitemShut {NoStop}%
\bibitem [{\citenamefont {Kresse}\ and\ \citenamefont
  {Hafner}(1993)}]{1993KRE}%
  \BibitemOpen
  \bibfield  {author} {\bibinfo {author} {\bibfnamefont {G.}~\bibnamefont
  {Kresse}}\ and\ \bibinfo {author} {\bibfnamefont {J.}~\bibnamefont
  {Hafner}},\ }\href@noop {} {\bibfield  {journal} {\bibinfo  {journal}
  {Physical Review B}\ }\textbf {\bibinfo {volume} {47}},\ \bibinfo {pages}
  {558} (\bibinfo {year} {1993})}\BibitemShut {NoStop}%
\bibitem [{\citenamefont {Kresse}\ and\ \citenamefont
  {Hafner}(1994)}]{1994KRE}%
  \BibitemOpen
  \bibfield  {author} {\bibinfo {author} {\bibfnamefont {G.}~\bibnamefont
  {Kresse}}\ and\ \bibinfo {author} {\bibfnamefont {J.}~\bibnamefont
  {Hafner}},\ }\href@noop {} {\bibfield  {journal} {\bibinfo  {journal}
  {Physical Review B}\ }\textbf {\bibinfo {volume} {49}},\ \bibinfo {pages}
  {14251} (\bibinfo {year} {1994})}\BibitemShut {NoStop}%
\bibitem [{\citenamefont {Kresse}\ and\ \citenamefont
  {Furthm{\"u}ller}(1996{\natexlab{a}})}]{1996KRE_a}%
  \BibitemOpen
  \bibfield  {author} {\bibinfo {author} {\bibfnamefont {G.}~\bibnamefont
  {Kresse}}\ and\ \bibinfo {author} {\bibfnamefont {J.}~\bibnamefont
  {Furthm{\"u}ller}},\ }\href@noop {} {\bibfield  {journal} {\bibinfo
  {journal} {Computational materials science}\ }\textbf {\bibinfo {volume}
  {6}},\ \bibinfo {pages} {15} (\bibinfo {year}
  {1996}{\natexlab{a}})}\BibitemShut {NoStop}%
\bibitem [{\citenamefont {Kresse}\ and\ \citenamefont
  {Furthm{\"u}ller}(1996{\natexlab{b}})}]{1996KRE_b}%
  \BibitemOpen
  \bibfield  {author} {\bibinfo {author} {\bibfnamefont {G.}~\bibnamefont
  {Kresse}}\ and\ \bibinfo {author} {\bibfnamefont {J.}~\bibnamefont
  {Furthm{\"u}ller}},\ }\href@noop {} {\bibfield  {journal} {\bibinfo
  {journal} {Physical review B}\ }\textbf {\bibinfo {volume} {54}},\ \bibinfo
  {pages} {11169} (\bibinfo {year} {1996}{\natexlab{b}})}\BibitemShut {NoStop}%
\bibitem [{\citenamefont {Chen}\ \emph
  {et~al.}(2015{\natexlab{b}})\citenamefont {Chen}, \citenamefont {He},
  \citenamefont {Benesty}, \citenamefont {Khotilovich}, \citenamefont {Tang},
  \citenamefont {Cho} \emph {et~al.}}]{2015CHE}%
  \BibitemOpen
  \bibfield  {author} {\bibinfo {author} {\bibfnamefont {T.}~\bibnamefont
  {Chen}}, \bibinfo {author} {\bibfnamefont {T.}~\bibnamefont {He}}, \bibinfo
  {author} {\bibfnamefont {M.}~\bibnamefont {Benesty}}, \bibinfo {author}
  {\bibfnamefont {V.}~\bibnamefont {Khotilovich}}, \bibinfo {author}
  {\bibfnamefont {Y.}~\bibnamefont {Tang}}, \bibinfo {author} {\bibfnamefont
  {H.}~\bibnamefont {Cho}},  \emph {et~al.},\ }\href@noop {} {\bibfield
  {journal} {\bibinfo  {journal} {R package version 0.4-2}\ }\textbf {\bibinfo
  {volume} {1}} (\bibinfo {year} {2015}{\natexlab{b}})}\BibitemShut {NoStop}%
\bibitem [{\citenamefont {Glass}\ \emph {et~al.}(2006)\citenamefont {Glass},
  \citenamefont {Oganov},\ and\ \citenamefont {Hansen}}]{2006GLA}%
  \BibitemOpen
  \bibfield  {author} {\bibinfo {author} {\bibfnamefont {C.~W.}\ \bibnamefont
  {Glass}}, \bibinfo {author} {\bibfnamefont {A.~R.}\ \bibnamefont {Oganov}}, \
  and\ \bibinfo {author} {\bibfnamefont {N.}~\bibnamefont {Hansen}},\
  }\href@noop {} {\bibfield  {journal} {\bibinfo  {journal} {Computer physics
  communications}\ }\textbf {\bibinfo {volume} {175}},\ \bibinfo {pages} {713}
  (\bibinfo {year} {2006})}\BibitemShut {NoStop}%
\bibitem [{\citenamefont {Flores-Livas}\ \emph {et~al.}(2020)\citenamefont
  {Flores-Livas}, \citenamefont {Boeri}, \citenamefont {Sanna}, \citenamefont
  {Profeta}, \citenamefont {Arita},\ and\ \citenamefont {Eremets}}]{2020FLO}%
  \BibitemOpen
  \bibfield  {author} {\bibinfo {author} {\bibfnamefont {J.~A.}\ \bibnamefont
  {Flores-Livas}}, \bibinfo {author} {\bibfnamefont {L.}~\bibnamefont {Boeri}},
  \bibinfo {author} {\bibfnamefont {A.}~\bibnamefont {Sanna}}, \bibinfo
  {author} {\bibfnamefont {G.}~\bibnamefont {Profeta}}, \bibinfo {author}
  {\bibfnamefont {R.}~\bibnamefont {Arita}}, \ and\ \bibinfo {author}
  {\bibfnamefont {M.}~\bibnamefont {Eremets}},\ }\href@noop {} {\bibfield
  {journal} {\bibinfo  {journal} {Physics Reports}\ }\textbf {\bibinfo {volume}
  {856}},\ \bibinfo {pages} {1 } (\bibinfo {year} {2020})},\ \bibinfo {note} {a
  perspective on conventional high-temperature superconductors at high
  pressure: Methods and materials}\BibitemShut {NoStop}%
\bibitem [{\citenamefont {Semenok}\ \emph
  {et~al.}(2020{\natexlab{b}})\citenamefont {Semenok}, \citenamefont
  {Kvashnin}, \citenamefont {Ivanova}, \citenamefont {Svitlyk}, \citenamefont
  {Fominski}, \citenamefont {Sadakov}, \citenamefont {Sobolevskiy},
  \citenamefont {Pudalov}, \citenamefont {Troyan},\ and\ \citenamefont
  {Oganov}}]{2020SEM_a}%
  \BibitemOpen
  \bibfield  {author} {\bibinfo {author} {\bibfnamefont {D.~V.}\ \bibnamefont
  {Semenok}}, \bibinfo {author} {\bibfnamefont {A.~G.}\ \bibnamefont
  {Kvashnin}}, \bibinfo {author} {\bibfnamefont {A.~G.}\ \bibnamefont
  {Ivanova}}, \bibinfo {author} {\bibfnamefont {V.}~\bibnamefont {Svitlyk}},
  \bibinfo {author} {\bibfnamefont {V.~Y.}\ \bibnamefont {Fominski}}, \bibinfo
  {author} {\bibfnamefont {A.~V.}\ \bibnamefont {Sadakov}}, \bibinfo {author}
  {\bibfnamefont {O.~A.}\ \bibnamefont {Sobolevskiy}}, \bibinfo {author}
  {\bibfnamefont {V.~M.}\ \bibnamefont {Pudalov}}, \bibinfo {author}
  {\bibfnamefont {I.~A.}\ \bibnamefont {Troyan}}, \ and\ \bibinfo {author}
  {\bibfnamefont {A.~R.}\ \bibnamefont {Oganov}},\ }\href@noop {} {\bibfield
  {journal} {\bibinfo  {journal} {Materials Today}\ }\textbf {\bibinfo {volume}
  {33}},\ \bibinfo {pages} {36} (\bibinfo {year}
  {2020}{\natexlab{b}})}\BibitemShut {NoStop}%
\bibitem [{\citenamefont {Boeri}\ and\ \citenamefont
  {Bachelet}(2019)}]{2019BOE}%
  \BibitemOpen
  \bibfield  {author} {\bibinfo {author} {\bibfnamefont {L.}~\bibnamefont
  {Boeri}}\ and\ \bibinfo {author} {\bibfnamefont {G.~B.}\ \bibnamefont
  {Bachelet}},\ }\href@noop {} {\bibfield  {journal} {\bibinfo  {journal}
  {Journal of Physics: Condensed Matter}\ }\textbf {\bibinfo {volume} {31}},\
  \bibinfo {pages} {234002} (\bibinfo {year} {2019})}\BibitemShut {NoStop}%
\bibitem [{\citenamefont {Flores-Livas}\ and\ \citenamefont
  {Arita}(2019)}]{2019FLO}%
  \BibitemOpen
  \bibfield  {author} {\bibinfo {author} {\bibfnamefont {J.~A.}\ \bibnamefont
  {Flores-Livas}}\ and\ \bibinfo {author} {\bibfnamefont {R.}~\bibnamefont
  {Arita}},\ }\href@noop {} {\bibfield  {journal} {\bibinfo  {journal}
  {Physics}\ }\textbf {\bibinfo {volume} {12}},\ \bibinfo {pages} {96}
  (\bibinfo {year} {2019})}\BibitemShut {NoStop}%
\bibitem [{\citenamefont {Wang}\ \emph {et~al.}(2018)\citenamefont {Wang},
  \citenamefont {Li}, \citenamefont {Gao}, \citenamefont {Li},\ and\
  \citenamefont {Ma}}]{2018WAN}%
  \BibitemOpen
  \bibfield  {author} {\bibinfo {author} {\bibfnamefont {H.}~\bibnamefont
  {Wang}}, \bibinfo {author} {\bibfnamefont {X.}~\bibnamefont {Li}}, \bibinfo
  {author} {\bibfnamefont {G.}~\bibnamefont {Gao}}, \bibinfo {author}
  {\bibfnamefont {Y.}~\bibnamefont {Li}}, \ and\ \bibinfo {author}
  {\bibfnamefont {Y.}~\bibnamefont {Ma}},\ }\href@noop {} {\bibfield  {journal}
  {\bibinfo  {journal} {Wiley Interdisciplinary Reviews: Computational
  Molecular Science}\ }\textbf {\bibinfo {volume} {8}},\ \bibinfo {pages}
  {e1330} (\bibinfo {year} {2018})}\BibitemShut {NoStop}%
\bibitem [{\citenamefont {Li}\ \emph {et~al.}(2019{\natexlab{c}})\citenamefont
  {Li}, \citenamefont {Miao}, \citenamefont {Ti}, \citenamefont {Liu},
  \citenamefont {Chen}, \citenamefont {Shi},\ and\ \citenamefont
  {Gregoryanz}}]{2019LI_a}%
  \BibitemOpen
  \bibfield  {author} {\bibinfo {author} {\bibfnamefont {B.}~\bibnamefont
  {Li}}, \bibinfo {author} {\bibfnamefont {Z.}~\bibnamefont {Miao}}, \bibinfo
  {author} {\bibfnamefont {L.}~\bibnamefont {Ti}}, \bibinfo {author}
  {\bibfnamefont {S.}~\bibnamefont {Liu}}, \bibinfo {author} {\bibfnamefont
  {J.}~\bibnamefont {Chen}}, \bibinfo {author} {\bibfnamefont {Z.}~\bibnamefont
  {Shi}}, \ and\ \bibinfo {author} {\bibfnamefont {E.}~\bibnamefont
  {Gregoryanz}},\ }\href@noop {} {\bibfield  {journal} {\bibinfo  {journal}
  {Journal of Applied Physics}\ }\textbf {\bibinfo {volume} {126}},\ \bibinfo
  {pages} {235901} (\bibinfo {year} {2019}{\natexlab{c}})}\BibitemShut
  {NoStop}%
\bibitem [{\citenamefont {Bernstein}\ \emph {et~al.}(2015)\citenamefont
  {Bernstein}, \citenamefont {Hellberg}, \citenamefont {Johannes},
  \citenamefont {Mazin},\ and\ \citenamefont {Mehl}}]{2015BER}%
  \BibitemOpen
  \bibfield  {author} {\bibinfo {author} {\bibfnamefont {N.}~\bibnamefont
  {Bernstein}}, \bibinfo {author} {\bibfnamefont {C.~S.}\ \bibnamefont
  {Hellberg}}, \bibinfo {author} {\bibfnamefont {M.}~\bibnamefont {Johannes}},
  \bibinfo {author} {\bibfnamefont {I.}~\bibnamefont {Mazin}}, \ and\ \bibinfo
  {author} {\bibfnamefont {M.}~\bibnamefont {Mehl}},\ }\href@noop {} {\bibfield
   {journal} {\bibinfo  {journal} {Physical Review B}\ }\textbf {\bibinfo
  {volume} {91}},\ \bibinfo {pages} {060511} (\bibinfo {year}
  {2015})}\BibitemShut {NoStop}%
\bibitem [{\citenamefont {Kruglov}\ \emph {et~al.}(2018)\citenamefont
  {Kruglov}, \citenamefont {Kvashnin}, \citenamefont {Goncharov}, \citenamefont
  {Oganov}, \citenamefont {Lobanov}, \citenamefont {Holtgrewe}, \citenamefont
  {Jiang}, \citenamefont {Prakapenka}, \citenamefont {Greenberg},\ and\
  \citenamefont {Yanilkin}}]{2017KRU}%
  \BibitemOpen
  \bibfield  {author} {\bibinfo {author} {\bibfnamefont {I.~A.}\ \bibnamefont
  {Kruglov}}, \bibinfo {author} {\bibfnamefont {A.~G.}\ \bibnamefont
  {Kvashnin}}, \bibinfo {author} {\bibfnamefont {A.~F.}\ \bibnamefont
  {Goncharov}}, \bibinfo {author} {\bibfnamefont {A.~R.}\ \bibnamefont
  {Oganov}}, \bibinfo {author} {\bibfnamefont {S.~S.}\ \bibnamefont {Lobanov}},
  \bibinfo {author} {\bibfnamefont {N.}~\bibnamefont {Holtgrewe}}, \bibinfo
  {author} {\bibfnamefont {S.}~\bibnamefont {Jiang}}, \bibinfo {author}
  {\bibfnamefont {V.~B.}\ \bibnamefont {Prakapenka}}, \bibinfo {author}
  {\bibfnamefont {E.}~\bibnamefont {Greenberg}}, \ and\ \bibinfo {author}
  {\bibfnamefont {A.~V.}\ \bibnamefont {Yanilkin}},\ }\href@noop {} {\bibfield
  {journal} {\bibinfo  {journal} {Science advances}\ }\textbf {\bibinfo
  {volume} {4}},\ \bibinfo {pages} {eaat9776} (\bibinfo {year}
  {2018})}\BibitemShut {NoStop}%
\bibitem [{\citenamefont {Matasov}\ and\ \citenamefont
  {Krasavina}(2020)}]{2020MAT}%
  \BibitemOpen
  \bibfield  {author} {\bibinfo {author} {\bibfnamefont {A.}~\bibnamefont
  {Matasov}}\ and\ \bibinfo {author} {\bibfnamefont {V.}~\bibnamefont
  {Krasavina}},\ }\href@noop {} {\bibfield  {journal} {\bibinfo  {journal} {SN
  Applied Sciences}\ }\textbf {\bibinfo {volume} {2}},\ \bibinfo {pages} {1}
  (\bibinfo {year} {2020})}\BibitemShut {NoStop}%
\bibitem [{\citenamefont {Pedregosa}\ \emph {et~al.}(2011)\citenamefont
  {Pedregosa}, \citenamefont {Varoquaux}, \citenamefont {Gramfort},
  \citenamefont {Michel}, \citenamefont {Thirion}, \citenamefont {Grisel},
  \citenamefont {Blondel}, \citenamefont {Prettenhofer}, \citenamefont {Weiss},
  \citenamefont {Dubourg} \emph {et~al.}}]{2011PED}%
  \BibitemOpen
  \bibfield  {author} {\bibinfo {author} {\bibfnamefont {F.}~\bibnamefont
  {Pedregosa}}, \bibinfo {author} {\bibfnamefont {G.}~\bibnamefont
  {Varoquaux}}, \bibinfo {author} {\bibfnamefont {A.}~\bibnamefont {Gramfort}},
  \bibinfo {author} {\bibfnamefont {V.}~\bibnamefont {Michel}}, \bibinfo
  {author} {\bibfnamefont {B.}~\bibnamefont {Thirion}}, \bibinfo {author}
  {\bibfnamefont {O.}~\bibnamefont {Grisel}}, \bibinfo {author} {\bibfnamefont
  {M.}~\bibnamefont {Blondel}}, \bibinfo {author} {\bibfnamefont
  {P.}~\bibnamefont {Prettenhofer}}, \bibinfo {author} {\bibfnamefont
  {R.}~\bibnamefont {Weiss}}, \bibinfo {author} {\bibfnamefont
  {V.}~\bibnamefont {Dubourg}},  \emph {et~al.},\ }\href@noop {} {\bibfield
  {journal} {\bibinfo  {journal} {the Journal of machine Learning research}\
  }\textbf {\bibinfo {volume} {12}},\ \bibinfo {pages} {2825} (\bibinfo {year}
  {2011})}\BibitemShut {NoStop}%
\bibitem [{\citenamefont {Hastie}\ \emph {et~al.}(2001)\citenamefont {Hastie},
  \citenamefont {Tibshirani},\ and\ \citenamefont {Friedman}}]{2001HAS}%
  \BibitemOpen
  \bibfield  {author} {\bibinfo {author} {\bibfnamefont {T.}~\bibnamefont
  {Hastie}}, \bibinfo {author} {\bibfnamefont {R.}~\bibnamefont {Tibshirani}},
  \ and\ \bibinfo {author} {\bibfnamefont {J.}~\bibnamefont {Friedman}},\
  }\href@noop {} {\emph {\bibinfo {title} {The Elements of Statistical
  Learning}}},\ Springer Series in Statistics\ (\bibinfo  {publisher} {Springer
  New York Inc.},\ \bibinfo {address} {New York, NY, USA},\ \bibinfo {year}
  {2001})\BibitemShut {NoStop}%
\bibitem [{\citenamefont {Bishop}(2006)}]{2006BIS}%
  \BibitemOpen
  \bibfield  {author} {\bibinfo {author} {\bibfnamefont {C.~M.}\ \bibnamefont
  {Bishop}},\ }\href@noop {} {\emph {\bibinfo {title} {Pattern Recognition and
  Machine Learning}}}\ (\bibinfo  {publisher} {Springer},\ \bibinfo {year}
  {2006})\BibitemShut {NoStop}%
\bibitem [{\citenamefont {Tibshirani}(1996)}]{1996TIB}%
  \BibitemOpen
  \bibfield  {author} {\bibinfo {author} {\bibfnamefont {R.}~\bibnamefont
  {Tibshirani}},\ }\href {http://www.jstor.org/stable/2346178} {\bibfield
  {journal} {\bibinfo  {journal} {Journal of the Royal Statistical Society.
  Series B (Methodological)}\ }\textbf {\bibinfo {volume} {58}},\ \bibinfo
  {pages} {267} (\bibinfo {year} {1996})}\BibitemShut {NoStop}%
\bibitem [{\citenamefont {Yoshida}\ \emph {et~al.}(2019)\citenamefont
  {Yoshida}, \citenamefont {Hongo},\ and\ \citenamefont {Maezono}}]{2019YOS}%
  \BibitemOpen
  \bibfield  {author} {\bibinfo {author} {\bibfnamefont {T.}~\bibnamefont
  {Yoshida}}, \bibinfo {author} {\bibfnamefont {K.}~\bibnamefont {Hongo}}, \
  and\ \bibinfo {author} {\bibfnamefont {R.}~\bibnamefont {Maezono}},\
  }\href@noop {} {\bibfield  {journal} {\bibinfo  {journal} {The Journal of
  Physical Chemistry C}\ }\textbf {\bibinfo {volume} {123}},\ \bibinfo {pages}
  {14126} (\bibinfo {year} {2019})}\BibitemShut {NoStop}%
\bibitem [{\citenamefont {Yoshida}\ \emph {et~al.}(2020)\citenamefont
  {Yoshida}, \citenamefont {Maezono},\ and\ \citenamefont {Hongo}}]{2020YOS}%
  \BibitemOpen
  \bibfield  {author} {\bibinfo {author} {\bibfnamefont {T.}~\bibnamefont
  {Yoshida}}, \bibinfo {author} {\bibfnamefont {R.}~\bibnamefont {Maezono}}, \
  and\ \bibinfo {author} {\bibfnamefont {K.}~\bibnamefont {Hongo}},\
  }\href@noop {} {\bibfield  {journal} {\bibinfo  {journal} {ACS omega}\
  }\textbf {\bibinfo {volume} {5}},\ \bibinfo {pages} {13403} (\bibinfo {year}
  {2020})}\BibitemShut {NoStop}%
\bibitem [{\citenamefont {Zou}\ and\ \citenamefont {Hastie}(2005)}]{2005ZOU}%
  \BibitemOpen
  \bibfield  {author} {\bibinfo {author} {\bibfnamefont {H.}~\bibnamefont
  {Zou}}\ and\ \bibinfo {author} {\bibfnamefont {T.}~\bibnamefont {Hastie}},\
  }\href {http://www.jstor.org/stable/3647580} {\bibfield  {journal} {\bibinfo
  {journal} {Journal of the Royal Statistical Society. Series B (Statistical
  Methodology)}\ }\textbf {\bibinfo {volume} {67}},\ \bibinfo {pages} {301}
  (\bibinfo {year} {2005})}\BibitemShut {NoStop}%
\bibitem [{\citenamefont {de~Castro}\ \emph {et~al.}(2020)\citenamefont
  {de~Castro}, \citenamefont {Terashima}, \citenamefont {Yamamoto},
  \citenamefont {Hou}, \citenamefont {Iwasaki}, \citenamefont {Matsumoto},
  \citenamefont {Adachi}, \citenamefont {Saito}, \citenamefont {Song},
  \citenamefont {Takeya} \emph {et~al.}}]{2020CAS}%
  \BibitemOpen
  \bibfield  {author} {\bibinfo {author} {\bibfnamefont {P.~B.}\ \bibnamefont
  {de~Castro}}, \bibinfo {author} {\bibfnamefont {K.}~\bibnamefont
  {Terashima}}, \bibinfo {author} {\bibfnamefont {T.~D.}\ \bibnamefont
  {Yamamoto}}, \bibinfo {author} {\bibfnamefont {Z.}~\bibnamefont {Hou}},
  \bibinfo {author} {\bibfnamefont {S.}~\bibnamefont {Iwasaki}}, \bibinfo
  {author} {\bibfnamefont {R.}~\bibnamefont {Matsumoto}}, \bibinfo {author}
  {\bibfnamefont {S.}~\bibnamefont {Adachi}}, \bibinfo {author} {\bibfnamefont
  {Y.}~\bibnamefont {Saito}}, \bibinfo {author} {\bibfnamefont
  {P.}~\bibnamefont {Song}}, \bibinfo {author} {\bibfnamefont {H.}~\bibnamefont
  {Takeya}},  \emph {et~al.},\ }\href@noop {} {\bibfield  {journal} {\bibinfo
  {journal} {NPG Asia Materials}\ }\textbf {\bibinfo {volume} {12}},\ \bibinfo
  {pages} {1} (\bibinfo {year} {2020})}\BibitemShut {NoStop}%
\bibitem [{\citenamefont {Yoshida}\ \emph {et~al.}(2021)\citenamefont
  {Yoshida}, \citenamefont {Maezono},\ and\ \citenamefont {Hongo}}]{2021YOS}%
  \BibitemOpen
  \bibfield  {author} {\bibinfo {author} {\bibfnamefont {T.}~\bibnamefont
  {Yoshida}}, \bibinfo {author} {\bibfnamefont {R.}~\bibnamefont {Maezono}}, \
  and\ \bibinfo {author} {\bibfnamefont {K.}~\bibnamefont {Hongo}},\ }\href
  {\doibase 10.1021/acsanm.0c03298} {\bibfield  {journal} {\bibinfo  {journal}
  {ACS Applied Nano Materials}\ }\textbf {\bibinfo {volume} {4}},\ \bibinfo
  {pages} {1932} (\bibinfo {year} {2021})},\ \Eprint
  {http://arxiv.org/abs/https://doi.org/10.1021/acsanm.0c03298}
  {https://doi.org/10.1021/acsanm.0c03298} \BibitemShut {NoStop}%
\bibitem [{\citenamefont {Bergstra}\ \emph {et~al.}(2015)\citenamefont
  {Bergstra}, \citenamefont {Komer}, \citenamefont {Eliasmith}, \citenamefont
  {Yamins},\ and\ \citenamefont {Cox}}]{2015BER_b}%
  \BibitemOpen
  \bibfield  {author} {\bibinfo {author} {\bibfnamefont {J.}~\bibnamefont
  {Bergstra}}, \bibinfo {author} {\bibfnamefont {B.}~\bibnamefont {Komer}},
  \bibinfo {author} {\bibfnamefont {C.}~\bibnamefont {Eliasmith}}, \bibinfo
  {author} {\bibfnamefont {D.}~\bibnamefont {Yamins}}, \ and\ \bibinfo {author}
  {\bibfnamefont {D.~D.}\ \bibnamefont {Cox}},\ }\href@noop {} {\bibfield
  {journal} {\bibinfo  {journal} {Computational Science \& Discovery}\ }\textbf
  {\bibinfo {volume} {8}},\ \bibinfo {pages} {014008} (\bibinfo {year}
  {2015})}\BibitemShut {NoStop}%
\bibitem [{\citenamefont {Perdew}\ \emph {et~al.}(1996)\citenamefont {Perdew},
  \citenamefont {Burke},\ and\ \citenamefont {Ernzerhof}}]{1996PER}%
  \BibitemOpen
  \bibfield  {author} {\bibinfo {author} {\bibfnamefont {J.~P.}\ \bibnamefont
  {Perdew}}, \bibinfo {author} {\bibfnamefont {K.}~\bibnamefont {Burke}}, \
  and\ \bibinfo {author} {\bibfnamefont {M.}~\bibnamefont {Ernzerhof}},\
  }\href@noop {} {\bibfield  {journal} {\bibinfo  {journal} {Physical review
  letters}\ }\textbf {\bibinfo {volume} {77}},\ \bibinfo {pages} {3865}
  (\bibinfo {year} {1996})}\BibitemShut {NoStop}%
\bibitem [{\citenamefont {Togo}\ and\ \citenamefont {Tanaka}(2015)}]{2015TOG}%
  \BibitemOpen
  \bibfield  {author} {\bibinfo {author} {\bibfnamefont {A.}~\bibnamefont
  {Togo}}\ and\ \bibinfo {author} {\bibfnamefont {I.}~\bibnamefont {Tanaka}},\
  }\href@noop {} {\bibfield  {journal} {\bibinfo  {journal} {Scr. Mater.}\
  }\textbf {\bibinfo {volume} {108}},\ \bibinfo {pages} {1} (\bibinfo {year}
  {2015})}\BibitemShut {NoStop}%
\bibitem [{\citenamefont {Nakano}\ \emph {et~al.}(2017)\citenamefont {Nakano},
  \citenamefont {Hongo},\ and\ \citenamefont {Maezono}}]{2017NAK}%
  \BibitemOpen
  \bibfield  {author} {\bibinfo {author} {\bibfnamefont {K.}~\bibnamefont
  {Nakano}}, \bibinfo {author} {\bibfnamefont {K.}~\bibnamefont {Hongo}}, \
  and\ \bibinfo {author} {\bibfnamefont {R.}~\bibnamefont {Maezono}},\
  }\href@noop {} {\bibfield  {journal} {\bibinfo  {journal} {Inorganic
  chemistry}\ }\textbf {\bibinfo {volume} {56}},\ \bibinfo {pages} {13732}
  (\bibinfo {year} {2017})}\BibitemShut {NoStop}%
\bibitem [{\citenamefont {Nakano}\ \emph {et~al.}(2016)\citenamefont {Nakano},
  \citenamefont {Hongo},\ and\ \citenamefont {Maezono}}]{2016NAK}%
  \BibitemOpen
  \bibfield  {author} {\bibinfo {author} {\bibfnamefont {K.}~\bibnamefont
  {Nakano}}, \bibinfo {author} {\bibfnamefont {K.}~\bibnamefont {Hongo}}, \
  and\ \bibinfo {author} {\bibfnamefont {R.}~\bibnamefont {Maezono}},\
  }\href@noop {} {\bibfield  {journal} {\bibinfo  {journal} {Scientific
  reports}\ }\textbf {\bibinfo {volume} {6}},\ \bibinfo {pages} {1} (\bibinfo
  {year} {2016})}\BibitemShut {NoStop}%
\bibitem [{\citenamefont {Marzari}\ \emph {et~al.}(1999)\citenamefont
  {Marzari}, \citenamefont {Vanderbilt}, \citenamefont {De~Vita},\ and\
  \citenamefont {Payne}}]{1999MAR}%
  \BibitemOpen
  \bibfield  {author} {\bibinfo {author} {\bibfnamefont {N.}~\bibnamefont
  {Marzari}}, \bibinfo {author} {\bibfnamefont {D.}~\bibnamefont {Vanderbilt}},
  \bibinfo {author} {\bibfnamefont {A.}~\bibnamefont {De~Vita}}, \ and\
  \bibinfo {author} {\bibfnamefont {M.}~\bibnamefont {Payne}},\ }\href@noop {}
  {\bibfield  {journal} {\bibinfo  {journal} {Physical review letters}\
  }\textbf {\bibinfo {volume} {82}},\ \bibinfo {pages} {3296} (\bibinfo {year}
  {1999})}\BibitemShut {NoStop}%
\bibitem [{\citenamefont {Baroni}\ \emph {et~al.}(2001)\citenamefont {Baroni},
  \citenamefont {De~Gironcoli}, \citenamefont {Dal~Corso},\ and\ \citenamefont
  {Giannozzi}}]{2001BAR}%
  \BibitemOpen
  \bibfield  {author} {\bibinfo {author} {\bibfnamefont {S.}~\bibnamefont
  {Baroni}}, \bibinfo {author} {\bibfnamefont {S.}~\bibnamefont
  {De~Gironcoli}}, \bibinfo {author} {\bibfnamefont {A.}~\bibnamefont
  {Dal~Corso}}, \ and\ \bibinfo {author} {\bibfnamefont {P.}~\bibnamefont
  {Giannozzi}},\ }\href@noop {} {\bibfield  {journal} {\bibinfo  {journal}
  {Reviews of modern Physics}\ }\textbf {\bibinfo {volume} {73}},\ \bibinfo
  {pages} {515} (\bibinfo {year} {2001})}\BibitemShut {NoStop}%
\bibitem [{\citenamefont {Giannozzi}\ \emph {et~al.}(2009)\citenamefont
  {Giannozzi}, \citenamefont {Baroni}, \citenamefont {Bonini}, \citenamefont
  {Calandra}, \citenamefont {Car}, \citenamefont {Cavazzoni}, \citenamefont
  {Ceresoli}, \citenamefont {Chiarotti}, \citenamefont {Cococcioni},
  \citenamefont {Dabo} \emph {et~al.}}]{2009GIA}%
  \BibitemOpen
  \bibfield  {author} {\bibinfo {author} {\bibfnamefont {P.}~\bibnamefont
  {Giannozzi}}, \bibinfo {author} {\bibfnamefont {S.}~\bibnamefont {Baroni}},
  \bibinfo {author} {\bibfnamefont {N.}~\bibnamefont {Bonini}}, \bibinfo
  {author} {\bibfnamefont {M.}~\bibnamefont {Calandra}}, \bibinfo {author}
  {\bibfnamefont {R.}~\bibnamefont {Car}}, \bibinfo {author} {\bibfnamefont
  {C.}~\bibnamefont {Cavazzoni}}, \bibinfo {author} {\bibfnamefont
  {D.}~\bibnamefont {Ceresoli}}, \bibinfo {author} {\bibfnamefont {G.~L.}\
  \bibnamefont {Chiarotti}}, \bibinfo {author} {\bibfnamefont {M.}~\bibnamefont
  {Cococcioni}}, \bibinfo {author} {\bibfnamefont {I.}~\bibnamefont {Dabo}},
  \emph {et~al.},\ }\href@noop {} {\bibfield  {journal} {\bibinfo  {journal}
  {J. Phys. Condens. Matter.}\ }\textbf {\bibinfo {volume} {21}},\ \bibinfo
  {pages} {395502} (\bibinfo {year} {2009})}\BibitemShut {NoStop}%
\end{thebibliography}%

\clearpage
\section{Supplemental Information}
\subsection{{\it Ab initio} calculations}
\label{computational}
The crystal structures predicted
for YKH$_{12}$ and LaKH$_{12}$ at each pressure
are given in Table.~\ref{table.crytal_struc}.
\begin{table*}
 \begin{center}
   \caption{
     Crystal structures of
     YKH$_{12}$ and LaKH$_{12}$
     predicted at each pressure~($P$).
     Lattice parameters ($a$, $b$ and $c$)
     are given in unit of $\AA$.
   }
     \label{table.crytal_struc}
\begin{tabular}{c|c|c|r|crrr}
\hline
& & & & \multicolumn{4}{l}{Atomic coordinates (fractional)}
\\
&&$P$~(GPa) && Atoms & $x$  & $y$  & $z$
  \\
\hline
YKH$_{12}$    & $C2/m$ & 240 & \begin{tabular}[c]{@{}l@{}}
  $a$ = 4.685
  \\
  $b$ = 4.959
  \\
  $c$ = 3.412
  \\
  $\alpha$= $\gamma$ = 90$^\circ$
  \\ $\beta$ = 94.640$^\circ$
  \end{tabular}
  & \begin{tabular}[c]{@{}r@{}}K(2$c$)
  \\
   Y(2$b$)
  \\
  H(8$j$)
  \\
  H(8$j$)
  \\
  H(8$j$)
  \end{tabular}
  & \begin{tabular}[c]{@{}r@{}}
  0.00000
  \\
  0.00000
  \\
  0.10546
  \\
  0.11840
  \\
  0.24652
  \end{tabular}
  & \begin{tabular}[c]{@{}r@{}}
  0.00000
  \\
   0.50000
  \\
  0.13491
  \\
  0.35886
  \\
  0.27559
  \end{tabular}
  & \begin{tabular}[c]{@{}r@{}}
  0.50000
  \\
  0.00000
  \\
  0.02245
  \\
  0.52932
  \\
  0.74033\end{tabular} \\
\hline
LaKH$_{12}$
& $R\bar{3}m$
& 140
& \begin{tabular}[c]{@{}r@{}}
$a$ = $b$ = 5.457
\\
$c$ = 5.844
\\
$\alpha$ = $\beta$ = 90$^\circ$
\\
$\gamma$ = 120$^\circ$
\end{tabular}
& \begin{tabular}[c]{@{}r@{}}
K(3$b$)
\\
La(3$a$)
\\
H(36$i$)
\end{tabular}
& \begin{tabular}[c]{@{}r@{}}
0.00000
\\
0.00000
\\
0.00522
\end{tabular}
& \begin{tabular}[c]{@{}r@{}}
0.00000
\\
0.00000
\\
0.29080\end{tabular}
& \begin{tabular}[c]{@{}r@{}}
0.50000
\\
0.00000
\\
0.25260\end{tabular}
\\
\hline
LaKH$_{12}$
& $C2/m$
& 250
& \begin{tabular}[c]{@{}r@{}}
$a$ = 8.126
\\
$b$ = 5.479
\\
$c$ = 4.379
\\
$\alpha$ = $\gamma$ = 90$^\circ$
\\
$\beta$ = 122.253$^\circ$
\end{tabular}
& \begin{tabular}[c]{@{}r@{}}
K(4$i$)
\\
 La(4$i$)
 \\
 H(8$j$)
 \\
 H(8$j$)
 \\
 H(8$j$)
 \\
 H(8$j$)
 \\
 H(8$j$)
 \\
 H(4$g$)
 \\
 H(4$h$)
 \end{tabular}
 & \begin{tabular}[c]{@{}r@{}}
 0.12534
 \\
 0.12649
 \\
 0.11445
 \\
 0.11651
 \\
 0.12613
 \\
0.13714
 \\
 0.24797
\\
 0.00000
 \\
0.00000
\end{tabular}
& \begin{tabular}[c]{@{}r@{}}
0.00000
\\
0.50000
 \\
 0.17656
 \\
 0.10279
 \\
 0.30763
 \\
0.36984
\\
0.21751
\\
0.21727
 \\
 0.26959
 \end{tabular}
& \begin{tabular}[c]{@{}r@{}}
 0.82783
 \\
 0.34168
 \\
 0.45546
 \\
 0.25125
 \\
 0.96974
 \\
 0.78838
 \\
 0.24331
 \\
 0.00000
 \\
 0.50000\end{tabular}  \\
\hline
\end{tabular}
 \end{center}
\end{table*}
For electronic structure calculations
(required to get pressure-dependent DOS)
and phonon calculations
(to evaluate dynamical stabilities),
we used VASP package
~\cite{1993KRE,1994KRE,1996KRE_a,1996KRE_b}
with GGA-PBE exchange-correlation functionals.
~\cite{1996PER}
Ionic cores are described by
ultrasoft pseudo potentials
provided in the package.
To assist the convergence in the
self-consistent field iterations,
we used Marzari-Vanderbilt smearing scheme.
~\cite{1999MAR}
Resolutions for the plane wave basis set
expansions [Energy cutoff ($E_{\rm cut}$)]
and the Brillouin-zone integration
[$k$-mesh] were determined
so that the resultant energy values
could converge within the required
accuracies, finally getting
$E_{\rm cut}$=75~Ry with
($8 \times 8 \times 8$) $k$-mesh
for the electronic Brillouin-zone.
Phonon calculations are
performed by
linear-response method
~\cite{2001BAR}
with ($4 \times 4 \times 4$) $q$-mesh.

\vspace{2mm}
To estimate $T_c$,
we used Allen-Dynes formula implemented
in Quantum Espresso package~\cite{2009GIA}
with the effective Coulomb interaction
$\mu^*$ being chosen 0.1 empirically.
Denser $k$-meshes, $16 \times 16 \times 16$,
were used for
the double-delta integrations in
electron-phonon calculations.
The estimated results are
summarized in Table~\ref{table.Pdep}.
\begin{table*}
 \begin{center}
   \caption{
     $T_c$ estimated by Allen-Dynes formula using
     {\it ab initio} phonon calculations
     for
     YKH$_{12}$ [$C2/m$] and KH$_{12}$ [$R\bar{3}m$]
     at each pressure.
     $\lambda$ and $\omega_{\rm log}$ are
     the parameters appearing in the formula.
   }
     \label{table.Pdep}
\begin{tabular}{rccrr}
\hline
 & $P$~[GPa] & $\lambda$ & $\omega_{\rm log}$~[K] & $T_c$~[K] \\
\hline
YKH$_{12}$& 180          & 1.398    &1289.465   &  137.1     \\
YKH$_{12}$ & 200         &1.400    & 1324.987      &  141.1     \\
YKH$_{12}$ & 240        &1.427     &1318.072        & 143.2      \\
YKH$_{12}$ & 260         &1.436     & 1127.344       & 123.2      \\
YKH$_{12}$ & 300          &1.587     & 911.696     &  109.4     \\
LaKH$_{12}$ & 140            &1.531     & 854.12     &  99.2     \\
LaKH$_{12}$ & 160          &1.665     & 789.38     & 98.8     \\
\hline
\end{tabular}
 \end{center}
\end{table*}

\subsection{Descriptors}
\label{descriptors}
Total 84 descriptors used for the regression are
listed in Table~\ref{table.descriptors},
corresponding to those
for (i) [space group], (ii) [pressure],
and (iii) [chemical composition].
A descriptor for a composition is
composed from that for each atomic species
in various way.
Denoting $\alpha$ as the index specifying
atomic species, and
$f_\alpha$ as the descriptor for $\alpha$,
\begin{eqnarray}
  f_{\rm ave} &=&
  \sum_\alpha W_{\alpha}^{*}\cdot f_{\alpha} \ ,
  \nonumber \\
  f_{\rm sum} &=&
  \sum_\alpha W_{\alpha}^{}\cdot f_{\alpha} \ ,
  \nonumber \\
  f_{\rm var}&=& \sum_\alpha W_{\alpha}^{*}\cdot (f_{\alpha} - f_{ave})^{2} \ ,
  \nonumber \\
  f_{\rm max} &=& \max_\alpha\left\{f_{\alpha}\right\}\ ,
  \nonumber \\
        f_{\rm min} &=& \min_\alpha \left\{f_\alpha\right\} \ ,
\label{weights}
\end{eqnarray}
are used, where $W_\alpha$ is
the number of $\alpha$-species
included in the composition,
and $W_\alpha^*$ is the normalized fraction.
\begin{table*}
 \begin{center}
   \caption{
     Total 84 descriptors used for the regression,
corresponding to those
for (i) [space group], (ii) [pressure],
and (iii) [chemical composition].
Each weighting scheme (ave/sum/var/max/min)
is defined in Eqs.(\ref{weights}).
     }
     \label{table.descriptors}
	\begin{tabular}{llll}
	  \hline
          Weighting scheme & Property & Label & Description
          \\
        \hline
        ave & (i) &
        atomic\_radius\_rahm
        & Atomic radius by Rahm {\it et al.}
        \\
 ave&(i) & boiling\_point
 & Boiling temperature
 \\
 ave&(i) & covalent\_radius\_cordero
 & Covalent radius by Cerdero {\it et al.}
 \\
 ave&(i) & covalent\_radius\_pyykko
 & Single bond covalent radius by Pyykko {\it et al.}
 \\
 ave/max&(i) & covalent\_radius\_slater
 & Covalent radius by Slater
 \\
 ave&(i) & en\_allen
  & Allen's scale of electronegativity
  \\
 ave/sum&(i) & en\_ghosh
 & Ghosh's scale of electronegativity
 \\
 ave/sum&(i) & first\_ion\_en
 & First ionisation energy
 \\
 ave&(i) & fusion\_enthalpy
 & Fusion heat
 \\
 ave/sum/var&(i) & gs\_bandgap
 & DFT bandgap energy of {\it T} = 0 K ground state
 \\
 ave/max&(i) & gs\_volume\_per
 & DFT volume per atom of {\it T} = 0 K ground state
 \\
 ave/sum/var/max/min&(i) & hhi\_p
 & Herfindahl-Hirschman Index (HHI) production values
 \\
 ave/max/min&(i) & hhi\_r
 & Herfindahl-Hirschman Index (HHI) reserves values
 \\
 ave/sum/var&(i) & heat\_capacity\_mass
 & Mass specific heat capacity
 \\
 ave&(i) & evaporation\_heat
 & Evaporation heat
 \\
 ave/var/max/min&(i) & lattice\_constant
 & Physical dimension of unit cells in a crystal lattice
 \\
 ave/min&(i) & mendeleev\_number
 & Atom number in mendeleev's periodic table
 \\
 ave/sum&(i) & melting\_point
 & Melting point
  \\
 ave/sum/max&(i) & molar\_volume
 & Molar volume
 \\
 ave/sum/max&(i) & num\_unfilled
 & Total unfilled electron
  \\
 ave/var/max&(i) & num\_valance
 & Total valance electron
  \\
 ave&(i) & period
 & Period in the periodic table
  \\
 ave&(i) & vdw\_radius
  & Van der Waals radius
   \\
 ave/max&(i) & vdw\_radius\_alvarez
 & Van der Waals radius according to Alvarez
  \\
 ave/max&(i) & vdw\_radius\_mm3
 & Van der Waals radius from the MM3 FF
  \\
 sum&(i) & atomic\_radius
 & Atomic radius
  \\
 sum&(i) & atomic\_volume
 & Atomic volume
  \\
 sum&(i) & c6\_gb
  & C\_6 dispersion coefficient in a.u
   \\
 sum/max&(i) & covalent\_radius\_pyykko\_triple
 & Triple bond covalent radius by Pyykko {\it et al.}
  \\
 sum/min&(i) & electron\_negativity
 & Tendency of an atom to attract a shared pair of electrons
 \\
 var/min&(i) & en\_pauling
 & Mulliken's scale of electronegativity
 \\
 sum/max&(i) & gs\_est\_bcc\_latcnt
 & Estimated BCC lattice parameter based on the DFT volume
 \\
 sum/var&(i) & num\_s\_unfilled
 & Unfilled electron in s shell
  \\
 sum&(i) & specific\_heat
 & Specific heat at 20oC
 \\
 var&(i) & icsd\_volume
 & Atom volume in ICSD database
 \\
 var/max&(i) & vdw\_radius\_uff
 & Van der Waals radius from the UFF
  \\
 max/min&(i) & bulk\_modulus
 & Bulk modulus
  \\
 max&(i) & num\_p\_unfilled
 & Unfilled electron in p shell
 \\
 ave/var/min&(ii) & s\_dos
 & Density of states of s electron at Fermi surface (state/eV/atom)
 \\
 ave/min&(ii) & Free\_energy
 & Pressure-related free energy
  \\
 var/max/sum&(ii) & p\_dos
 & Density of states of p electron at Fermi surface (state/eV/atom)
 \\
 max/sum&(ii) & element\_dos
 & Density of states at Fermi surface (state/eV/atom)
 \\
 &(iii) & spg number
 & space group number
 \\
	\hline
\end{tabular}
 \end{center}
\end{table*}

\subsection{List of estimated and training data}
Total 426 training data used to
construct the regression are listed
in Table~\ref{table.training}.
By using the XGBoost regression,
we estimated $T_c$ for a set of
chemical compositions as shown in
Table~\ref{table.estData}.
\begin{table*}
 \begin{center}
   \caption{
     28 chemical compositions     
     to estimate $T_c$ by using
     the XGBoost regression learned with
     the 426 training data.
     }
     \label{table.estData}
\begin{tabular}{r|rr||r|rr}
\hline
No. &     & $T_c$~[K]&No. &  & $T_c$~[K] \\
\hline
1  & YKH$_{12}$   & 168.93 &15 & LaSH$_{12}$  & 74.94  \\   
2  & LaKH$_{12}$  & 162.85 &16 & LaArH$_{12}$ & 74.91  \\   
3  & YClH$_{12}$  & 137.60 &17 & YBiH$_{12}$  & 72.54  \\   
4  & LaMgH$_{12}$ & 135.33 &18 & YFeH$_{12}$  & 64.58 \\    
5  & LaClH$_{12}$ & 132.70 &19 & LaPH$_{12}$  & 60.59\\     
6  & LaAcH$_{12}$ & 128.99 &20 & YTaH$_{12}$ & 58.93\\      
7  & YSeH$_{12}$  & 127.01 &21 & YTiH$_{12}$  & 57.21 \\    
8  & YSH$_{12}$   & 123.04 &22 & LaBiH$_{12}$ & 56.59\\     
9  & YArH$_{12}$  & 99.59  &23 & LaSbH$_{12}$ & 52.38\\     
10 & YGeH$_{12}$  & 98.60  &24 & LaSiH$_{12}$ & 50.53  \\   
11 & LaSe$_{12}$  & 98.25  &25 & LaTaH$_{12}$ & 42.25  \\   
12 & YInH$_{12}$  & 93.81  &26 & LaTiH$_{12}$ & 39.85\\     
13 & YPH$_{12}$   & 92.78  &27 & LaFeH$_{12}$ & 24.48 \\    
14 & LaTeH$_{12}$ & 77.93  &28 & LaGeH$_{12}$ & 18.47\\     
\hline
\end{tabular}
 \end{center}
\end{table*}

\clearpage
\begin{longtable*}{r|r|rl|lll}
   \caption{
Total 533 training data~\cite{2008CHE,
2009KIM,2009TSE,2010JIN,2010LI,2010GAO,2010DUA,2011GAO,
2011KIM,2011ABE,2012ZHO_a,2011ZHA,2012ZHO_b,2013GAO,2013ABE, 2013HOO,2013LON,
2013HU,2014XIE,2014YU,2014LI,
2014WAN,2014CHE,2014DUA,
2015YAN,2015HOU,2015DUA,2015ZHA_a,
2015LIU_a,2015SHA,2015LIU_b,2015CHE_a,2015FEN,
2015ZHA_b,2015LIU,2015YU,2015CHE_b,2015ZHA,
2015ERR,2016ISH,2016MAH,2016ZHA,2016FU,2016SHA,2016LIU,
2016LI_a,2016ZHO,2016LI_b,2017RAH,2017MA,
2017MA_b,2017KOK,2017LI_a,2017ZHU,2017LIU,2017ZEN,2017ISH,
2017LI_b,2017PEN,2017DAV,2017MAJ,2018ZAR,2018SEM,
2018LI,2018YE,2018KVA_a,2018DUR,2018ZHE,2018LIU,
2018ABE,2018NAK,2018ZHU,2018KVA_b,2018WU,2019LIA_a,2019SUN,2019HEI,2019GRI,
2019CHE,2019LI_b,2019SHA,2019XIE,2019LIA_b,2019YUA,2019WAN,2019AMS,
2019YAN,2019XI,2019ABE,2019DU,2020GUO,2020LV,2020XIE,2020ZHA_a,2020ZHA_b,2020LI,
2020HAO_a,2020CUI,2015MA,2020TRO,2020DIC,2019LI_c,
2018KAN,2020HAO_b}
used to construct our regressiors (see the main text).
The label 'Spg' abbreviates
'Space group' with the numbering.
$\mu$ is the parameter for the
effective Coulomb interactions
appearing in Allen-Dynes formula.
   }
     \label{table.training}
     \\
\hline
  No.   &          & Spg. &   & $P$~(GPa) & $T_c$~(K)   & $\mu$   \\
  \hline
  0   & SiH$_{4}$              & 64         & $Cmca$      & 60.0          & 75.0    & 0.1  \\
1   & SiH$_{4}$              & 64         & $Cmca$      & 150.0         & 20.0    & 0.1  \\
2   & SiH$_{4}$              & 64         & $Cmca$      & 200.0         & 30.0    & 0.1  \\
3   & SiH$_{4}$              & 64         & $Cmca$      & 250.0         & 50.0    & 0.1  \\
4   & YH$_{3}$               & 225        & $Fm\bar{3}m$     & 17.7          & 40.0    & 0.1  \\
5   & YH$_{3}$               & 225        & $Fm\bar{3}m$     & 28.0          & 9.0     & 0.1  \\
\end{longtable*}

\end{document}